%% file: aar.tex
\documentclass[smallextended]{springer/svjour3} 
\pdfoutput=1
\usepackage{nomencl}
\makenomenclature

\usepackage{natbib}
\bibpunct{(}{)}{;}{a}{}{,}

\usepackage{graphicx}
\graphicspath{{figs/}}
\DeclareGraphicsExtensions{.pdf,.png,.jpg}
\usepackage{sidecap} 
\usepackage{amsmath,amssymb,amsfonts,textcomp,aas_macros}
\usepackage{url}
\smartqed 

\journalname{Astronomy Astrophysics Review}

\def\lesssim{\mathrel{\hbox{\rlap{\hbox{\lower4pt\hbox{$\sim$}}}\hbox{$<$}}}}
\def\gtrsim{\mathrel{\hbox{\rlap{\hbox{\lower4pt\hbox{$\sim$}}}\hbox{$>$}}}}
\def\arcsec{\hbox{$^{\prime\prime}$}}
\def\arcmin{\hbox{$^{\prime}$}}
\def\deg{\hbox{${^\circ}$}}
\def\gr{$^{\circ}$}

\def\ts{\thinspace}
\def\hi{H{\ts{$\scriptstyle {\rm I}$}} }
\def\hei{He{\ts{$\scriptstyle {\rm I}$}} }
\def\mgx{Mg\ts{$\scriptstyle {\rm X}$} }
\def\mgi{Mg\ts{$\scriptstyle {\rm I}$} }
\def\mgii{Mg\ts{$\scriptstyle {\rm II}$} }
\def\oiv{O\ts{$\scriptstyle {\rm IV}$} }
\def\ov{O\ts{$\scriptstyle {\rm V}$} }
\def\feii{Fe\ts{$\scriptstyle {\rm II}$} }
\def\ci{C\ts{$\scriptstyle {\rm I}$}}
\def\cii{C\ts{$\scriptstyle {\rm II}$}}
\def\caii{Ca\ts{$\scriptstyle {\rm II}$} }
\def\ni{N\ts{$\scriptstyle {\rm I}$} }
\def\nii{N\ts{$\scriptstyle {\rm II}$}}
\def\siii{Si\ts{$\scriptstyle {\rm II}$} }
\def\siv{Si\ts{$\scriptstyle {\rm IV}$}}
\def\ne8{Ne\ts{$\scriptstyle {\rm VIII}$} }

\begin{document}

\input{aar_def}

\title{Spectroradiometry with Space Telescopes} 
\author{Anuschka~Pauluhn 
\and Martin~C.E.~Huber 
\and Peter L. Smith 
\and Luis Colina 
}

\institute{Anuschka Pauluhn 
\and 
Martin C.E. Huber 
\at Paul Scherrer Institut, 5232 Villigen PSI, 
Switzerland 
\email{anuschka.pauluhn@psi.ch} 
\and
Peter L. Smith \at  Harvard-Smithsonian Center for Astrophysics, 
Cambridge MA 02138, USA 
\and
Luis Colina \at 
Centro de Astrobiolog\'{i}a (CSIC/INTA) 
28850  Torrej\'{o}n de Ardoz, Madrid, Spain 
}

\date{accepted for publication in AARev 29 Sept 2015}
\maketitle
\markboth{Spectroradiometry with Space Telescopes  (Pauluhn et al.)}{}

\setcounter{tocdepth}{2}
\setcounter{secnumdepth}{2}
\tableofcontents    


\vspace{1cm}
\input{abstract0}

\input{intro1}

\input{laboratory2}

\input{celestial3} 
\input{examples4}

\input{concl5}

\input{abbrev_cal}

\begin{acknowledgements}

The authors sincerely thank K. Bennett, C. T. Bingham, 
R. C. Bohlin,  
D. J. Coletti, T. Dudok de Wit, C. Fr\"ohlich, 
L. D. Gardner, A. Gottwald, U. Grothkopf, 
J. B. Holberg, J. Hollandt, C. V. H. Huber-Ott, 
F. Jansen, 
C. Jones, E. M. Kellogg, R. M. Klein, J. L. Kohl, 
J. W. Kruk, M. K\"{u}hne, X. Liu, R. Paladini, 
W. H. Parkinson, G. Schmidtke, J. G. Timothy,  G. Ulm, 
 K. Wilhelm, C. Winkler and B. J. Wargelin for illuminating 
discussions, reference material, both published and unpublished, and other 
assistance. PLS thanks the Space Science Department of ESA for travel support. 
He was also supported in part by NASA Grants NAGW-1596, and SOHO, and SPARTAN grants.

\end{acknowledgements}

\printnomenclature

\bibliography{aarbib}   
\end{document}

%% file: aar_def
\def\3{\ss}                                                         
\def\cm{\vspace{1cm}}
\def\hcm{\vspace{0.5cm}}                                               

\def\qm{\phantom{-}} 
\def\q0{\phantom{1}}                                                        
\def\la{\langle}                                                        
\def\ra{\rangle}                                                        
\def\lb{\lbrack}                                                        
\def\rb{\rbrack}                                                        

\def\lra{$\Longrightarrow$}
\def\rar{$\Rightarrow$}

\def\al{\alpha}                                                         
\def\be{\beta}                                                         
\def\si{\sigma}                                                         
\def\Si{\Sigma}                                                         
\def\de{\delta}                                                         
\def\De{\Delta}                                                         
\def\lam{\lambda}                                                         
\def\ga{\gamma}                                                         
\def\ka{\kappa}                                                         
\def\th{\theta}                                                         
\def\om{\omega}                                                         
\def\Om{\Omega}                                                         
\def\eps{\epsilon}                                                      
\def\vare{\varepsilon} 
                          
\def\nj{$\tilde{\mbox{n}}$}
\def\gr{$^{\circ}$}
\def\arcsec{\hbox{$^{\prime\prime}$}}
\def\arcmin{\hbox{$^{\prime}$}}
\def\deg{\hbox{${^\circ}$}}
\def\half{$\frac{1}{2} $}

\hyphenation{spec-tro-radio-me-try}
\hyphenation{spec-tro-radio-me-tric}
\hyphenation{fi-gure}
\hyphenation{re-con-cile}
\hyphenation{ca-li-bra-tion}
\hyphenation{ca-li-bra-ted}
\hyphenation{in-te-gra-ting}
\hyphenation{re-sul-ting}
\hyphenation{res-ponse}
\hyphenation{re-la-tively}
\hyphenation{li-be-ra-ted}
\hyphenation{theo-rem}
\hyphenation{ana-lysed}
\hyphenation{ana-ly-ti-cal-ly}
\hyphenation{tem-pe-ra-ture}
\hyphenation{high-tem-pe-ra-ture}
\hyphenation{dia-meter}
\hyphenation{litho-graphy}
\hyphenation{signi-fi-cant}
\hyphenation{hand-ling}
\hyphenation{tem-pe-ra-ture-dependent}
\hyphenation{an-iso-tropy}
\hyphenation{an-iso-tropies}
\hyphenation{re-flec-ti-vity}
\hyphenation{va-li-da-ting}
\hyphenation{me-di-um}
\hyphenation{photo-poly-meri-sation}
\hyphenation{de-li-ve-red}
\hyphenation{geo-metric}
\hyphenation{geo-metry}
\hyphenation{nor-ma-lized}
\hyphenation{arc-minutes}
\hyphenation{inter-fero-metric}
\hyphenation{inter-fero-meter}
\hyphenation{inter-fero-meters}
\hyphenation{cor-res-ponding}
\hyphenation{sche-ma-tic}
\hyphenation{qua-li-ta-tive}
\hyphenation{far-in-fra-red}
\hyphenation{un-biased}
\hyphenation{astro-no-my}
\hyphenation{long-wave-length}
\hyphenation{re-so-lu-tion}
\hyphenation{thres-hold}
\hyphenation{neu-tra-li-zation}
\hyphenation{ca-vi-ty}
\hyphenation{ca-pa-bi-li-ty}
\hyphenation{arc-minute}
\hyphenation{arc-minutes}
\hyphenation{astro-nomi-cal}
\hyphenation{extra-va-gant}
\hyphenation{be-treffen-den}
\hyphenation{high-tem-pe-ra-ture}
\hyphenation{sys-te-ma-tic}
\hyphenation{Ka-zakh-stan}
\hyphenation{gra-ting}
\hyphenation{know-ledge}
\hyphenation{com-bi-ning}
\hyphenation{com-pa-ra-tive}
\hyphenation{com-pa-ri-sons}
\hyphenation{he-te-ro-dyne}
\hyphenation{interfero-metry}
\hyphenation{de-fi-nition}
\hyphenation{para-meters}
\hyphenation{par-ti-cu-lar}
\hyphenation{par-ti-ci-pa-ting}
\hyphenation{co-herence}
\hyphenation{proper-ties}
\hyphenation{sensi-tivity}
\hyphenation{techno-logy}
\hyphenation{ac-celerator}
\hyphenation{opera-ted}
\hyphenation{Lei-cester}
\hyphenation{simu-lated}
\hyphenation{re-solving}
\hyphenation{ele-ment}
\hyphenation{qua-lity}
\hyphenation{manu-facture}
\hyphenation{stres-sing}
\hyphenation{pro-viding}
\hyphenation{coup-ling}
\hyphenation{seve-ral}
\hyphenation{rela-tive}
\hyphenation{po-lari-zation}
\hyphenation{dia-gnostics}
\hyphenation{maxi-mize}
\hyphenation{maxi-mum}
\hyphenation{li-quid}
\hyphenation{specifi-cally}
\hyphenation{se-pa-rate}
\hyphenation{se-pa-ra-te-ly}
\hyphenation{geo-phy-si-cal}
\hyphenation{phy-si-cal}
\hyphenation{ima-ging}
\hyphenation{spec-tro-meter}
\hyphenation{para-meter}
\hyphenation{mo-del}
\hyphenation{in-ter-action}
\hyphenation{cha-rac-te-ristic}
\hyphenation{re-fe-ren-ces}
\hyphenation{signi-fi-cant-ly}
\hyphenation{ca-me-ra}
\hyphenation{a-stig-ma-tism}
\hyphenation{li-be-ra-ted}
\hyphenation{pola-ri-sation}
\hyphenation{pola-ri-zation}
\hyphenation{re-alizing}
\hyphenation{li-quid}
\hyphenation{atmo-sphere}
\hyphenation{pre-eminent}
\hyphenation{wave-guide} 
\hyphenation{sensi-ti-vi-ty}
\hyphenation{navi-ga-tion}
\hyphenation{modi-fi-cations}
\hyphenation{ga-la-xies}
\hyphenation{re-fe-rence}
\hyphenation{pro-du-cing}
\hyphenation{cryo-coolers}
\hyphenation{se-pa-rate}
\hyphenation{vi-sible}
\hyphenation{si-li-con}
\hyphenation{cha-rac-te-ris-tics}
\hyphenation{elec-tro-nics}
\hyphenation{micro-roughness}
\hyphenation{en-vi-saged}
\hyphenation{never-the-less}
\hyphenation{ultra-violett-empfindlicher}
\hyphenation{ex-pe-ri-men-tal}
\hyphenation{si-mi-larly}
\hyphenation{par-ti-cu-larly}
\hyphenation{par-ti-cu-late}
\hyphenation{se-con-da-ry}
\hyphenation{se-cond}
\hyphenation{nano-technology}
\hyphenation{mo-dest}
\hyphenation{em-pi-ri-cal}
\hyphenation{em-pi-ri-cally}
\hyphenation{sta-bi-li-ty}
\hyphenation{astig-ma-tism-cor-rec-ted}
\hyphenation{open-struc-ture}
\hyphenation{gra-vi-ty}
\hyphenation{dis-co-ve-ry}
\hyphenation{pho-to-me-try}
\hyphenation{con-ti-nu-ous}

\def\nsp{\hspace{-.08cm}}
\def\nsps{\hspace{-.07cm}}

\def\mb[#1]{\makebox[0.15cm][l]{#1}}
\def\pu{{\mb[\"u]}}


\def\gapprox{{_>\atop{^\sim}}} 
\def\lapprox{{_<\atop{^\sim}}}

\newcommand*\avl{<\!\!}  
\newcommand*\avr{\!\!>}

\newcommand{\greeksym}[1]{{\usefont{U}{psy}{m}{n}#1}}
\newcommand{\udelta}{\mbox{\greeksym{d}}}
\newcommand{\uDelta}{\mbox{\greeksym{D}}}
\newcommand{\uOmega}{\mbox{\greeksym{W}}}

\def\ga{\rlap{\lower 2.5pt\hbox{$\sim$}}\raise 1.5pt\hbox{$>$}\;} 

\def\ts{\thinspace}
\def\hei{He{\ts{$\scriptstyle {\rm I}$}} }
\def\heii{He{\ts{$\scriptstyle {\rm II}$}} }
\def\mgx{Mg\ts{$\scriptstyle {\rm X}$} }
\def\mgi{Mg\ts{$\scriptstyle {\rm I}$} }
\def\mgii{Mg\ts{$\scriptstyle {\rm II}$} }
\def\oiv{O\ts{$\scriptstyle {\rm IV}$} }
\def\ov{O\ts{$\scriptstyle {\rm V}$} }
\def\feii{Fe\ts{$\scriptstyle {\rm II}$} }
\def\ci{C\ts{$\scriptstyle {\rm I}$}}
\def\cii{C\ts{$\scriptstyle {\rm II}$}}
\def\caii{Ca\ts{$\scriptstyle {\rm II}$} }
\def\ni{N\ts{$\scriptstyle {\rm I}$} }
\def\nii{N\ts{$\scriptstyle {\rm II}$}}
\def\siii{Si\ts{$\scriptstyle {\rm II}$} }
\def\siv{Si\ts{$\scriptstyle {\rm IV}$}}
\def\ne8{Ne\ts{$\scriptstyle {\rm VIII}$} }


\newcommand*\m{\mathrm{m}}                     
\newcommand*\s{\mathrm{s}}                     
\newcommand*\K{\mathrm{K}}                     
\newcommand*\kg{\mathrm{kg}}                   
\newcommand*\A{\mathrm{A}}                     
\newcommand*\mol{\mathrm{mol}}                 
\newcommand*\cd{\mathrm{cd}}                   


\newcommand*\da{\mathrm{da}}                   
\newcommand*\h{\mathrm{h}}                     
\newcommand*\kilo{\mathrm{k}}                  
\newcommand*\M{\mathrm{M}}                     
\newcommand*\G{\mathrm{G}}                     
\newcommand*\peta{\mathrm{P}}                  
\newcommand*\E{\mathrm{E}}                     
\newcommand*\Z{\mathrm{Z}}                     
\newcommand*\Y{\mathrm{Y}}                     
\newcommand*\deci{\mathrm{d}}                  
\newcommand*\centi{\mathrm{c}}                 
\newcommand*\n{\mathrm{n}}                     
\newcommand*\p{\mathrm{p}}                     
\newcommand*\f{\mathrm{f}}                     
\newcommand*\z{\mathrm{z}}                     
\newcommand*\y{\mathrm{y}}                     


\newcommand*\rad{\mathrm{rad}}                 
\newcommand*\sr{\mathrm{sr}}                   
\newcommand*\Hz{\mathrm{Hz}}                   
\newcommand*\henry{\mathrm{H}}                 
\newcommand*\N{\mathrm{N}}                     
\newcommand*\Pa{\mathrm{Pa}}                   
\newcommand*\Bq{\mathrm{Bq}}                   
\newcommand*\J{\mathrm{J}}                     
\newcommand*\W{\mathrm{W}}                     
\newcommand*\C{\mathrm{C}}                     
\newcommand*\degC{\mathrm{\deg{C}}}            
\newcommand*\V{\mathrm{V}}                     
\newcommand*\F{\mathrm{F}}                     
\newcommand*\T{\mathrm{T}}                     
\newcommand*\Ohm{\mathrm{\Omega}}              


\newcommand*\nm{\mathrm{nm}}                   
\newcommand*\ua{\mathrm{ua}}                   
\newcommand*\lit{\mathrm{L}}                   
\newcommand*\Lap{\mathrm{\nabla^2}}            
\newcommand*\dAl{\mathrm{\Box}}                
\newcommand*\an{\mathrm{a}}                    


%
	


%% file: abstract0.tex
\begin{abstract}

Radiometry, i.e., measuring the power of electromagnetic radiation -- 
hitherto often referred to as ``photometry'' -- is of fundamental 
importance in astronomy. 
We provide an overview of how to achieve a valid 
laboratory calibration of space telescopes 
and discuss ways to reliably extend this 
calibration to the spectroscopic telescope's 
performance in space. A lot of effort has been, and still 
is going into radiometric ``calibration'' of telescopes once they 
are in space; these methods use celestial primary and transfer 
standards and are based in part on stellar models. 
The history of the calibration of the Hubble Space Telescope serves 
as a platform to review these methods.
However, we insist that a true calibration of spectroscopic space 
telescopes must directly be based on and traceable to 
laboratory standards, and thus be independent of the observations. 

This has recently become a well-supported aim, following the 
discovery of the acceleration of the cosmic expansion by use of type-Ia 
supernovae, and has led to plans for launching calibration 
rockets for the visible and infrared spectral range. This is timely, 
too, because an adequate exploitation of data from 
present space missions, such as {\em Gaia}, and from many current 
astronomical projects like {\em Euclid} and {\em WFIRST} demand higher 
radiometric accuracy than is generally available today.

A survey of the calibration of instruments observing from 
the X-ray to the infrared spectral domains that includes 
instrument- or mission-specific estimates of radiometric 
accuracies rounds off this review.


\end{abstract}

%% file: intro1.tex
\section{Introduction}                                        
\label{sec1}

The advent of spectroscopy in astronomy led, about 100 years ago, to the 
development of astrophysics.  Today, our knowledge of the Universe is, to a 
significant extent, based on spectroradiometric\footnote
	{In astronomy, the term {\em photometry} is often used when
	dealing with broadband light-level 
       measurements; those
	with higher spectral
	resolution are called {\em spectrophotometry}. 
	However, in general  terminology of radiation measurements, 
        photometry 
	refers to intensity determinations that are relevant 
	to human vision.  Therefore, we have chosen to use 
	the terms {\em radiometry} or {\em spectroradiometry},  
 which apply to measurements in the entire electromagnetic spectrum, in 
 this paper.  
To avoid confusion, we use the term ``(astronomical) photometry'' 
when discussing results of the photometry method of astronomy.}
observations, i.e., on the determination of the spectral irradiance or 
spectral radiance.\footnote
	{If an object is not spatially resolved, irradiance, $I$, the
	detected power per unit area (with unit symbol W m$^{-2}$; often,
	loosely, called radiative flux) is measured. {\em Spectral} irradiance 
	refers to the irradiance per energy (or wavelength) interval at 
	a given energy (or wavelength).
	Radiance, $R$, is the power per unit area per unit solid angle,
	with unit symbol W m$^{-2}$ sr$^{-1}$ 
(cf., The International System of Units (SI), Brochure, 8th edition 
(2006, updated in 2014), Bureau International des Poids et Mesures (BIPM)  
\url{http://www.bipm.org/en/publications/si-brochure}). 
{\it Spectral}
	radiance refers to the
	radiance per energy (or wavelength) interval at a given energy (or
	wavelength), with unit symbol W m$^{-2}$ sr$^{-1}$ eV$^{-1}$ (or
	W m$^{-2}$ sr$^{-1}$ nm$^{-1}$). 
	If the distance, $d$, to a 
	uniformly emitting object 
	of area, $S$, is known, then the irradiance is related 
	to the radiance by $I = R (S/d^{2})$.  }
Such measurements require determining the number as well as the spectral and 
spatial (or rather directional) distributions of photons that 
arrive at an observer's telescope from an 
object. When combined with atomic and molecular data and correct distances, 
spectroradiometric observations provide the essential evidence
for deriving temperatures, densities, gravity, element abundances, ionisation
stages, and flow and turbulent velocities in objects that are in the gaseous or 
plasma state.  In this way, we elucidate their physical structures, 
chemical compositions, and the processes that cause them to emit 
 \citep[see, e.g.,][]{WilFr13}. 
Such observations also provide similar inferences about  dust and 
solid bodies in interplanetary, interstellar, and intergalactic
space. 

The history of astronomy shows that observations with an improved 
accuracy of measurement frequently uncover flaws in our knowledge 
of celestial objects.  Usually, this process leads to a more
detailed and comprehensive understanding of the object under study. 
In some instances -- the discovery by 
\cite{1966ApJ...146..666P} 
of the cosmic microwave background, for
example -- radiometric measurements of improved accuracy uncover
hitherto hidden phenomena in backgrounds, foregrounds, and 
surroundings that are 
also observed.



Findings from astronomical observations thus critically  depend upon 
knowledge of the calibration of the  data with respect to primary 
laboratory radiance standards  (cf., Sect.~2). 
Because rigourous science requires rigourous measurements, astronomers should 
ensure that their spectroradiometric observations are traceable to such 
standards through frequent, thorough, spectroradiometric calibration.
This implies judging the uncertainty of 
the effective area\footnote{The effective area of a 
spectroradiometric instrument is the collecting area 
of an instrument with loss-free optical elements, i.e., with 
perfect reflections, diffraction or dispersion 
efficiency as well as detector efficiency.} 
of the telescope-spectrometer 
combination as a function of wavelength or energy of their observation.  

The need for precise spectroradiometric calibration has 
been demonstrated  
once more by the fact 
that the uncertainties in the 
spectral irradiance (often called ``flux distribution'' in the references) 
of stellar standards 
are the dominant systematic error in 
measuring relative irradiances (often called ``relative fluxes'' 
in the literature) 
of redshifted supernovae Ia and, 
thus, in determining the nature of the
dark energy that is driving the observed accelerating cosmic 
expansion 
\citep{2011ApJ...737..102S, 2014PASP..126..711B}. 

\begin{figure}[ht]
\centering \includegraphics[width=0.8\textwidth]{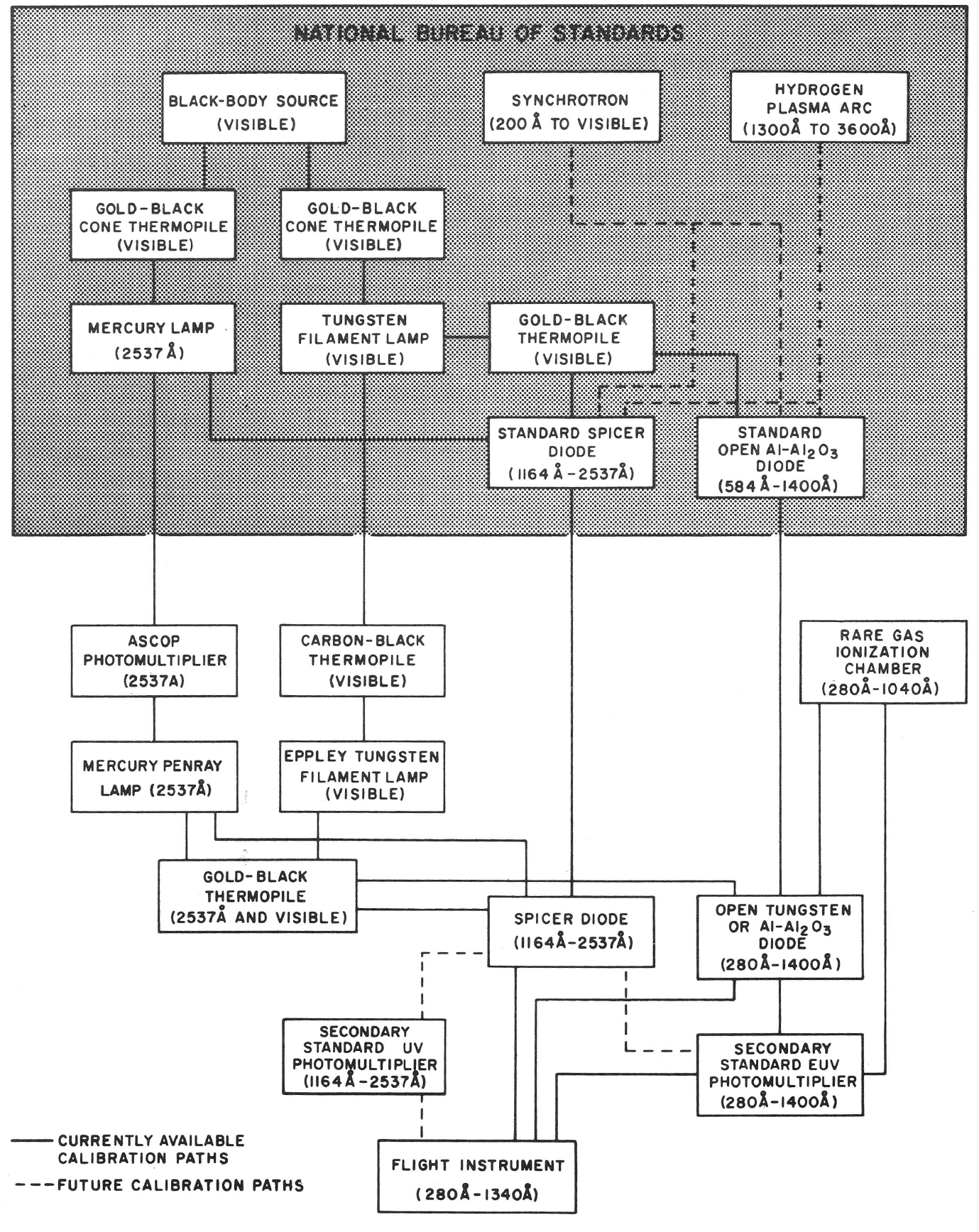} 
\caption{Pedigree of the calibration 
of the photoelectric Harvard S-055 spectroheliometer 
on the Apollo Telescope Mount (ATM) of the 
{\em Skylab} mission \citep{1974spop.conf...33H}.}
\label{Pedigree}
\end{figure}

For spectrometric telescopes in space, current spectroradiometric calibrations 
are based on a wide variety of 
techniques whose relationships to 
primary laboratory 
standards are not always evident.  In this paper we elucidate and critically 
assess the pedigrees of calibration methods for the X-ray through the
sub-millimetre wavelength ranges.\footnote{We leave out the 
gamma-ray and radio regimes which have their own, markedly different 
calibration methods 
 \citep[see, e.g.,][]{Kan13, SchKan13}.} 
%
An example of a historical pedigree of the calibration 
of the spectroheliometer on the Apollo Telescope Mount (ATM) of the 
{\em Skylab} mission is given in Figure~\ref{Pedigree}. 
%
%
%

\begin{figure}[ht]
\centering \includegraphics[width=0.8\textwidth]{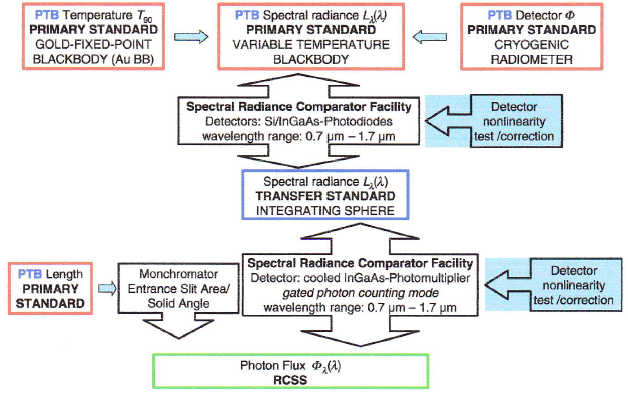} 
\caption{
Calibration pedigree for the radiometric calibration spectral source 
(RCSS), a very low photon flux radiation source 
operated under vacuum and cryogenic conditions. It will be 
used for testing and calibrating the near-infrared multi-object 
dispersive spectrograph NIRSpec, to be flown on the 
{\em JWST}, in the wavelength range 
0.7~\textmu m to 5~\textmu m 
\citep{2009Metro..46S.207T}.}
\label{nirspec_pedigree}
\end{figure}

The calibration pedigree for the NIRSpec instrument on the 
James Webb Space Telescope ({\em JWST})  in 
Figure~\ref{nirspec_pedigree} looks less complex, as it can 
rely directly on a primary standard. In the vacuum ultraviolet domain 
the primary standard would now be synchrotron radiation from a storage 
ring, which at the time of the calibration presented in Figure~\ref{Pedigree} 
had not yet been available.
%

The nomenclatures for wavelength bands are often arbitrary and 
not consistent from author to author; we therefore list  in Table~\ref{tab1}
our preferences, 
and some of the representative space missions flown more recently 
carrying instruments for the corresponding bands. 
%
\begin{table}
\caption{Wavelength bands as used in this article, together with example missions.}
\label{tab1}
\resizebox{\textwidth}{!} {
\begin{tabular}{llll} 
\hline\noalign{\smallskip} 
 {Name}          &         &  {nominal range}     & {mission} \\
\noalign{\smallskip}\hline\noalign{\smallskip}
 X-ray                     &         & 0.1~keV to 10~keV      & {\em XMM-Newton, Chandra, Rhessi, Yohkoh, Suzaku}       \\
 \hspace*{4mm}soft X-ray   &         & 0.1~keV to 3 keV       &      {\em  Yohkoh, Hinode}                      \\
 extreme ultraviolet       & EUV     & 10 nm to 90 nm         & {\em EUVE, SOHO, TRACE, STEREO, Hinode, SDO, IRIS}                \\
 far ultraviolet           & FUV     & 90 nm to 120 nm          & {\em FUSE}                \\

 vacuum UV                 & VUV     & 120 nm to 200 nm         & {\em IUE, HST}            \\
 ultraviolet               & UV      & 200 nm to 450 nm         & {\em HST}                 \\
 visible                   &         &450 nm to 750 nm          & {\em HST, SOHO, STEREO}                 \\ 
 near infrared             & NIR     & 750 nm to 5~\textmu m     & {\em HST},  {\em AKARI}        \\ 
 mid infrared              &   & 5~\textmu m to 30~\textmu m &  {\em JWST}                \\
 far infrared              & FIR     & 30 \textmu m to 100 \textmu m & {\em ISO},  {\em Spitzer},  {\em AKARI, Herschel}   \\
 sub-millimetre            & sub-mm  & 100 \textmu m to 1 mm       & {\em COBE, WMAP, Herschel, Planck}	      \\
\noalign{\smallskip}\hline 
\end{tabular}}
\end{table}
%
%
Although the focus of this review is on spectrometric telescopes, we include a 
number of  solar irradiance measurements, both spectral and total.  
These are among the best calibrated
measurements from spacecraft and, therefore, provide a model for  accurate 
astronomical spectroradiometric observations.

We consider the laboratory bases for radiometric calibration, discuss the 
transfer of the calibration to satellite instruments through measurements as 
well as the establishment of ``celestial standards'',\footnote
	{These should not be confused with the primary laboratory
	standards described in Sect.~2.1.}  and point to the 
spectroradiometric calibration of some selected astronomical spacecraft 
and their instruments. 

Because recent and future space 
missions will provide data on the Cosmos to an exquisite accuracy, 
the need for thorough calibration before, and continued monitoring after, 
launch of spectroscopic instrumentation has been receiving increased 
attention for several missions lately. 


The {\em Gaia} mission,\footnote{{\em Gaia} was launched on 19 December 2013 
and began its five-year science phase on 29 July 2014. For its progress 
follow \url{http://sci.esa.int/gaia}.}  
for example,  
will provide a global stellar census of our Galaxy, which will 
result in the measurement of distances for more than 20 million stars 
up to a few kiloparsec away from Earth to one-percent 
accuracy \citep{2008IAUS..248..217L}.  
A one- to three-percent radiometric accuracy 
would be compatible 
with measurements on such stars 
\citep[cf.,][]{2012MNRAS.426.1767P}. 
%


 

For a duration of six years, starting ca. 2020,  
the {\em Euclid} mission will map at least 
35.6~\% of the entire sky to derive the geometry and evolution of the 
dark universe. This will place stringent constraints on dark energy, 
dark matter, gravity, and cosmic initial conditions \citep{2013LRR....16....6A}.  
Unprecedented radiometric precision and accuracy will be required: 
the relative radiometric uncertainty in 
imaging with its Near-infrared Spectrograph 
and Photometer (NISP), for example, shall be less than 0.5~\%, and it is a 
mission goal to obtain an absolute radiometric calibration better than 3~\%  
\citep{2012SPIE.8442E..0TL}. 

\begin{figure}[t]
\centering 
\includegraphics[width=0.8\textwidth]{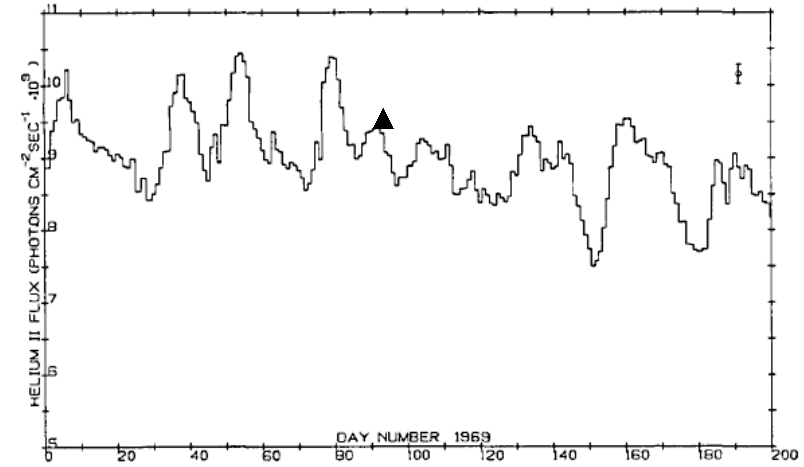} 
\caption{The long-term solar irradiance variations in the 
\heii Lyman-$\alpha$ line as measured in 1968 by a grazing-incidence 
spectrometer on the {\em OSO-4} satellite. A comparison with the
irradiance measured on day 93 by two calibration-rocket spectrometers 
(black triangle) 
showed that the responsivity of the satellite spectrometer had 
not changed \citep{1971ApOpt..10...28B}. 
The variations in the 
signal stem from the 28-day solar rotation, which makes 
active regions appear and disappear on the solar disk. The active 
regions are also waxing and waning, on different time scales. 
}
\label{oso}
\end{figure}

\begin{SCfigure} 
\includegraphics[width=0.4\textwidth]{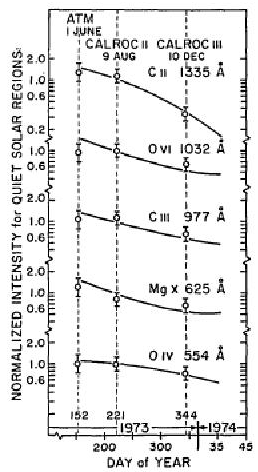} 
\caption{Comparison of the apparent relative calibration 
changes of the Harvard S-055 spectroheliometer 
on the Apollo Telescope Mount (ATM)
observed in orbit and the results of the preflight 
and rocket measurements. (From \citet{1977ApOpt..16..849R}).
}
\label{calroc}
\end{SCfigure}

Moreover, 
it will be a major challenge to maintain a stable radiometric response on 
ESA's {\em Solar Orbiter} (to be launched in 2018), which will observe the 
Sun from about 0.15~au, i.e., from half the distance between Mercury and 
the Sun. 
In the 1960s already it was noted that the effective area of the 
normal-incidence vacuum-ultraviolet (VUV) spectroheliometer on \mbox{\em OSO-4} 
changed during laboratory calibration and again before early 
observations in space \citep{1970ApOpt...9.1201R}. 
Degradation, with spectrally different rates, was registered during 
further exposure in orbit on {\em OSO-4} and on {\em OSO-6} 
\citep{1973ApJ...183..291H}. 
In contrast, no degradation of the responsivity on grazing-incidence 
spectrometers was found, except in some cases, where the reduction could 
be ascribed to loss of responsivity 
of open-structure detectors 
\citep[cf.,][]{1969SoPh....6..183N, 1969SoPh....6..175H}. 
Moreover, comparisons 
between the responsivity of the grazing-incidence spectrometer on 
{\em OSO-4} and a 
sounding rocket measurement by the University College group  
\citep{1971ApOpt..10...28B} 
had shown perfect 
agreement of the measured irradiance by the satellite and rocket 
instruments (cf., Figure~\ref{oso}).\footnote{In spite of grazing-incidence optics being much 
less sensitive to contamination, and therefore having a better 
radiometric stability, normal-incidence instrumentation is usually 
preferred -- at least as long as surfaces with sufficient 
reflectivity are available, because the optical design 
of normal-incidence optics is considerably less difficult.}
The responsivity of the follow-on normal-incidence instrument S-055 on 
the Apollo Telescope Mount on {\em Skylab} \citep{1977ApOpt..16..837R} 
was therefore 
monitored by use of a ``celestial standard'', namely a ``quiet area'' in 
the centre of the solar disk, and checked by three calibration-rocket 
flights \citep{1975SSI.....1...23T}, 
which simultaneously observed the same ``quiet area''. 
The results \citep{1977ApOpt..16..849R} 
are shown in Figure~\ref{calroc}. This comparison 
at the same time gives an indication on the errors one may have to contend 
with, when using a ``quiet'' area in the centre of the solar disk as 
``celestial standard'' to monitor instrument responsivity. 
%
%
It was generally concluded that the main reason for the loss of sensitivity 
of normal-incidence VUV instruments was polymerisation of minute layers 
of volatile substances on the optical surfaces. Based on the experiences 
with other early solar space observatories, a special effort was undertaken 
to maintain an extensive cleanliness programme for the {\em SOHO} instruments 
and spacecraft \citep{2002ISSIR...2.....P}. 
More recently, these kind of studies 
have been taken up again by \citet{2013SoPh..288..389B}     
who comprehensively 
analysed the behaviour of numerous instruments designed to observe the Sun 
in space and drew renewed attention to this problem. 


Missions whose primary aim is solar physics can (and will) insist 
on clean platforms. Monitoring the solar EUV irradiance for use in 
modelling the thermosphere and the ionosphere, however, is  
an operational issue that 
cannot wait until a clean platform becomes available: real-time measurements 
of the solar VUV irradiance provide data needed for ionospheric modelling 
that in turn permits correcting propagation delays of navigation signals 
from space to Earth.\footnote{By adding airglow and auroral-emission 
monitoring, the impact of space weather on the terrestrial 
thermosphere/ionosphere can be studied as well, and used to investigate 
real-time space weather effects. This in turn helps to derive detailed 
correction procedures for the evaluation of global navigation satellite 
system signals.}
In an aper\c{c}u on the history of VUV solar-irradiance measurements since 
1946, \citet{2015HistGeoSpSc}   
points out that it was only at the end of the 
20th century that enough progress had been made to solve the serious 
problem of degradation. The use of a primary standard in orbit now 
permits measuring the solar VUV irradiance with sufficient 
radiometric accuracy. As a consequence, a data set covering a full 
11-year solar cycle is now available for the first time. 

In their report about measurements over eight years of the VUV solar 
irradiance by SolACES on the International Space Station ({\em ISS}) 
\citet{2014SoPh..289.1863S} 
describe ``a very strong up-and-down variability of 
the spectrometric efficiency by orders of magnitude'', which is assumed 
to stem  at least in part from contamination by supply missions. 
The use of an in-orbit absolute standard nevertheless led to the 
reliable results now available. 

Finally, we should draw attention to the summarising report by 
\citet{2010JGRA..11512313S},    
in which the environments of several spacecraft -- 
and in particular of {\em Rosetta}  -- 
during their trajectories and operation in space have been studied and 
interpreted in detail.  
The authors conclude that ``the most important process of outgassing 
during the first 200 days of a mission is desorption of water from exposed 
surfaces of the spacecraft.'' For a spacecraft at 1~au, the half‐life of 
the density associated with desorption was observed to be ca. 21~days. 
After several years in space, decomposition of the material, mostly 
attributed to solar UV radiation, begins to play a role. 
The contamination of sensitive surfaces may vary considerably between 
different spacecraft, and between instruments on the same spacecraft; 
it can also vary with time, as the spacecraft attitude changes and 
payload operations take place. In general, shadowed parts contain mostly 
water, whereas parts that are mostly in the Sun may exhibit a large 
contribution from decomposition of, for example, lubricants. 
In spite of strong efforts to avoid contamination by a careful design and 
selection of materials, there remains a risk of measurements falsified
 by contamination of optical surfaces. In the long range, as more and more 
accurate spectroradiometry will be needed to keep pace with progress in 
astronomy, the use of primary standards in orbit will be required. Source 
standards will then probably become the preferred kind of standard.

%% file: laboratory2.tex
\section{Laboratory basis for spectroradiometry and its transfer 
to space}  
\label{sec2}

Spectroradiometry that is performed on an astronomical object to 
derive physical  
properties must be based on primary laboratory standards,  
i.e., on absolute radiation sources or detectors.\footnote{A brief summary 
of Section~\ref{sec2} has been given earlier by \citet{2013opsg.book.....H}.}
     The criterion for primary standards is that they
     give results directly in Syst\`{e}me International (SI) units without
     the need for any calibration relative to the quantity being measured.
     One must be able to write the complete operating equation for the
     standard without any unknown or empirically determined constants or
     functions that depend upon the quantity being measured 
     \citep{1997Metro..34...61Q}.  
These standards are realised in the laboratory; 
\citet{1994Cookbook} 
has discussed the fundamental necessity for this approach.

In this section we discuss 
the current primary standards for radiometry, 
ways of 
applying spectroradiometric calibration to space instruments, and methods for 
monitoring changes in these calibrations in orbit. 
In Section~\ref{sec3} we will then discuss celestial 
standards for astronomical spectroradiometry and the spectra of model stellar 
atmospheres for white dwarfs.  
These modelled spectra are used to extend 
calibration capability beyond the wavelength and irradiance ranges of the 
astronomical standards.

\subsection{Primary radiometric standards}

Due to the large ranges of intensity and 
energy (or wavelength) of electromagnetic radiation,  
practical considerations and current technology 
have resulted in four primary standards for absolute spectral 
radiometry: two emission or source standards, viz., black bodies, 
and electron (or positron) storage rings, and two detector 
standards, viz., cryogenic radiometers and double-ionisation chambers. 

Black bodies \citep[cf.,][]{1984Kaase} 
are emission standards 
based on thermodynamics: the radiation from a black body is related to its 
temperature by the Planck law, where the operational temperatures are normally 
those of melting points of appropriate metals 
\citep{2009Hartmann}. 

Electron (or positron) storage rings \citep[cf.,][]{1992Madden}
are standards based on electrodynamics.  The \citet{1949Schwinger}
formula 
is used to 
calculate the synchrotron radiation emitted by the
accelerated charged particles. Inputs required for the calculation are the 
energy and current of the stored beam as well as the magnetic induction at the 
point where the radiation is emitted; small corrections are required to account 
for the finite vertical extent and divergence of the particle beam. 



Cryogenic electrical substitution radiometers 
 \citep[ESR, cf.,][]{1991Moestl} are 
detector standards that are based on accurate current and voltage measurements. 
The temperature increase of an irradiated cavity cooled to liquid helium 
temperatures is compared by a null method with the temperature increase caused 
by deposition of accurately measured electrical power into an equivalent cavity. 
Although corrections for reflections, scattering and diffraction must be made, 
cryogenic radiometers are accepted as the most accurate radiometric standards 
\citep{Fox:90, Foukal:90, HoytFoukal1991, Quinn:91b, Quinn:91a}. 

Rare-gas double-ionisation chambers are detector standards based on the fact 
that the photo-ionisation yield of rare gases is unity over the wavelength 
range 30~nm to 102~nm \citep{Samson:64,Samson:74,SamsonEderer:2000}. 
Ionisation chambers 
are the only primary standards that have been used in space to date 
\citep{Carlson:84, 2014SoPh..289.1863S}, 
though non-cryogenic ESRs, 
which are less sensitive and accurate than 
the cryogenic variety, have been used to measure and monitor the 
solar constant (see Sect.~4).

The state of agreement between the radiometric scales realised by the first 
three primary standards mentioned above was discussed by 
\citet{1993Metro..30..439S}. 
Their general uncertainty analysis indicated that  
black bodies and cryogenic radiometers produced and detected spectral radiant 
power with relative 
uncertainties of 0.08~\% and 0.007~\%, respectively.\footnote
	{Unless otherwise indicated, all uncertainties in this paper are given
	in terms of one standard deviation, i.e., 68~\% confidence limit  
        (coverage factor $k=1$).} 
\citet{1996Metro..32..459T} 
measured emission from an electron storage 
ring with a cryogenic radiometer and found agreement between the two primary
standards within 0.3~\% at a photon energy of 15~keV and 0.08~\% for visible 
radiation.\footnote{
The Bureau International des Poids et Mesures 
(BIPM, \url{http://www.bipm.fr}) has the 
mandate to provide the basis for a single, coherent system of measurements 
-- traceable to the International System of Units (SI) -- throughout the 
world. The Bureau was set up by the Convention of the Metre and operates 
under the exclusive supervision of the Comit\'{e} International des Poids 
et Mesures (CIPM). The Committee's principal task is to 
ensure worldwide uniformity in units of measurement, particularly
between national measurement standards, but the CIPM also 
takes on the more fundamental task to arrange for and monitor comparisons 
that determine the accuracies with which the individual primary 
standards are realised.}
\citet{2005ExMPS..41..213H} 
have summarised further such comparisons; 
these included a comparison between synchrotron radiation from the
 electron storage ring BESSY~I with radioactive standards 
\citep{1994NIMPA.339...43A}. 

\subsection{Metrology light sources} 
\begin{figure}[t,h]
\centering \includegraphics[width=0.6\textwidth]{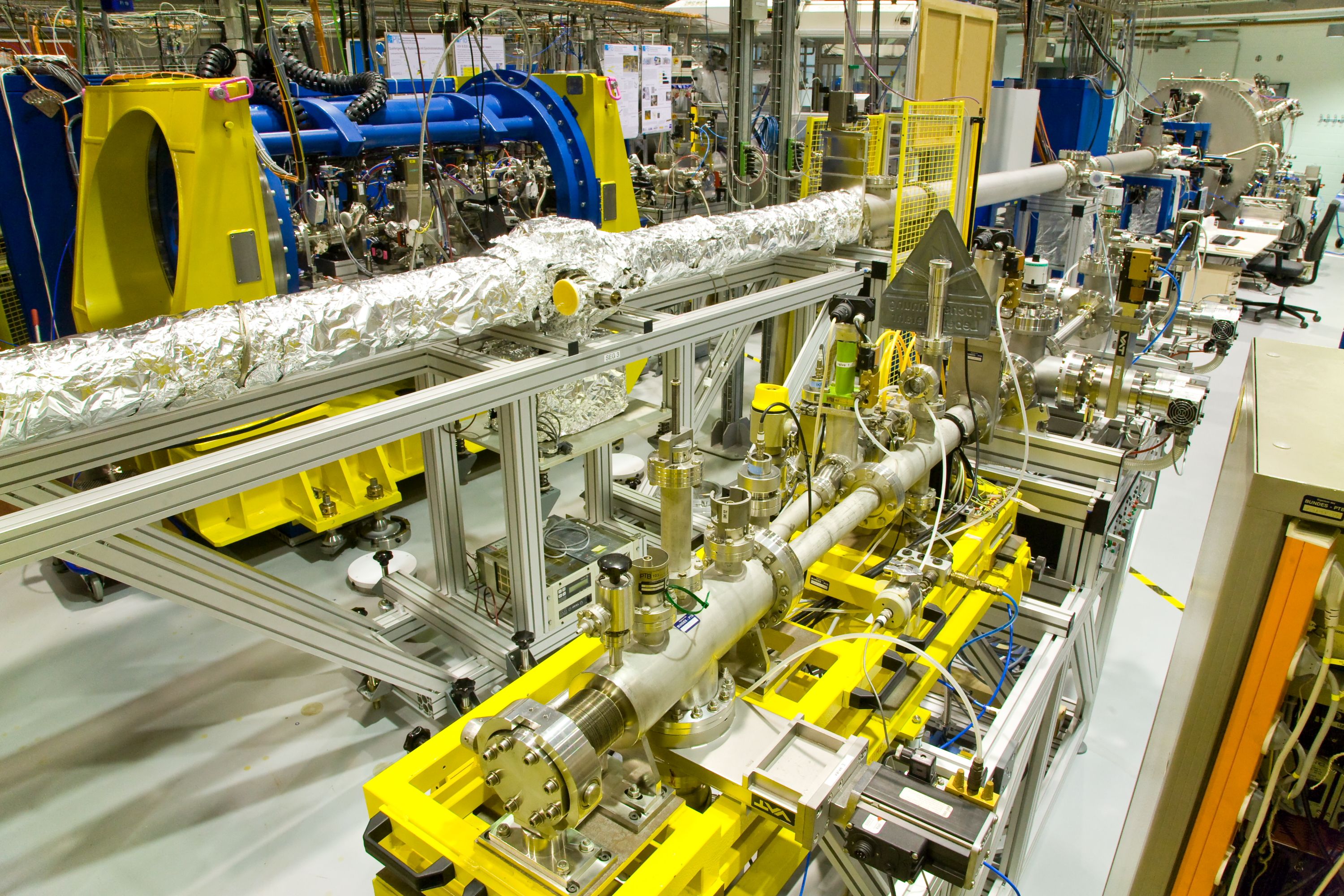} 
\caption{View from the storage ring downstream at the metrology beamline 
of the MLS at PTB. The vacuum tank for calibration (also shown in the 
next figure) 
is seen at the right upper corner of the image. 
A transportable VUV transfer source based on a window-less hollow-cathode 
discharge 
that has been developed for the calibration of {\em SOHO} instruments in 
the home 
laboratories of the Principle Investigators, can also be brought into the beam; 
it is seen on the lower right side. 
(From \citet{2014Gottwald},  
with permission by PTB).}
\label{MLS_beamline}
\end{figure}

\begin{figure}[t,h]
\centering \includegraphics[width=0.6\textwidth]{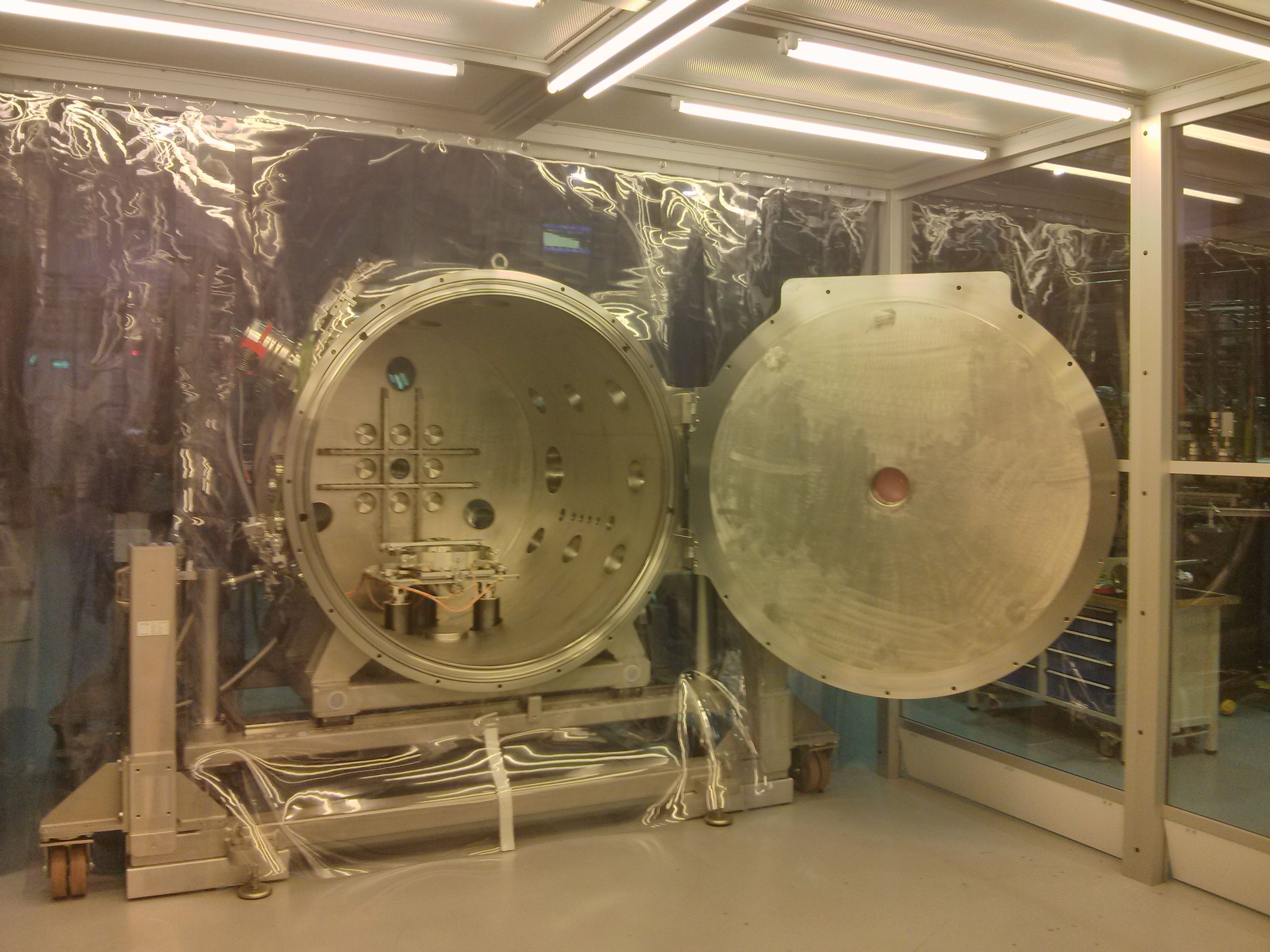} 
\caption{The vacuum tank for calibration of 
space instruments at the metrology beamline 
at the MLS. The tank is embedded in a cleanroom environment to enable the handling 
of flight hardware directly at the beamline. 
(From \citet{2014Gottwald},  
with permission by PTB).}
\label{MLS_tank}
\end{figure}

Today, user facilities, such as the Metrology Light Source (MLS) at the PTB in 
Berlin\footnote{\url{http://www.ptb.de/mls/index.html}} 
\citep{2009Metro..46S.266K} 
or the Synchrotron Ultraviolet Radiation Facility 
SURF-III\footnote{\url{http://www.nist.gov/pml/div685/grp07/surffacility.cfm}}   
at NIST\footnote{the US National Institute of Science and Technology 
(earlier National Bureau of Standards, NBS)} 
\citep{2002RScI...73.1674A} 
which can accommodate space hardware for radiometric calibrations 
with synchrotron radiation, are in operation. 
A dedicated beamline at the MLS  in Berlin 
and a corresponding calibration chamber in a cleanroom -- 
both part of installations 
that can be used by customers -- are shown in 
Figures~\ref{MLS_beamline} and~\ref{MLS_tank}.  


%

\subsection{Transfer standards} 

It is often practical to use simple, stable 
sources and detectors, known as transfer standards, that are themselves 
calibrated in the laboratory by comparison with the primary standards. Examples 
include argon arcs \citep{1977ApOpt..16..367B}, 
bare metal photoemissive diodes \citep{1978ApOpt..17.1489S}, 
D$_2$ lamps \citep{1988JRes...93...21K}, 
tungsten-filament incandescent lamps 
\citep{1988ntm..rept...64W},        
hollow-cathode \citep{1988ApOpt..27.4947D, 1994ApOpt..33...68H}
and Penning \citep{1994ApOpt..33.5111H} 
discharges, and silicon trap detectors  \citep{1996ApOpt..35.4404G} 
and photodiodes \citep{1998Metro..35..355K,1998Metro..35..329C}. 

\subsection{Radiometric calibration of spectroscopic space telescopes}
\label{sec2rcal}
\setlength{\parskip}{0mm}
The procedures used for radiometric calibration of spectrometric space telescopes 
may be divided into four categories: 

\noindent \hangindent=7mm \hangafter=1 ($i$) \, Pre-flight calibration in the 
laboratory at the component and/or subsystem levels or in end-to-end (i.e., 
from entrance aperture to detector output) tests using primary or transfer 
standards.  
This approach was followed  for the spectroheliometers on 
the ESA/NASA Solar and Heliospheric Observatory {\em SOHO}
\citep{1995SoPh..162.....F} 
and for the X-ray missions {\em Chandra} \citep{2000SPIE.4012....2W} 
and {\em XMM-Newton} \citep{2001A&A...365L...1J}. 

\noindent \hangindent=7mm \hangafter=1 ($ii$) Calibration in orbit by celestial 
standards once these are established through observations or by theoretical 
models \citep{2014PASP..126..711B}. 
For example, the calibration of the Hubble Space Telescope {\em HST}  
\citep{2007ASPC..364..315B} 
or of the solar Interface Region Imaging Spectrograph
{\em IRIS} have been largely done this way \citep{2014SoPh..289.2733D}. 
Several instruments on {\em SOHO}, in particular SUMER and UVCS, have used 
stars and planets for monitoring the radiometric calibration or 
to extend the calibration range to longer wavelengths where no ground 
calibration was available \citep{2000ApOpt..39..418S, Lemaire02,2002ISSIR...2..161G}. 
%


\noindent \hangindent=7mm \hangafter=1 ($iii$) Calibration of an orbiting 
spectrometric telescope by comparing its observations with simultaneous ones 
made by a similar ``underflight" instrument on a sub-orbital rocket or 
the Space 
Shuttle.  The latter instrument is then recalibrated in the laboratory 
after use 
in space to minimise uncertainties stemming from changes in effective area. 
As mentioned above, 
the radiometric calibration of the ATM spectroheliometer on {\em Skylab} 
has been checked three times with this method 
\citep{1975SSI.....1...23T,1977ApOpt..16..849R}, 
see also Figure~\ref{calroc}. 
%
 The
radiometric accuracy of observations with the Hopkins Ultraviolet Telescope 
 \citep[HUT,][]{1992ApJ...392..264D}  
was established in this way and they were then used
to validate modelled spectra used as celestial standards 
\citep{1997ApJ...482..546K}.  
The Solar Extreme Ultraviolet Experiment (SEE) on NASA's {\em TIMED} 
satellite also used three sounding-rocket flights to follow changes 
in its radiometric calibration. 
Rocket underflights have also been used extensively for monitoring 
the calibration and performance of the 
{\em SOHO} VUV instruments. 
In particular, they have been extremely useful to track changes after 
{\em SOHO}'s temporary loss of attitude in 1998, 
 see, e.g.,  \citet{2006AdSpR..37..225W} 
and references therein. 

\noindent \hangindent=7mm \hangafter=1 ($iv$) Calibration by use 
of a primary or 
transfer standard that is carried into and operated in orbit as part of the 
scientific payload.\footnote
	{An additional concept, viz., having emission transfer standards on
	the International Space Station ({\em ISS}), or on a small calibration 
        satellite
	orbiting the Earth,  with the purpose of calibrating 
co-orbiting satellites in the extreme-ultraviolet range has been suggested
	\citep{1991eua..coll..390S}.   
} 
Standards carried into orbit can seldom be used to assure an end-to-end 
calibration,
but are employed for monitoring the stability of a crucial part of an 
instrument, such as a detector.  
An exception is the Midcourse Space Experiment (MSX), which performed 
in-orbit infrared radiometry by releasing calibrated 2~cm diameter 
spheres, whose size and emissivity had been precisely measured in 
the laboratory, and whose temperature could be determined to $\pm 1$~K 
 \citep{2004AJ....128..889P}. 
 Instruments devoted to studying total solar irradiance, such as the 
Variability of Solar Irradiance and Gravity Oscillations (VIRGO) 
experiment on {\em SOHO} and the Spectral Irradiance Monitor (SIM) on 
the  {\em SORCE} satellite do use electrical substitution radiometers  
to directly calibrate their measurements in space.\footnote{In contrast 
to the cryogenic electrical substitution radiometers, 
where corrections covering non-ideal performance
 nearly vanish owing to the low operating temperature, 
ESRs used at higher temperatures require a careful evaluation 
of outside influences, such as heating of baffles. 
Besides, for ESRs to be used in space, such tests are complex.} 
Moreover, rare-gas 
ionisation chambers, which are primary radiometric detector standards, 
are operating in the SolACES experiment 
\citep{2014SoPh..289.1863S}. 
%
%
%

Nowadays, a combination of the described  techniques is often used.  
However, in view 
of: $(i)$, the multitude of observing modes and the concomitant large 
parameter space that should be covered during a calibration, 
$(ii)$,  concerns about the differences between conditions 
during ground testing 
and those of in-orbit operations, and $(iii)$, the inevitable changes in 
performance over time, it has become 
common practice to employ instrument models that use well-characterised 
properties of optical components, coatings, and contaminants to interpolate 
between measured benchmark values of instrument performance and to predict time 
behaviour 
\citep[cf.,][and discussion of {\em Chandra} and {\em XMM-Newton} in 
Sect.~4]{1997ASPC..125..411R,1997A&AS..126..563B}. 

\subsection{Monitoring spectoradiometric response in space}

Changes in the effective areas of spectrometric telescopes on spacecraft can be 
caused by exposure to contaminating pre-launch 
environments and material outgassed from the instrument or spacecraft itself. 
Unless extreme care is exercised, degradation can take particularly dramatic 
forms for solar telescopes, which are directly exposed to the harsh 
electromagnetic and energetic particle emission of the Sun. 

It is difficult to detect whether changes in spectroradiometric efficiency have 
occurred between laboratory measurements and first light in orbit unless an 
on-board calibration source that tests the overall system from end to end or 
reliable celestial standards are available.  In order to monitor changes that 
occur in space, users of spaceborne spectrometers often define an appropriate 
ensemble of non-variable objects that are observed at regular intervals. 
However, this procedure is not without complications for solar and X-ray 
telescopes where the objects observed are variable; on-board calibration 
systems or underflights remain the best options in these cases. 
Regular measurements of common targets with several instruments in common 
wavelength ranges in specifically designed campaigns have proven 
extremely useful as well.  A recent account of in-flight 
monitoring of the performance 
of solar instrumentation has been given by 
\citet{2013SoPh..288..389B}. 





%% file: celestial3.tex
\section{Model-supported celestial standards and beyond}       
\label{sec3}

Spectroradiometry -- from the VUV through the NIR -- of astronomical night-sky objects 
relies most commonly on the radiometric calibration of the spectra of a small set 
of stars, the so-called celestial primary standards -- with Vega ($\alpha$~Lyr) 
as the original primary standard. 
The state-of-the-art calibration of these celestial standards is based on direct, 
ground-based comparisons of 
the irradiance of Vega with that of laboratory standards, viz. black bodies or calibrated lamps, 
at the wavelength of 555.6~nm  
\citep{1970ApJ...161.1015O,1975ApJ...197..593H,1977A&A....61..679T,
1985IAUS..111..225H,1995A&A...296..771M}.\footnote{Spectroradiometric calibrations -- at the 
time called determining the ``spectral energy distribution'' (SED) of stars -- had been performed  
starting in the 1910s with visual and photographic comparisons with assumed stellar models. 
Direct comparisons with calibrated standard lamps started at the end of the 1930s at 
Ann Arbor and continued in the 1940s with extensions to the infrared, and by going to 
higher altitude, for example at Jungfraujoch, into the ultraviolet as well. The results were usually 
expressed as colour temperatures and in  magnitudes 
\citep[see, for example,][]{1960stat.conf...50C}.
}  
The calibration was  then extended to 
other wavelength domains by stellar atmosphere models. This was rather successful 
when going to shorter wavelengths by use of model spectra of hot white dwarfs (WD) 
with pure hydrogen atmospheres. 
%
Calibration in the infrared to 35~\textmu m was established 
in a series of 14 papers in the Astronomical Journal that was initiated by Martin Cohen in 1992  
\citep{1992AJ....104.1650C}. 
This effort involved measurements from the ground, by the Kuiper Airborne Observatory, 
by the Low Resolution Spectrometer on the {\em IRAS} satellite, and by 
stellar atmosphere programmes 
\citep[see, e.g.,][]{1992AJ....104.2030C, 2003AJ....126.1090C}.  

Three additional papers in this series, by \citet{2004AJ....128..889P}, 
  \citet{2006AJ....132.1445E},  
and \citet{2010AJ....140.1919E}  
reported a physics-related calibration performed in orbit 
(the details of which will 
be discussed below) and associated improvements 
to the catalogue of standard stars. 
 \citet{2010AJ....140.1919E} 
extended the spectral range covered by their standard stars to shorter 
wavelengths, i.e., to  350~nm.  
\citet{2008AJ....135.2245R} 
published another extensive report on 
calibration in the infrared; they had determined an absolute radiometric 
calibration for the 24~\textmu m band of the Multiband Imaging Photometer on 
{\em Spitzer}, and recommended adjustments to calibrations of the Infrared Array 
Camera on {\em Spitzer} \citep{2004ApJS..154...10F}, 
%
to the (ground-based) Two Micron All Sky Survey (2MASS), and  to the 
{\em IRAS} (astro)photometry. 
Rieke et al also showed that integrating measurements of the Sun 
with those of solar-type stars led to an accurate estimate of 
the solar irradiance spectrum from 1~\textmu m to 30~\textmu m that appears 
to agree with theoretical solar models.

Blackwell and his group at Oxford, who had been performing accurate 
astronomy-related
 measurements in the 1970s  -- originally with measurements 
of oscillator-strengths 
for abundance determinations and concomitant testing of stellar 
atmosphere models 
and nuclear astrophysics\footnote{The Oxford group had  been working on 
precision laboratory astrophysics early on.  In the 1970s  they experimentally 
determined transition probabilities with one-percent accuracies, 
 when such measurements normally had uncertainties of $\pm 10~\%$ or more. 
Not surprisingly then, Blackwell chose the topic of 
``Uncertainty in Astronomy'' for his Presidential Address to the 
Royal Astronomical  Society \citep{1975QJRAS..16..361B}.} -- 
%
later  determined Vega's irradiance at the Observatorio del Teide  
out to the infrared 
at $\lambda = $ 4.92~\textmu m by comparing it with that of a furnace 
\citep[cf., for example,][]{1985A&A...151..399M, 1989A&A...218..167B}. 

Observations of Vega by the {\em IRAS} satellite \citep{1984ApJ...278L...1N} 
that went further into the infrared 
have shown that Vega's spectrum exhibits an infrared 
excess beyond 12~\textmu m,  
because  solid particles  form a warm surrounding of this star 
\citep{1984ApJ...278L..23A}.  
Moreover, a decade ago, an interferometric 
study revealed that $\alpha$~Lyrae is a fast rotator seen almost pole-on 
\citep{2006Natur.440..896P}.  
Such deviations from model assumptions 
of the  environment of Vega and of the 
behaviour of the star itself  brought home the 
general danger of trying to rely on a stellar model and underlined the care 
that should be exercised in choosing a standard star if one wanted 
to extrapolate radiometric properties to hitherto uncalibrated wavelength regions.

\citet{2010AJ....140.1919E} 
have described the calibration history of (astro) photometry 
that has been associated with Vega.\footnote{In reviewing the literature 
relating to a potential variability of Vega, \citet{2014AJ....147..127B}  
came to the conclusion that there was not enough support for this claim. 
Nevertheless, \citet{2011AN....332..956B} and \citet{2014BCrAO.110...80B} 
did not exclude 
a long-term variability of Vega. The conclusion was 
that a (very minor) 21-year variability is ``most probable''. 
\citet{2015A&A...577A..64B} have recently observed starspots on Vega. 
}
They analysed the voluminous work that had gone into testing Vega's 
use as a celestial primary standard in the wavelength range between 
350~nm and 35.0~\textmu m and concluded that Vega should be replaced 
as primary celestial standard for (astro)photometry by 109~Vir in the visible 
and by Sirius ($\alpha$~CMa) in the infrared. Engelke et al also used results 
of a direct calibration of infrared irradiances in space by the 
Midcourse Space Experiment. 
As part of the MSX, \citet{2004AJ....128..889P} 
have compared 
the irradiance of a number of infrared standard stars with that of five 
emissive reference spheres that were ejected at various times during the 
mission. They describe the calibration properties 
of the released spheres as follows: ``The physical properties of the 2~cm 
diameter spheres, such as size and emissivity, had been precisely measured 
in the laboratory. The energy balance equation between the total flux 
absorbed and that emitted by the sphere is solved to obtain the time-dependent 
temperature of the sphere under the assumption that the sphere radiates as a 
black body with the measured wavelength-dependent emissivity.''
This resulted in absolutely calibrated values with a claimed 
standard uncertainty in the one-per-cent range for five wavelengths 
between 4.3~\textmu m and 21.3~\textmu m. 

Based on a comparison of data from MSX measurements with data from the 
Short Wavelength Spectrometer of ESA's Infrared Space Telescope 
\citep[{\em ISO},][]{1996A&A...315L..27K}, 
 \citet{2006AJ....132.1445E}   
created new spectra for stars that had earlier 
been selected as standards.  These new spectra led to model temperatures 
and stellar angular diameters that, on the one hand, compared favourably 
with independent measurements \citep[cf., e.g.,][]{2003AJ....126.2502M}, 
but, on the other hand, led to irradiances that were lower by 4~\% to 7~\% 
at wavelengths between 1~\textmu  m and 4~\textmu m than those found earlier by Cohen.  

\begin{figure}[th]
\centering \includegraphics[width=0.8\textwidth]{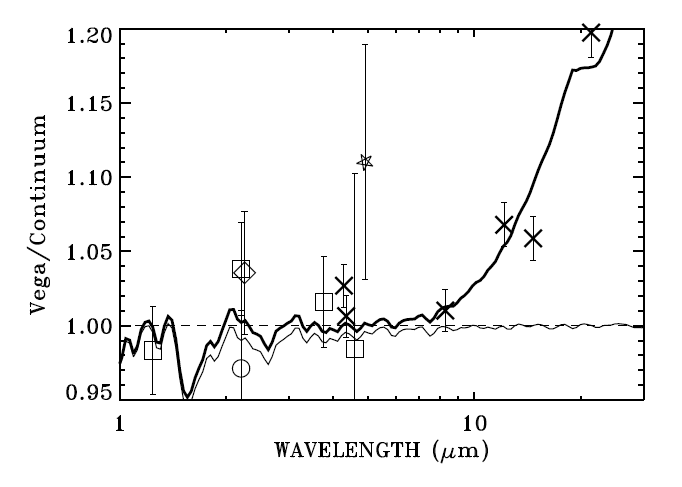} 
\caption{
Comparison of measured infrared absolute irradiances with the modelled 
stellar energy distributions derived by  \citet{2014AJ....147..127B}. 
The {\em light line} represents a model with a photospheric temperature of 9400~K at a 
resolution of R = 20; the {\em heavy line} includes the emission from the dust ring.
 The physics-based measurements obtained on the ground by \citet{1983MNRAS.203..795S} 
are represented by a {\em circle}, those by \citet{1989A&A...218..167B},  
by   \citet{1985A&A...151..399M}       
and by  \citet{1983MNRAS.205..897B}  
by a {\em diamond}, a {\em star} and {\em squares}, respectively. 
The crosses display measures obtained 20 years later in space by \citet{2004AJ....128..889P} 
from the MSX. 
(After Figure~9 in  \citet{2014PASP..126..711B},    
with permission.)}
\label{BohlinetalVEGA}
\end{figure}

Figure~\ref{BohlinetalVEGA} 
shows a comparison between the physics-related stellar irradiances 
mentioned above with a model-derived stellar energy distribution (SED). 
The uncertainties of the ground- and space-based measurements are shown. 
Those of the tracings of the SED, although not shown, are assumed to be smaller 
than those of the direct measurements. Clearly, using a stellar atmosphere 
model to interpolate between physics-related measurements is a pragmatic 
approach, but inherently retains the danger of circular conclusions 
if other models are later to be compared with this model. 
%
%

%

In order to put the calibration of stars on a solid, physics-based procedure, 
NIST 
has developed a Telescope Calibration Facility (TCF) with the aim of 
achieving SI-traceable spectroradiometric measurements of stellar irradiances with an accuracy 
of 0.5~\% (so-called ``SI stars'') for ground-based and space-borne observations  
  \citep{2009Metro..46S.219S}.      
This approach has been made possible not only by improvements in radiometry 
itself but also by advances in measuring and modelling the interfering effects 
of the terrestrial atmosphere. 

The TCF measurements make use of a redundant system of calibrated detectors and sources. 
Initially, a reference telescope-spectrometer system is calibrated in the Telescope 
Calibration Facility. This reference telescope is taken to a large astronomical 
telescope, and both these telescopes then observe a distant calibrated source.
 If, in a next step, both telescopes measure a given star's irradiance, they 
should also obtain the same measured value.

This value is, however, still only the ``ground-level irradiance'' rather than the 
value that prevails above the atmosphere. A further step is then taken to assess 
and subsequently eliminate the influence of the atmosphere above the telescopes. 
To determine the transmittance of the atmosphere one uses inputs from both   
LIDAR and atmospheric modelling. 
Assuming stable stars, repeated measurements under different 
atmospheric conditions will then help to assess the uncertainty caused by correcting 
for atmospheric transmittance. In this way, it is hoped that the ``above-the-atmosphere'' 
irradiance of standard stars  prevailing in space can be established with sub-percent 
accuracy 
for 1~nm resolution elements over the wavelength interval 350~nm to 1.05~\textmu m 
\citep{2012SPIE.8450E..1SM}.  

In preparation for the US National Academy's ``Astro2000'' Decadal Survey, 
 \citet{2009astro2010S.155K}       
wrote a White Paper 
entitled ``Photometric Calibrations for 21st Century Science'', where they 
criticised the traditional spectroradiometry in astronomy: ``Our traditional 
standard star systems, while sufficient until now, need to be 
improved and extended in order to serve future astrophysics experiments.'' 
They therefore advocated ``a program to improve upon and expand the current 
networks of spectrophotometrically\footnote{``spectroradiometrically'' in our terminology} 
calibrated stars to provide precise calibration with an accuracy of equal 
to and better than 1~\% in the ultraviolet, visible, and near-infrared 
portions of the spectrum with excellent sky coverage and large dynamic range.'' 

A sub-orbital programme, ``ACCESS: Absolute 
Color Calibration Experiment for Standard Stars'' 
\citep{2008SPIE.7014E..5YK},  
has subsequently been approved. This programme aims to transfer irradiance represented by
absolute detector standards to additional standard stars with 
better than 1~\% precision over the spectral range 350~nm to 1.7~\textmu m 
with a spectral resolving power of ca. 500. It is planned to repeatedly 
fly the ACCESS 
to directly measure the irradiance of six target stars (including Sirius und Vega), 
whereby each of the target stars is to be observed by two separate rocket 
flights to ensure repeatability \citep{2014SPIE.9143E..4YK}. 
Comparing the results of TCF (aim $\pm$ 0.5~\%) with those obtained by 
ACCESS (aim $\pm$ 1~\%) will hopefully lead to a reliable system 
of standard stars. 


\subsection{Vega, Sirius and the process of radiometrically calibrating  
 the Hubble Space Telescope}

\citet{2014PASP..126..711B}    
have presented a well-documented description of the elaborate  
procedures 
that have been used to calibrate {\em HST}. These have led to today's 
most advanced 
database for spectroradiometry from the ultraviolet to the 
mid-infrared wavelength 
range (110~nm to 35~\textmu m), i.e., 
to the CALSPEC database, which will, in future, support the 
operation of the James Webb Space Telescope \citep{2014AJ....147..127B}.  

\begin{figure}[!htb]
\centering \includegraphics[width=0.8\textwidth]{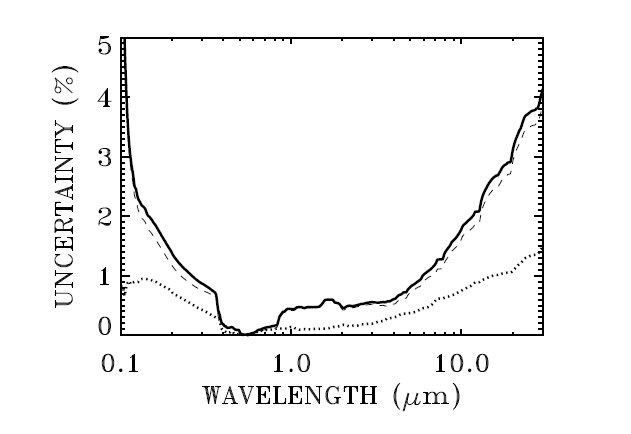} 
\caption{
The uncertainty of the irradiance (beyond that owing to the ground-based 
comparison with a laboratory-based standard at 555.6~nm) for various  
model assumptions. {\em Dotted line}: r.m.s. uncertainty in the white-dwarf 
irradiance scale from formal errors in the effective temperature used in the 
original model. 
{\em Dashed line}: The r.m.s. uncertainty as determined from differences 
between three pairs of each two independently calculated sets of 
pure hydrogen NLTE models,  where the three pairs of models
have the same $T_{\rm eff}$ and log~$g$. Differences arise from 
errors in one or both sets of calculations that should give the same results. 
{\em Heavy solid line}: Combination in quadrature of 
these two uncertainties. Owing to the definition of the scale, the relative 
uncertainty at 555.6 nm is zero by definition. 
(After Figure~14 in \citet{2014PASP..126..711B},   
with permission).}
\label{BohlinetalUncert}
\end{figure}

The calibration started with the above-mentioned ground-based comparisons 
of Vega's 
irradiance at the wavelength of 555.6~nm with standard light sources
and was extended to shorter and longer wavelengths by stellar models.   
Figure~\ref{BohlinetalUncert} illustrates the current r.m.s. 
uncertainty in Vega's irradiance resulting from the use of models over 
the wavelength range 100~nm to 25~\textmu m.\footnote{As pointed out 
before, Vega has now been replaced by the 
primary standard stars 109~Vir in the visible and by Sirius 
($\alpha$~CMa) in the infrared, cf., 
 \citet{2010AJ....140.1919E}.}   
%
%
%

As a complete ground-based end-to-end calibration of the {\em HST} telescope 
and its instruments was not available, Bohlin and his collaborators 
were facing a gigantic task in spectroradiometrically calibrating this 
arguably most important space telescope. They chose an in-orbit calibration 
supported by use of atmosphere models of selected standard stars, supported 
by physics-related measurements that were available.


This meant compounding the difficulties that one has to contend 
with after launch of a perfectly calibrated spectrometric telescope, namely 
\begin{itemize}
 \item a changing responsivity during space-operations, resulting from  
  \begin{itemize}
       \item a changing reflection of optical surfaces, 
       \item   changes in detection efficiencies and in non-linearities, as well as 
       \item a degrading of charge-transfer efficiency (CTE) in CCDs, which 
              may also be complicated by a change of performance across the CCD
              (i.e., a ``changing flat-field''),
   \end{itemize}
\end{itemize}
by the problems inherent in spectroradiometrically calibrating a spectroscopic 
telescope that had not been fully calibrated before launch by relying on 
standard stars, namely 
\begin{itemize}
 \item the complications arising, when the spectral resolution of a standard 
    star's spectral energy distribution differs from that of an unknown star, 
  \item partially unknown non-linearities in detectors, and 
  \item scattered light in the spectrometers. 
\end{itemize}

\begin{figure}[t]
\centering \includegraphics[width=0.6\textwidth]{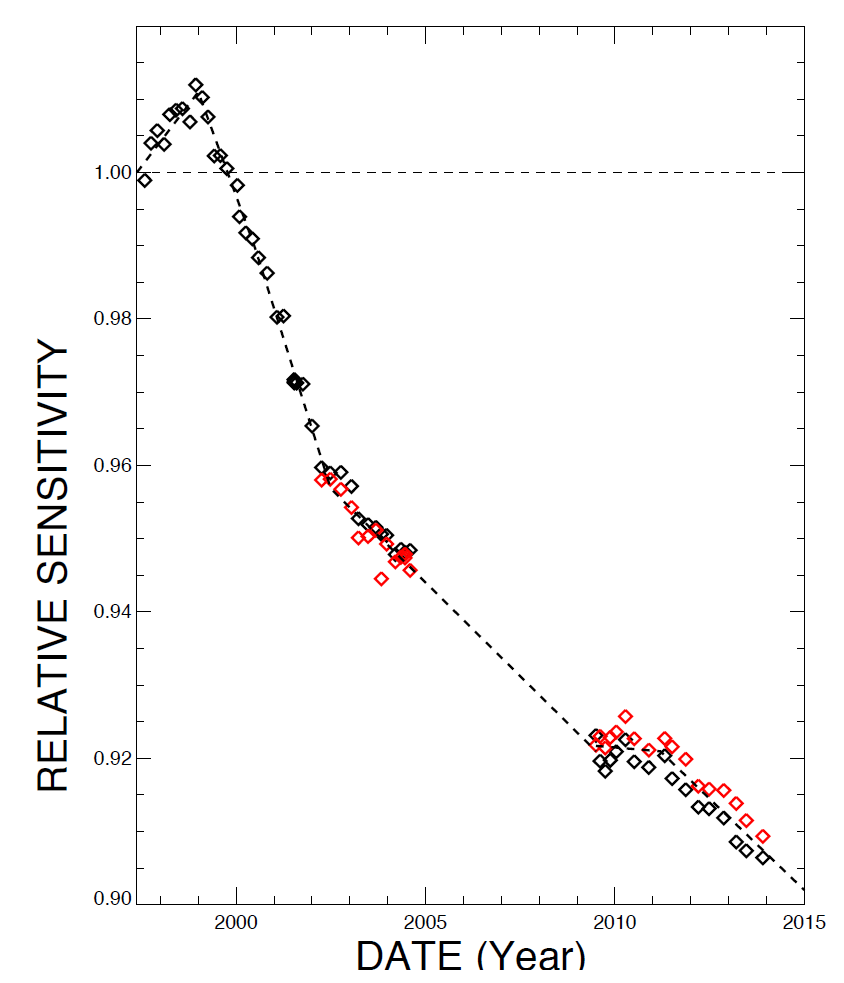} 
\caption{
The time-behaviour of the 
responsivity of the Space Telescope Imaging Spectrograph (STIS) 
during observations in space as observed by monitoring a standard star. 
The black diamonds are data -- relating 
to the wavelength range 200~nm to 300~nm and corrected for charge transfer 
efficiency losses -- from the centre of the CCD. The red diamonds 
correspond to measurements taken at a position near the read-out amplifier. 
The five-year gap in the data corresponds to the waiting time until STIS 
could be repaired during SM4. (After Figure~1 in \citet{2014PASP..126..711B},  
with permission.) 
}
\label{HSTstis}
\end{figure}

 A peculiarity of {\em HST} was that its instruments could be replaced, 
exchanged or even repaired in orbit. Indeed, astronauts visited the 
spacecraft on the occasion of five
Servicing Missions (SM1 (1993), SM2 (1997), SM3A (1999), SM3B (2002), 
and SM4 (2009)). 
Obviously, this brought enormous benefits to the mission, not the least being 
the replacement of one of the instruments by optics 
\citep{1993ApOpt..32.1768B}  
that corrected the initial spherical aberration of the primary mirror. 
Other benefits were the repair of the Advanced Camera for Surveys (ACS) 
and of the Space Telescope Imaging Spectrograph (STIS). On the other hand 
this also required additional efforts for the in-orbit calibration of the 
newly-installed instruments, and prevented continuous monitoring of the 
radiometric performance of defective instruments. Figure~\ref{HSTstis}
illustrates the 
change with time in responsivity that the {\em HST} instrument STIS has 
experienced in orbit: an initial improvement of efficiency and later decay 
of efficiencies, as well as the occurrence of a data 
gap.\footnote{cf.,  
 \url{http://hubblesite.org/the_telescope/team_hubble/servicing_missions.php}, 
as well as the caption of Figure~\ref{HSTstis}. 
 Note, however, that {\em HST} was not the first scientific satellite to be 
repaired in orbit. 
At launch in 1973 the {\em Skylab} space station had lost a so-called 
micrometeorid shield 
during launch. This shield would also have been a thermal shield of the 
astronauts' 
living space. Upon their arrival at {\em Skylab} the first crew was able 
to mount and 
deploy a heat shade, which saved the mission. In 1984 the Solar Maximum Mission 
also underwent repair in orbit, when it was visited by the space shuttle Challenger.
Another example of a data gap stems from the {\em SOHO} mission, which 
experienced an intermediate time-out as well. Contact with the spacecraft was 
lost in August 1998 after a sequence of incorrect commands during what should 
have been a routine manoeuvre. Four weeks later a powerful radar signal from 
Earth produced a faint echo from the spacecraft indicating that {\em SOHO} 
had not 
drifted away from its position in its L1-halo orbit after loss of contact. It 
was slowly rotating and angled in such a way that sunlight was going to fall 
on its solar cells during the following months. Normal operations could then be 
resumed after an extended turn-on and test period. The responsivity of many 
instruments 
had changed -- out-baking in the absence of thermal control had, in fact, 
improved 
the responsivity of some instruments \citep[cf.,][]{2002ISSIR...2.....P}. 
In this context we recall the earlier mentioned reference 
about the outgassing of a spacecraft after launch  
\citep{2010JGRA..11512313S}.}
%

The description by \citet{2014PASP..126..711B}  
of the {\em HST} calibration begins with an 
in-depth discussion of the comparison of stars to ``Laboratory Flux Standards'', 
i.e. ``laboratory-pedigreed'' stellar irradiances. The initial wavelength range 
being discussed is the ultraviolet below the atmospheric cut-off at 330~nm; 
calibration attempts for this wavelength range  started in the late 
1960s, mostly by rocket flights. 

Space astronomers at that time -- many of them being experimental physicists -- were keen 
on working with instruments that had a calibration traceable to laboratory standards. 
\citet{1976ApJ...203..410B}  
judged the initial calibration 
of the Wisconsin equipment package \citep{1970ApJ...161..377C} 
on NASA's Orbiting Astronomical Observatory ({\em OAO-2}) 
to be rather insecure ``because of the long time (18 months) required by spacecraft 
testing between instrument calibration and launch, and because of the feeble far-ultraviolet 
light sources available {to them} in the mid-sixties.'' 
They therefore measured the irradiance of $\alpha$ Vir, $\eta$ UMa, 
and $\alpha$ Leo in several 
pass bands within the wavelength range 137~nm to 292~nm during the 
flight of an Aerobee sounding rocket. 

In a pioneering effort,    \citet{1976ApJ...203..410B}                 
had calibrated the telescope 
for (astro)photometry on this sounding rocket by use of synchrotron radiation as 
standard. The electron storage ring at the physics department of the University of 
Wisconsin with electron-energy of 240~MeV generated a spectrum resembling that of a B5-star, i.e., 
the kind of stars to be calibrated. This largely eliminated problems with scattered light. 
Bless et al performed complete absolute calibrations before and after the flight. 
In addition, they had prepared ``field calibration units'', which provided checks of the rocket 
instrument's responsivity whenever judged beneficial. 
Field checks could be made three days before, and the day after the flight.

\begin{figure}[th]
\centering \includegraphics[width=0.8\textwidth]{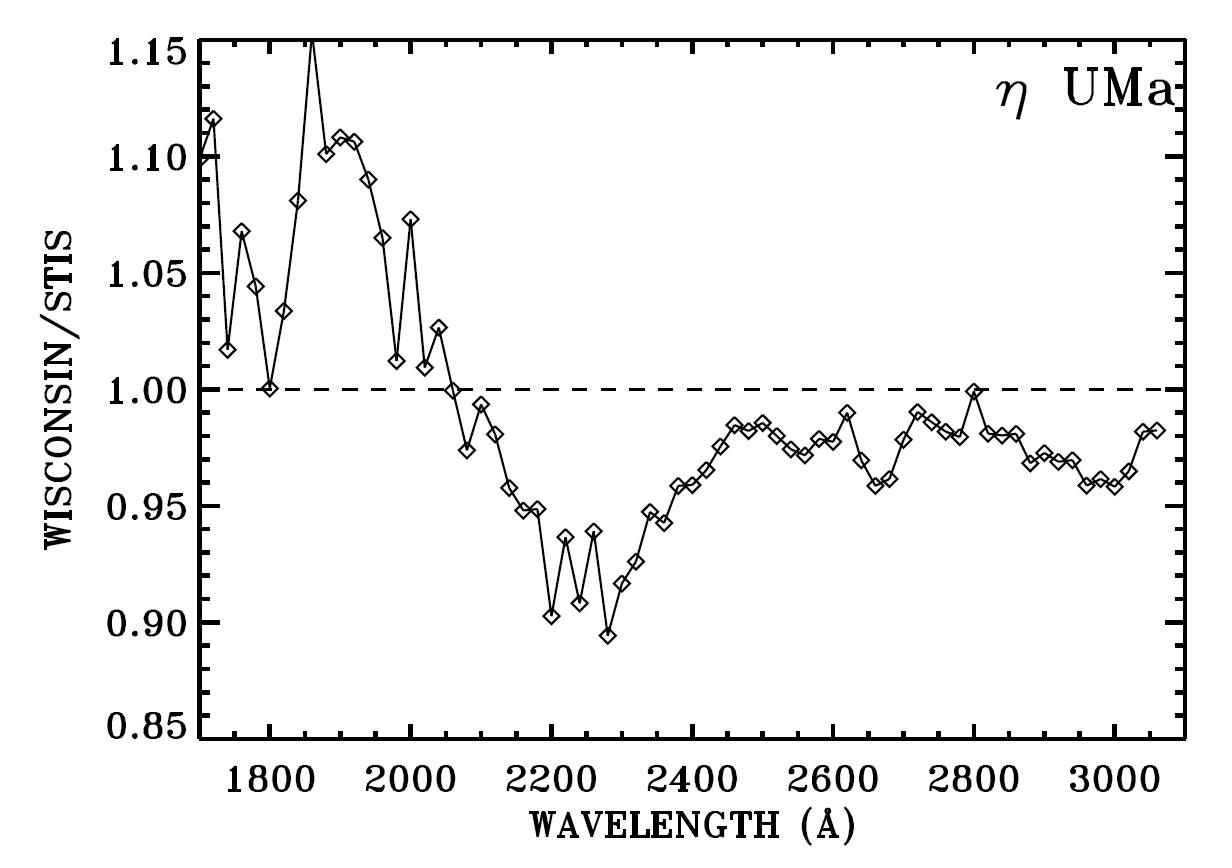} 
\caption{
The spectral stellar irradiance of  $\eta$~UMa  according to the calibration of 
{\em OAO-2} with a rocket flight carrying an (astro)photometric telescope 
that had been calibrated by use of synchrotron radiation is 
compared here with data obtained with {\em HST} using the CALSPEC 
database as calibration file. 
(After Figure~2 in \citet{2014PASP..126..711B}, 
with permission.) 
}
\label{STIS_Uma}
\end{figure}

Bless et al then revised the responsivity of the {\em OAO-2} (astro)photometers 
according to the results of the rocket flight. Figure~\ref{STIS_Uma} 
compares the spectral 
stellar irradiance of  $\eta$~UMa      
obtained in this way with that of the 
modern CALSPEC database. The agreement within roughly $\pm 10~\%$ is remarkable, 
particularly as it also substantiates the original uncertainty estimate. 

In 1975, a series of laboratory-pedigreed, extraterrestrial measurements 
of ultraviolet stellar irradiances that came from groups working in Belgium, 
the Netherlands, the UK, and the US had been available for a comparison. 

\citet{1975SSI.....1..343A}    
at the University in Groningen had calibrated the Astronomical Netherlands 
Satellite ({\em ANS}) based on a detector standard, namely an EMI phototube  
that had been maintained at the University of Wisconsin. 
Their estimated calibration uncertainty was $\pm 30~\%$. 

Calibrations of the Sky Survey instrument on ESA's {\em TD-1} satellite by 
\citet{1976A&A....49..389H}   
 had been performed independently in Edinburgh and Li\`ege, based on 
fundamentally different laboratory standards. The uncertainty 
estimate of the overall calibration was   $\pm 20~\%$. 

A Johns Hopkins instrument  \citep{1975ApJ...201..613H}   
mounted in the service module of {\em Apollo~17}, observed low-resolution 
ultraviolet 
spectra from lunar orbit. It had been calibrated  by use of photodiodes 
provided by the NBS.\footnote{The U.S. National Bureau of Standards, 
now called National Institute of Standards and Technology (NIST).}
The estimate of the calibration uncertainty was   $\pm 10~\%$. 
There was a decline of the responsivity owing to the substantial 
flux of ultraviolet radiation reflected from the lunar surface, 
which could, however, be corrected.

At Colorado, \citet{1974A&A....30..127B}   
had measured the spectral irradiance of $\alpha$~Lyr, $\eta$~UMa,  
and $\zeta$~Oph in the wavelength 
range 170~nm to 340~nm by a rocket flight in 1972. 
The standards used in this 
case were a tungsten standard lamp and two caesium-telluride photodiodes,  
both calibrated by the NBS. \citet{1979PASP...91..205S}         
later revisited and corrected these results, and estimated that the 
uncertainties of the corrected results ranged from \mbox{$+27~\%/-17~\%$} at 175~nm 
to \mbox{$+15~\%/-12~\%$} at 335~nm.

\begin{figure}[!htb]
\centering \includegraphics[width=0.6\textwidth]{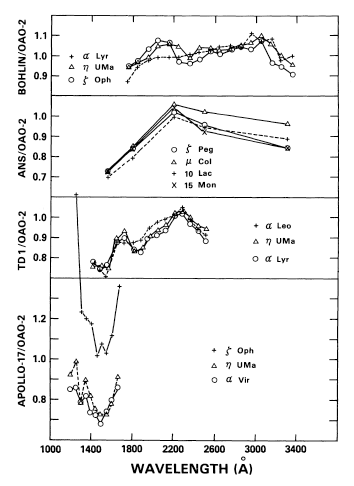} 
\caption{
A comparison of absolute ultraviolet irradiances of four independent 
space telescopes with those derived from the {\em OAO-2} satellite \citep{1976ApJ...203..410B}.    
Bohlin:  \citet{1979PASP...91..205S}; 
ANS:   \citet{1975SSI.....1..343A};  
TD-1: \citet{1976A&A....49..389H};   
Apollo-17:  \citet{1975ApJ...201..613H}. 
Note that the estimated uncertainties of the various measurements, 
with the exception of those by Henry et al turned out to be realistic. 
Note however, that the latter had singled out the data points relating to 
 $\zeta$~Oph as being particularly uncertain. 
(After Figure~4 in   \citet{1979PASP...91..205S}, 
with permission.) 
}
\label{oao2}
\end{figure}

The four graphs of Figure~\ref{oao2} show ratios 
between the absolute ultraviolet 
irradiances of stars measured by the four experiments described above 
and the corresponding ones obtained by {\em OAO-2}. Taken together, 
these early measurements indicate that the uncertainty estimates 
were realistic, with one exception, but, unfortunately, still rather sizeable. 

This discouraging situation began to improve after observers had studied 
white dwarfs with {\em IUE}. \citet{1979ApJ...229L.141G}    
recognised from studies with the {\em IUE} spectroscopic telescope that 
effective temperatures of degenerate stars -- determined by fitting line profiles of the 
Ly-$\alpha$ absorption -- matched those obtained by (astro)photometry in the visible, 
and also those obtained from the {\em ANS} in the ultraviolet,  
rather well. They thus concluded that an adaptation of the {{\em IUE} calibration was needed. 
They presaged, ``If we accept that theoretical predictions should be correct, 
corrections to the absolute {\em IUE} calibration derived are an upward shift of (3 to 5)~\%, 
with irregular residuals attaining $\pm 7~\%$.''\footnote{The idea that ``if we accept 
that theoretical predictions should be correct ...'' was later followed, and has been 
guiding the calibration of {\em IUE} and {\em HST} up to today. Improvements of the models 
(improved gravity values, non-LTE calculations, for example), which have taken place in the 
mean time will be mentioned below.}   
%
%
\citet{1984NASCP2349..277F}    
later proposed a change of the responsivity 
of {\em IUE} by about ten per cent, after they, too, had found that effective 
temperatures of white dwarfs obtained from visual magnitude measurements 
and ultraviolet irradiances measured by {\em IUE} were incompatible. 
\citet{1986ApJ...306..629H}  
confirmed these earlier conclusions. 
They had determined temperatures of the atmospheres of white 
dwarfs by fitting Ly-$\alpha$ profiles obtained from {\em IUE} observations and 
derived from them a correction of the spectroradiometric calibration of {\em IUE}. 
Following a further analysis of the spectra of seven hot DA white dwarfs, 
\citet{1990ApJ...359..483F}    
suggested that white-dwarf calibration should 
serve as the calibration for  {\em IUE} (and  {\em HST}). They thus affirmed 
Greenstein and Oke's 1979 ``conjecture'', namely that if one accepted 
``that theoretical predictions should be correct'', this would be 
an expedient for the absolute calibration of  {\em IUE}.

In preparation for the launch of {\em HST},  \citet{1990ApJS...73..413B} 
presented 37 ``standard stars'', whose absolute ultraviolet spectral irradiance 
distributions for the wavelength range 115~nm to 330~nm had been derived by use 
of the latest in-orbit calibration of {\em IUE}. There was agreement with the 
{\em OAO-2}, 
{\mbox{\em TD-1}} and {\em ANS} data to a few per cent for stars in common. The 
transition to ground-based data near the atmospheric cut-off between 
325~nm and 340~nm 
\citep{1975ApJ...197..593H, 1983ApJ...266..713O}  
implied an uncertainty of about 3~\%. These 37 standard stars, 
which had an appropriate 
distribution across the sky, a range of $10^{4}$ in brightness 
and a good coverage 
of spectral types, defined the first ultraviolet calibration 
of {\em HST}. At the time, a comparison with white-dwarf model atmospheres 
indicated an agreement of about 10~\% to 15~\%.

At the 1993 {\em HST} Calibration Workshop, which took place shortly before the 
originally intended imaging performance of {\em HST} was to be restored during 
{\em HST} Servicing Mission~1, a consensus had developed that one should set 
the radiometric calibrations of {\em IUE} and {\em HST} onto the same scale, 
and also that spectra of further white dwarfs should be obtained for 
the purpose of improving the absolute calibration with the {\em HST}  
Faint Object Spectrograph   
\citep{1994BAAS...26Q1212B}.  

\citet{1995AJ....110.1316B}    
then performed such observations on three additional white dwarfs with the 
Faint Object Spectrograph of {\em HST}, and showed that the spectral irradiance 
of these three newly observed stars agreed with the model spectra within 
about 2~\%. 
As the (astro)photometry in the blue and visual agreed within better than 
1~\% in the 
average as well, these authors recommended that four white dwarfs be primary 
reference standard stars, and recommended their use for all ultraviolet and 
optical absolute calibrations over the wavelength range from 100~nm 
to 1~\textmu m.\footnote{This recommendation, with an extension into the 
infrared, was made in view of the STIS, 
which covered the wavelength range 115~nm to 1~\textmu m and was eventually 
installed on board {\em HST} during SM~2 in 1997. The 
data for the longer wavelengths were based on observations by 
\citet{1990AJ.....99.1621O}    
with the 5.1~m Hale telescope on Palomar Mountain that were later 
slightly corrected by use of {\em HST} FOS observations  
\citep{1994AJ....108.1931C}.  
} 
The standard stars in question are G191-B2B, GD~71, GD~153 and HZ~43, whereby 
ground-based observers should note that HZ~43 has a red 
companion about $3\arcsec$ away.

In 1997, Kruk et al 
published reassuring results: they had observed the hot dwarf star G191-B2B with the 
Hopkins Ultraviolet Telescope during the Astro-1 space shuttle mission in 1990. 
HUT's postflight calibration was based on synchrotron radiation and 
NBS photodiodes 
as laboratory standards, and showed that the synthetic spectrum of G191-B2B and 
the effective area of HUT gave results that were consistent within the laboratory
 measurement uncertainties. Kruk et al thus concluded, ``These results validate 
the use of white-dwarf models as absolute flux standards in the far-ultraviolet, 
especially for wavelengths shortward of Ly-$\alpha$, where previous experiments 
often disagreed by large factors.''

The calibration for the Astro-2 mission of HUT confirmed these findings: 
synthetic spectra -- based on model parameters derived from fits to 
ground-based spectra --
differed from the observations interpreted by use of the laboratory calibration by 
less than 3~\% at all wavelengths (except at the cores of the Lyman lines). 
In a consistency check, the spectrum of an additional star (BD +75\gr\ 325) measured 
by HUT was found to differ from that observed by {\em HST}'s 
Faint Object Spectrograph 
by at most 5~\% in the wavelength region covered by both observations. 
\citet{1999ApJS..122..299K}  
therefore declared that their results being ``consistent within 
the laboratory measurement 
uncertainties, demonstrate that pure hydrogen white dwarfs ... may be used as 
primary flux standards down to the galactic Lyman edge.'' 

Another experimental confirmation of the white-dwarf calibration quality has been 
reported by \citet{2013ccfu.book..191S}:  
 the Solar Stellar Irradiance Comparison Experiment (SOLSTICE) on the 
{\em SORCE} (Solar Radiation and Climate Experiment) spacecraft is an instrument 
\citep{2005SoPh..230..225M}   
that is aimed at both solar and stellar irradiance measurements. It observes the 
Sun and stars by the same optics and detectors by changing the size of the 
entrance aperture and of the exit slit. SOLSTICE was calibrated over the wavelength 
range 115~nm to 320~nm on the SURF-III storage ring with an uncertainty of 
about 3~\% and transferred 
this laboratory calibration in an early on-orbit observing campaign to 18 stars, before 
any observations of the Sun would initiate a degradation of the responsivity 
\citep{2005SoPh..230..259M}.   
Figure~\ref{solstice_stis} shows a comparison of absolute irradiances measured by 
SOLSTICE for three common stars in the CALSPEC database, and again supports both 
the white-dwarf calibration procedure and the realistic uncertainty estimate of 
the SOLSTICE laboratory calibration. 

\begin{figure}[ht]
\centering \includegraphics[width=0.8\textwidth]{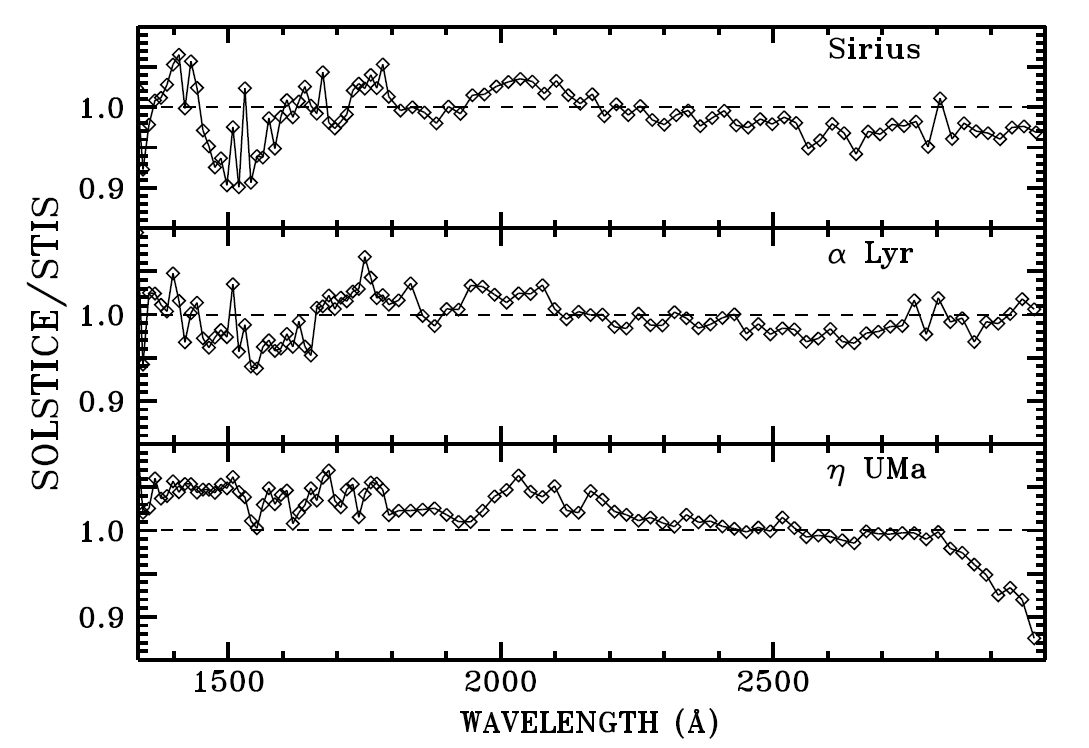} 
\caption{
Comparison of SOLSTICE irradiances for three stars with CALSPEC 
baseline data (STIS/{\em HST} irradiances for 
$\lambda > 170$~nm and {\em IUE} 
irradiances at shorter wavelengths) demonstrates that the values 
of 
the synchrotron-based irradiances of SOLSTICE and those of STIS 
rarely differ by more than $\pm 6~\%$, i.e., 
2~$\sigma$ of the uncertainty of the laboratory calibration of SOLSTICE. 
(After Figure~6 in    \citet{2014PASP..126..711B},        
with permission.) }
\label{solstice_stis}
\end{figure}

In the above text we have addressed the ultraviolet wavelength range in detail, 
because it was the first domain to be spectroradiometrically calibrated and 
also 
because it may be the experimentally most demanding area, when one tries to 
calibrate 
the responsivity of traditional space telescopes, i.e., telescopes giving 
access to 
the ultraviolet, visible, and infrared ranges of the electromagnetic spectrum. 
The calibration of vacuum-ultraviolet instrumentation was initially 
also the most 
uncertain, not the least because contamination of the surfaces of 
optical elements 
and detectors was (and actually still is) difficult to control. 
The visible part of 
the spectrum and its extension to ca. 1~\textmu m will soon have a
 more accurate 
calibration when data from the NIST TCF \label{TCFpage} 
and from the ACCESS 
will become available. 
As mentioned in the introductory paragraph, spectroradiometry 
in the infrared out to 
35~\textmu m seems to be settled at an uncertainty level for the irradiance of 
standard stars around two to three per cent thanks to the untiring efforts 
of the community  \citep{2008AJ....135.2245R}  
and particularly the series 
of 14 papers initiated by Martin Cohen   
\citep{1992AJ....104.1650C}.\footnote{Note that the Astrophysics Data System ADS 
gives access to all papers in a series, whenever one calls up one of the 
individual papers.}
Realistically, however, the physics-related measurements on the Midcourse Space 
Experiment reported in the last three papers of the series 
\citep{2004AJ....128..889P, 2006AJ....132.1445E, 2010AJ....140.1919E}  
 tied down the calibration scale to the two-per-cent uncertainty level. 

Rather than describing the extended efforts that went into modelling stellar 
atmospheres, particularly those of white dwarfs with hydrogen 
atmospheres -- exploiting non-LTE effects, determining effective temperatures 
of stellar atmospheres from line-profile analysis, and deriving stellar gravity 
via stellar-diameter measurements -- we refer to  \citet{2014PASP..126..711B}. 

The outcome of these efforts is the CALSPEC archive \citep{2014AJ....147..127B},  
which presents a standard calibration scale for use with {\em HST} and future {\em JWST} 
observations and which reaches from the vacuum ultraviolet far into the infrared. 
The calibration scale is backed up by an extensive modelling effort. The spectrum of Sirius 
calculated by 
\citet{1979ApJS...40....1K, 1993yCat.6039....0K,2013ascl.soft03024K}\footnote{cf., \url{http://kurucz.harvard.edu/stars/}}
for a model with \mbox{$T_{\rm eff}$ = 9850~K},  \mbox{log $g$ = 4.3} and an abundance ratio 
 \mbox{[M/H] = $+ 0.4$} 
agrees extremely well with the spectrum of Sirius as shown by Bohlin. 
Remarkably, because the spectrum was calculated long before CALSPEC was established, 
the original 1993 Kurucz model agrees with the 2013 update within better than 1~\% 
longward of 180~nm. At shorter wavelengths, the update removes some emission lines and, 
as a consequence, now also fits the observed irradiance significantly better. 

As a reminder of the difficulty of radiometric calibration, we may recall in conclusion 
the optimistic view that Bless et al  expressed in 1967, ``It appears likely that 
within the next few years the absolute energy distribution of stars will be as well 
known in the ultraviolet as in the visual region of the spectrum.'' Although the 
path to knowing the absolute energy distribution in the visible on the 0.5~\% or at least 1~\% level 
is still ahead of us, the way to a somewhat satisfactory knowledge of the ultraviolet and 
infrared calibration was much longer, and not principally guided by laboratory measurements. 
Until a calibration based exclusively on laboratory standards is available, we must remember 
that the calibration of space telescopes is still resting to a considerable amount on 
the use of stellar model atmospheres, and, as observations are often supposed to test 
new models, we must be aware that one has arrived at the hitherto derived calibrations 
by what, in effect, is a giant bootstrap procedure. This inevitably will lead to testing of 
models by models, which may lead to circular conclusions.




%% file: examples4.tex
\section{Examples of calibrated instruments aboard spacecraft}        

\subsection{X-ray instruments} 


High-resolution X-ray spectroscopy of cosmic sources is the 
key to understanding physical processes in hot astrophysical plasmas: it 
provides the essential tools for determining temperatures, densities, element 
abundances and ionisation stages, and flow and turbulent velocities.  
Concomitant studies of absorption features provide information about 
line-of-sight interstellar gas and dust.  The absolute radiometric 
responsivities of 
X-ray telescopes and spectrometers cannot be unambiguously determined by use 
of celestial standards because absorption by interstellar gas confounds 
comparisons with model spectra, because the models themselves are 
compromised by trace amounts of helium and metals in the real atmospheres, 
and, finally, and perhaps most fundamentally,  
because most astronomical sources of EUV and X-rays are variable. 

As a consequence of such complications, pre-launch spectroradiometric 
responsivity  
characterisation and effective prevention of contamination of the 
optical elements of the flight models 
are essential for astronomical EUV and X-ray missions.  
The accuracy of such determinations directly limits the 
astrophysical analyses made from them: in order 
to derive the differential emission measure or the elemental 
abundances and ionisation stages, relative line and continuum 
intensities must be known to within a few percent.  
As a further 
example, the 
 studies 
of the Sunyaev-Zel'dovich effect require $\pm$1~\% radiometric 
accuracy.  Such constraints drive the calibration requirements.

Instruments are designed to image far-away sources through the 
entire entrance aperture. 
Thus, ideally,  
spectroradiometric calibrations of telescope-spectrometer combinations 
would involve irradiation of the full telescope mirror 
with a parallel beam of radiation with known spectral content.  
In the absence of an appropriate X-ray collimator this is 
difficult to achieve in practice, and leads to a  finite 
source-to-telescope distance in the 
laboratory with the corresponding geometries. 
In general, 
the calibration is performed 
at long-beam X-ray facilities, i.e., with a limited source 
distance, and the on-ground calibration has to be extrapolated to in-orbit conditions. 
%
%

Moreover, after launch, the radiometric calibration of high-energy astronomy missions 
quite often differs from that measured on the ground -- not unlike the situation 
found in other wavelength ranges. As a rule, further changes then occur in the 
course of the mission, as frequently observed during science operations. 
In 2006, therefore, an International Astronomical Consortium for High Energy
 Calibration (IACHEC)\footnote{Updated general information on IACHEC is available 
on the web site \url{http://web.mit.edu/iachec/}. Results of the IACHEC collaboration 
are published as refereed papers and made accessible through 
\url{http://web.mit.edu/iachec/papers/}.} 
was founded with the aim of harmonising the 
calibration of high-energy astrophysics missions. This forum is  
endorsed and led by the user group of the two major missions,  
{\em XMM-Newton} and 
{\em Chandra}, and is funded by the experimenters of participating missions. 
In the framework of IACHEC, a sustained effort is undertaken to 
inter-calibrate high-energy missions by use of a set of ``standard candles''
and, in this way, to minimise calibration uncertainties 
\citep{2010AIPC.1248..593S}. 
%
%
%
In the following, 
we will trace the development of instrumentation and calibration methods 
in the course of time by describing
some example missions with their calibration measures and status.


{\it Uhuru}  
\citep[1970 to 1973;][]{1971ApJ...165L..27G}
was the first satellite launched specifically for the purpose of 
X-ray astronomy, using 
two sets of proportional counters and observing in the range from 2~keV to 20~keV.  
Information on this and other previous missions can be found on 
the webpage of the 
High Energy Astrophysics Science Archive Research Center (HEASARC) which is the primary 
archive of  NASA's 
(and other space agencies') missions studying electromagnetic radiation 
from extremely energetic cosmic phenomena.\footnote{\url{http://heasarc.gsfc.nasa.gov/}}  
In particular its database  
stores information 
(such as targets and sequences) and files  
for calibration and cross-calibration 
of high-energy astronomical 
instrumentation.\footnote{\url{http://heasarc.gsfc.nasa.gov/docs/heasarc/caldb/caldb_intro.html} 
and \url{http://heasarc.gsfc.nasa.gov/docs/heasarc/caldb/caldb_xcal.html}}


ESA's {\it EXOSAT} 
\citep[1983 to 1986;][]{1981SSRv...30..479T}\footnote{\url{http://www.cosmos.esa.int/web/exosat/home}} 
covered the energy range  between 0.05~keV  and 50~keV. It was followed by 
the {\it ROSAT}  mission 
\citep[R\"ontgensatellit, 1990 to 1999;][]{1981SSRv...30..569A},  
carrying a German-built imaging X-ray Telescope (XRT) with 
three focal-plane instruments, two 
Position Sensitive Proportional Counters (PSPC), 
and the US-supplied High Resolution Imager (HRI). The X-ray mirror assembly was a 
grazing-incidence four-fold nested Wolter~I telescope with an 84~cm diameter aperture and 240~cm 
focal length. 
The XRT assembly was sensitive to X-rays between 0.1~keV and 2~keV.
In addition, a British-supplied extreme ultraviolet telescope,  
the Wide Field Camera (WFC), was coaligned with the XRT and covered 
the energy band from 0.042~keV to 0.21~keV. 
Calibration 
had, to a large part, been performed at the 
PANTER facility\footnote{\url{http://www.mpe.mpg.de/heg/panter}} 
of the Max-Planck-Institut f\"ur extraterrestrische Physik (MPE) in Garching 
\citep{2005ExA....20..405F},   
documents are available at the websites of the 
MPE\footnote{
\url{http://www.mpe.mpg.de/xray/wave/rosat/doc/calibration/index.php}} 
and  
HEASARC.\footnote{\url{http://heasarc.gsfc.nasa.gov/docs/heasarc/caldb/caldb_docs_rosat.html}} 
 The {\em ROSAT} Users' Handbook 
\citep{rosatbook}\footnote{\url{http://www.mpe.mpg.de/xray/wave/rosat/doc/ruh/rosathandbook.php}} 
 contains additional, more detailed information. 

The Italian-Dutch {\it BeppoSAX}  mission  (1996 to 2003)  comprised five 
instruments to cover the range from 0.1~keV to 300~keV, 
four X-ray telescopes working in conjunction with one of the following detectors: 
the Low Energy Concentrator Spectrometer (LECS), the Medium Energy Concentrator 
Spectrometer (MECS), 
the High Pressure Gas Scintillator Proportional Counter (HPGSPC), and the Phoswich 
Detection System (PDS),  
as well as two units of a  Wide Field Camera (WFC).\footnote{Additionally, it carried a 
Gamma-Ray Burst Monitor (GRBM) of four CsI(Na) scintillators
that were also used as active lateral shields of the PDS experiment. Its calibration 
is described by \citet{1997SPIE.3114..176A}. 
}
%
Ground calibration of the LECS and MECS had been done at beamlines of the BESSY and PANTER  
facilities, component by component, and at the instrument level 
 \citep{1997A&AS..122..309P,1997A&AS..122..327B}. 
The HPGSPC and the PDS had been calibrated by use of radioactive sources; 
the PDS included a movable calibration source  Co$^{57}$    
\citep{1997SPIE.3114..216S}. 
The Crab nebula and supernova remnants were 
used for flux cross-calibration 
in orbit.\footnote{\url{http://www.asdc.asi.it/bepposax/calibration.html},  
\url{https://heasarc.gsfc.nasa.gov/docs/sax/sax.html}}
%

The Rossi X-ray Timing Explorer, 
{\em RXTE} 
\citep[1995 to 2012;][]{1993A&AS...97..355B,1998ApJ...496..538R} 
measured in the energy range between  2~keV and  250~keV, by use of  
a Proportional Counter Array (PCA), the High Energy X-ray Timing Experiment (HEXTE), and 
an All-Sky Monitor (ASM). 
The HEXTE consisted of two clusters of four phoswich scintillation detectors, fast switching 
between background and targets with a sampling time of eight microseconds.  
Automatic gain control was provided by comparison with a radioactive source ($^{241}$Am)  
mounted in each detector's field of view. 

	


{{\em Suzaku} (formerly {\em ASTRO-EII}) is a Japanese X-ray astronomy satellite launched in 2005. 
%
It is carrying high-resolution, 
wide-band instruments for 
detecting signals ranging from soft X-rays up to gamma-rays 
(0.3~keV to 600~keV).\footnote{\url{http://www.astro.isas.jaxa.jp/suzaku/}}
The payload systems are the 
    X-ray Spectrometer (XRS), an X-ray calorimeter,  
   the  X-ray Telescope (XRT), the 
    X-ray Imaging Spectrometer (XIS), and the 
    Hard X-ray Detector (HXD). 
The X-ray calorimeter of {\em Suzaku}  was the first such instrument 
to be flown in space and unfortunately stopped operation soon after 
launch due to failure of the cooling system.
Description and database of the calibration are given at the 
webpages.\footnote{\url{http://www.astro.isas.ac.jp/suzaku/caldb/}, 
\url{http://www.astro.isas.jaxa.jp/suzaku/process/caveats/}} 



The X-ray Multi-Mirror Mission \citep[{\em XMM-Newton};][]{2001A&A...365L...1J}
was launched  in 
1999 as the second ``cornerstone'' of ESA's 
Horizons 2000 Programme \citep{ESA94}. 
This space observatory performs 
high-throughput spectroscopy of cosmic X-ray sources. Its effective 
areas -- over 
nearly its entire spectral domain, 0.1~keV to 10~keV (12.5~nm to 0.125~nm) -- 
are approximately 1000~cm$^2$ and 
100~cm$^2$, respectively, for 
spectrometric imaging and high-resolution spectroscopy.  
{\em XMM-Newton} was  calibrated by a combination of the three 
calibration methods, ($i$) through ($iii$),  
enumerated in Section~\ref{sec2rcal}. There were pre-flight calibrations at the component and 
sub-system level and in end-to-end 
configurations. 
%
Specifically, the three telescope assemblies, each consisting of 58 Wolter type-I mirrors 
in a nested co-axial and confocal configuration, were tested in the 
PANTER facility of the MPE. 
Comprehensive numerical modelling was used to 
generate a calibration database from which the laboratory tests, taken with a 
finite source distance, could be extrapolated to in-orbit conditions 
\citep{1996SPIE.2808..390G,1998SPIE.3444..290G}. 
%
The spectroradiometric calibration 
of {\em XMM-Newton} is  monitored in orbit by regularly observing a selected 
set of sources and cross-calibration is carried out with contemporary X-ray observatories. 

All three telescope assemblies have CCD arrays -- the European Photon 
Imaging Cameras (EPIC) -- in their focal plane. Two of them are equipped 
with MOS-CCDs \citep{2001A&A...365L..27T}, 
and one uses pn-CCDs \citep{2001A&A...365L..18S}. 
These cameras had also been 
calibrated in several steps. First, their quantum efficiencies were determined by use of 
synchrotron light sources. Then, the effective area of the telescope-camera combination 
was calibrated in the PANTER facility. However, to minimise contamination of the flight mirrors, 
the end-to-end calibrations were carried out with the qualification-model rather than with the 
actual flight-model mirror assemblies. 

The two Reflection Grating Spectrometers on board {\em XMM-Newton} 
\citep{2001A&A...365L...7D}  
are objective spectrometers, i.e., they have no slits. Each of them 
collects radiation from 
half the aperture of the two telescope assemblies that feed 
the MOS-CCDs. 
The third telescope assembly is unobstructed by gratings and feeds the 
pn-CCD camera. Each spectrometer uses an array of 182 plane gratings with 
variable ruling-frequency. Their calibration was performed in three stages. First, the individual 
gratings and the CCD detectors were characterised. Then 
measurements were made of the assemblies of the two reflection grating arrays 
with their focal-plane cameras. Finally, the effective area was determined -- 
again in the PANTER facility -- in end-to-end tests. With the finite source distance, the mirror 
assemblies and the reflection-grating array were under-illuminated, and, consequently, the measured 
effective area had to be corrected by use of ray-tracing calculations. The results on the overall 
effective area (measured on axis at six energies, but at one energy only off axis) agreed with 
the model calculations to within 5~\% in the first and second grating order 
\citep{2001A&A...365L...7D}. 
%
%
%
{\em XMM-Newton} also carries internal X-ray sources ($^{244}$Cm).  
Although 
these do not permit an end-to-end calibration in orbit, they are 
useful for monitoring 
changes in the stability of the instrument responsivities. In particular 
they can be used for absolute energy calibrations, to establish the system gain at the 
time, and under the conditions of, a given observation. In principle, the internal sources also permit 
the measurement of radiation damage in orbit, manifesting itself in a loss in 
charge-transfer efficiency. 
However, the damage monitoring turned out to be considerably more 
complicated in the {\em XMM} case 
because the calibration source is also cool and thus a sink for contaminants.  
%

An overview of the recent calibration status is given by 
\citet{2015A&A...573A.128D} 
and gives an uncertainty of 10~\% for 
the  effective area in the energy range 0.5~keV  
to 1~keV, somewhat larger for higher-energy values, and up to  
40~\% at the low-energy end. 
%
Water vapour contamination, 
resulting in 
a stable oxygen edge in the instrument response, has been observed, 
most probably due to 
the long period between purging the instrument, sealing, and the launch. 
Additionally, loss of effective area at longer wavelengths 
increasing over time, attributed to out-gassing of hydrocarbons by
the carbon-fiber-reinforced  
structures of the telescope tube, has been found. 

Cross-calibration between the different {\em XMM-Newton} 
instruments and between 
{\em XMM-Newton} and 
various other observatories such as 
{\em Chandra, Suzaku, Swift, BeppoSAX, RXTE},   
are being performed on a regular basis (see, e.g., the 
{\em XMM} calibration portal and the documents found 
there).\footnote{\url{http://xmm2.esac.esa.int/external/xmm_sw_cal/calib/}} 
%
Simultaneous observations of a regular X-ray burster, GS 1826-238, with 
{\em Chandra} 
and {\em RXTE} 
resulted in 
agreement of the fluxes measured by  PCA/{\em RXTE} and {\em Chandra}  
within their formal uncertainties;  
{\em XMM-Newton}'s EPIC-pn measured (14.0$\pm$0.3)~\% less flux than 
the PCA/{\em RXTE} \citep{2015arXiv150105330G}. 



The {\em Chandra} X-ray Observatory \citep[{\em CXO}, launched 1999;][]{2002PASP..114....1W}, 
formerly known as the Advanced X-ray Astrophysics
Facility {\em AXAF}},  
carries a High Resolution Mirror Assembly (HRMA), two focal-plane instruments, the High 
Resolution Camera (HRC) and the Advanced CCD Imaging Spectrometer (ACIS), and two sets 
of transmission gratings, the High Energy Transmission Grating (HETG) and the Low Energy 
Transmission Grating (LETG).  The gratings can be moved in and out of the optical paths 
to both detectors, permitting both spectroscopy and imaging to be performed.  
The 
combination of high resolution, large collecting area, and sensitivity 
to higher-energy 
X-rays makes it possible for {\em Chandra} to study extremely faint 
sources 
in crowded fields whose lower-energy 
radiation is sometimes strongly absorbed. Since transmission -- 
away from absorption edges -- increases with increasing energy, 
higher-energy X-rays are less affected by obscuring material so 
that the irradiance of the sources can be determined. At the same time the 
hypothesis that some of these sources may be ``standard candles'' can 
be tested. And, if such ``standard candles'' are found and calibrated, 
distances to nearby galaxies can be determined. 
These distances are a crucial step in the derivation of the Hubble Constant.

{\em Chandra} has been calibrated at the X-ray and Cryogenics Facility 
(XRCF)\footnote{\url{https://optics.msfc.nasa.gov/}}  
at the NASA Marshall Space Flight Center. 
The determination of the effective areas was a major undertaking.  
In order to meet its science objectives, the {\em CXO} project started with a calibration 
goal of determining the effective areas of the various combinations of telescope mirror 
and focal-plane science instruments with overall uncertainties at the $\pm$1~\% level 
\citep{1997SPIE.3113..515K} 
through an extensive programme of sub-system and end-to-end measurements 
and modeling similar to that discussed above for {\em XMM}.  The effective area of 
the HRMA is given by \citet{2000SPIE.4012...28S}; 
that for the ACIS and HRC focal plane detectors is given by 
\citet{ 2000SPIE.4012...53B}  
and \citet{2000SPIE.4012...68M},  
respectively.
The website provides links to references and the calibration 
status, as well as to the most recent calibration workshops.\footnote{\url{http://cxc.harvard.edu/}} 



The NASA Small-Explorer (SMEX) mission {\em NuSTAR} 
\citep[Nuclear Spectroscopic Telescope Array, 
launched in 2012,][]{2013ApJ...770..103H}\footnote{\url{http://www.nustar.caltech.edu/}}  
is, for the first time, using focusing optics in the  high-energy X-ray domain  
(3~keV to 79~keV) 
 instead of coded apertures. 
It consists of two co-aligned grazing-incidence telescopes 
(of ten metre focal length) with specially coated optics 
and newly developed (hybrid pixel) detectors. 
For the ground calibration of the optics, the Rainwater Memorial 
Calibration Facility (RaMCaF) has been built, providing a long (175~m) 
high-energy beamline in order to approximate the situation in space 
\citep{2011SPIE.8147E..0IB, 2012SPIE.8443E..1YB}. 
In-flight calibration has largely been using the Crab nebula 
\citep[][and references therein]{2014SPIE.9144E..1PM}. 


\subsection{EUV and FUV instruments} 

Although primarily designed for observation of the atmospheres of solar system planets, 
the {\em Voyager} spacecraft 
\citep[launched 1977;][]{1977SSRv...21..183B} 
made many of the first 
observations of hot stars at EUV wavelengths.  Spectroradiometric 
accuracy was a 
major goal of this work, but uncertainties in calibrations -- based on sounding 
rocket observations -- plagued the results 
\citep[see][and discussion therein]{1991ApJ...375..716H}. 

Before the Hopkins Ultraviolet Telescope HUT (operated on two 
Space Shuttle missions in December 1990 and March 1995, 
cf., Section~\ref{sec3}), 
however, there were no other useful 
spectroradiometric observations at these wavelengths from spacecraft.  The 
calibrations of the {\em ORFEUS~I} and~{\em II} missions\footnote{Orbiting 
Retrievable Far and Extreme Ultraviolet Spectrometers, 
1993 and 1996 on the Astro-SPAS, a reusable shuttle-launched space platform 
\citep{1998ESASP.413..757G}},   
for example, were not based on direct calibration 
but on comparison to modelled spectra of G191-B2B and HZ~43, respectively 
\citep{1998ApJ...500L...1H, 1999A&AS..134..561B}. 
A comparison of fluxes from 
targets common to both  {\em ORFEUS} 
missions indicated that the calibration of the Berkeley 
spectrometers differed by about 10~\% \citep{1998ApJ...500L...1H}.   
The  HUT observations  
\citep{1997ApJ...482..546K,2013PASP..125..431D} 
showed that the final {\em Voyager} 
calibration was accurate to $\pm$10~\% over the wavelength ranges 91~nm  
to 118~nm. 



The performance of the Extreme Ultraviolet Explorer 
\citep[{\em EUVE},\footnote{\url{http://www.ssl.berkeley.edu/euve/index.html}} 1992 to 2001;][]{1991AdSpR..11..205B}, 
which comprised four 
grazing-incidence telescopes with a number of thin, metal-film filters that 
defined six, partially overlapping, passbands between 6~nm and 75~nm, is discussed 
by \citet{1997ApJS..110..347S}.  
The pre-launch effective area of {\em EUVE} had been determined 
from laboratory measurements at discrete wavelengths by use of a transfer-standard 
photodiode.  Effective-area values between the calibration points were modelled 
from measured values of mirror reflectivities, filter transmission, and detector responsivity.   
The differences between the measured and modelled effective areas were about $\pm$30~\%.  
The {\em EUVE} data are archived at HEASARC and MAST.\footnote{\url{http://archive.stsci.edu/euve/}} 
%
%
In-orbit calibration using WD models HZ~43 and GD~153 
\citep{2002ASPC..264...57S} has been shown to be consistent 
with 10~\% uncertainties 
except at the longest wavelengths ($\lambda > 60$~nm). 
%

The calibration of the   Far Ultraviolet Spectroscopic Explorer 
\citep[{\em FUSE}, 1999 to 2007,][]{2000SPIE.4139..131S} 
is based on WD models only 
\citep{2000SPIE.4013..334S}. 
Originally, uncertainties had been calculated to be less than 10~\%, 
however, 
repeated observations of standard white-dwarf stars showed that, while the 
instrument's responsivity had been roughly constant for the first two years of the 
mission, it later declined at a rate of 5~\% to 10~\% per 
year, \citep[see also][]{2000SPIE.4139..186C}.\footnote{cf., \url{http://fuse.pha.jhu.edu/analysis/calfuse_wp0.html; http://fuse.pha.jhu.edu/analysis/calfuse_wp1.html}} 




\subsection{VUV and UV instruments}

The International Ultraviolet Explorer 
\citep[{\em IUE}, 1978 to 1996;][]{1978Natur.275..372B}
was a common project 
of NASA, the UK Science Research Council and the European Space Agency (ESA). 
As  mentioned in Section~\ref{sec3}, 
all archived data were 
recalibrated by use of the pure hydrogen WD 
G191-B2B.\footnote{IUE data are available from the Mikulski Archive for 
Space Telescopes (MAST), 
\url{http://archive.stsci.edu/iue}.} 
The {\em IUE} flux reference deviates by 6~\% from the current 
{\em HST} CALSPEC reference 
\citep[see][and references therein]{2014PASP..126..711B}. 

The Hubble Space Telescope 
{\em HST},\footnote{\url{http://www.stsci.edu/hst/}; \url{http://www.spacetelescope.org/}} 
a NASA/ESA mission, 
has been launched in 1990, is still operating as of 2015, and may last until 2020. 
Its  main instruments observe in the  
ultraviolet, visible, and near infrared. 
Five subsequent Space Shuttle missions repaired, upgraded, and replaced systems 
on the telescope, the last  
mission having taken place in 2009. 



The current set of science instruments includes three cameras and two spectrographs. 
The calibration of {\em HST} has been described in  Section~\ref{sec3}. 
An overview of 
the standards used in the {\em HST} CALSPEC procedures is given by \citet{2014PASP..126..711B}. 
Electronic versions of the UV spectra of all {\em HST} calibration standards as well 
as overviews over the instruments and their documentations and the proceedings 
from the calibration workshops can be found on the STScI Web pages.\footnote{The 
two web sites \url{http://www.nasa.gov/mission_pages/hubble/main/index.html} 
and  \url{http://www.spacetelescope.org/about/general/instruments/} 
give an in-depth overview of {\em HST}, its history and its instruments.}



\subsection{Visible and near-, mid-, and far-infrared instruments}

The absolute calibration of visible and NIR spectrometric telescopes in space, 
follows a process 
similar to that 
used for the calibration of ground-based visible telescopes: A primary 
celestial 
standard is used to set the absolute flux scale and a small set of secondary 
standards is used in the calibration of the different instruments over 
their corresponding wavelength and responsivity ranges (cf., Section~3). 
%


The infrared astronomy satellite  {\em AKARI} 
\citep[][before launch called {\em Astro-F}]{2007PASJ...59S.369M, 2010yCat.2298....0Y}\footnote{\url{http://www.ir.isas.jaxa.jp/ASTRO-F/Outreach/souti_e.html}}
 has been in operation 
from 2006 to 2011. 
It has been 
developed by  the 
Japan Aerospace Exploration Agency JAXA, in cooperation with institutes from  
Europe and Korea. 
Its primary mission was to survey the entire sky in near-, mid-, and 
far-infrared, from 1.7~\textmu m to 180~\textmu m, 
through its 68.5~cm aperture telescope and  
%
%
with two kinds of instruments: the 
FIS (Far-Infrared Surveyor) for far-infrared observations 
and the IRC (InfraRed Camera) for near and mid-infrared observations.
The 
calibration of the {\em AKARI} Far-Infrared Imaging Fourier Transform Spectrometer 
 (FIS-FTS) has been described by \citet{2010PASJ...62.1155M}. 
It is based on bright astronomical sources, such as 
stars, asteroids and planets. An internal blackbody source has been used for 
monitoring the detector responsivity over time. 
The relative uncertainty of the calibration of the continuum has been estimated 
to be $\pm$ 20~\% or better over the entire wavelengh range, 
and the absolute uncertainty is estimated to be 
better than 
$+40/-60$~\%.   
For the IRC, the (astro)photometric accuracies using stars 
have been found to be better than 
20~\% \citep{2012PASJ...64..126T,2014PASJ...66...47A}. 


ESA's {\em Herschel} Space Observatory 
\citep[originally named {\em FIRST}, Far InfraRed and Sub-millimetre Telescope;][]{2008SPIE.7010E..02P} 
has been launched in 2009, together 
with ESA's {\em Planck} mission. It has been equipped with  the largest 
-- i.e., 3.5~m diameter -- 
single mirror 
built so far for a space telescope. Its operational phase ended in 
2013 after the helium supply had been exhausted. 
%
%
%
%
{\em Herschel}'s three instruments are 
the HIFI (Heterodyne Instrument for the Far Infrared), 
the PACS (Photodetector Array Camera and Spectrometer), 
and the SPIRE (Spectral and Photometric Imaging Receiver), covering 
the  wavelength range from 55~\textmu m to 672~\textmu m, combining 
spectrometry, imaging spectrometry, and imaging photometry. 
For calibration, three sets of celestial standards and associated models 
have been used:
planetary models, stellar models, and asteroid models. 
Data in the models have been compiled from flyby missions, 
near-infrared and sub-millimetre ground-based 
observations, space-based observations, 
radar measurements, and known planetary atmosphere 
and stellar atmosphere conditions. 
The set of asteroids used for calibration fills the gap between the
sub-millimetre and millimetre range covered by Mars, Uranus, and Neptune, 
and the mid-IR range covered by bright calibration stars. 
The thermophysical model predictions 
agree within 5~\% with the available (and
independently calibrated) {\em Herschel} 
measurements \citep{2014ExA....37..253M}. 
%

The HIFI high-resolution heterodyne spectrometer observes in the  
two bands:  157~\textmu m to 212~\textmu m and 240~\textmu m 
to 625~\textmu m.  
The instrumentation for imaging in the infrared, in particular using  
superconducting mixers as detectors, 
has been described for example by  \citet{Wild13} and \citet{ SwinWild13}. 
%
HIFI uses internal loads to determine sensitivities for each frequency setting, 
and very accurate frequencies can be established by the local oscillators. 
Mainly planets (Mars and Neptune) are  
used for 
calibration.

PACS consists of an imaging photometer and a medium-resolution grating 
spectrometer \citep{2010A&A...518L...2P}. 
%
%
%
%
%
%
%
From a set of ``fiducial stars'' selected for calibration, 
five have been used 
for the final analysis: \mbox{$\beta$~And}, \mbox{$\alpha$~Cet}, 
\mbox{$\alpha$~Tau}, 
\mbox{$\alpha$~Boo}, and \mbox{$\gamma$~Dra} \citep{2014ExA....37..129B}. 
%
%
%
For the special observing mode of ``chop-nod'',\footnote{chopping and 
nodding in perpendicular direction, used to reduce background signal in particular for 
point-source photometry} very high 
accuracy can be reached. 
The relative (astro)photometric calibration precision (repeatability) has been 
found to be 1~\% in the blue and green bands and up to 5~\% in 
the red band.\footnote{These bands are centred around the wavelengths 
56~\textmu m, 100~\textmu m and 160~\textmu m, respectively. An  
in-depth description of the filter 
bandpasses and how they have to be used in comparison with other instruments, 
can be found on the PACS calibration web sites 
\url{http://herschel.esac.esa.int/twiki/bin/view/Public/PacsCalibrationWeb}.}
At this precision, inconsistencies among the stellar calibration models 
become noticeable, and  the absolute calibration accuracy
is mainly limited by the model uncertainty, which is 5~\% 
for all three bands \citep{2013ExA....36..631N}. 
%
For the principal observation mode (scan map, 
with a boustrophedon pattern) 
the relative accuracy 
varies from 0.5~\% in the blue band to 2~\% in the red. 
The systematic difference in flux calibration to the chop-nod mode amounts to 
less than 6~\%.  

The SPIRE instrument is an imaging photometer and imaging Fourier 
transform spectrometer 
\citep[FTS, see also][]{GriffinAde13}, 
both operating over a frequency range of approximately 450~GHz to 1550~GHz 
(670~\textmu m to 190~\textmu m). 
%
After correction for pointing errors, the absolute flux calibration has been found 
to be uncertain to less than 6~\%, and for extended sources where mapping 
is required   
to at least 7~\%, with a precision of about 1~\% 
\citep{2014MNRAS.440.3658S}. 
Cross-calibration programmes have shown that the agreement between the 
spectrometers, 
although based on different  prime calibrators, 
e.g., PACS, using stellar models, 
and SPIRE, using  Uranus and Neptune, 
is  better than 20~\%.  
Evaluation of the data as well as 
update and improvement of the theoretical models is ongoing 
\citep{2014ExA....37..381F,2014SPIE.9143E..2DN,2014ExA....37..253M}.\footnote{\url{http://www.cosmos.esa.int/web/herschel/home}}  

\subsection{Infrared and microwave instruments}

Ground-based absolute IR flux measurements with direct reference to laboratory 
standards, 
such as those 
by \citet{1983MNRAS.203..795S}, \citet{1983MNRAS.205..897B}, \citet{1985A&A...151..399M}, and \citet{1989A&A...218..167B},  
have been introduced in Section~3 together with the space-based absolute
 mid-infrared calibrations by
%
the Midcourse Space Experiment.  
The latter were carried out by use of
emissive reference spheres  that were
ejected and observed as point sources with calibrations based on 
laboratory data and basic physics \citep{2001AJ....121.1180C, 2004AJ....128..889P}. 

Proceeding to the microwave region:  
studies of the Cosmic Microwave Background (CMB) 
and its polarisation require extremely high precision 
\citep[cf.,][]{LaDo13}. 
%
%
%
%
In this wavelength range  
the {\em Planck} mission (2009 to 
2013),\footnote{\url{http://www.cosmos.esa.int/web/planck}} 
refined the measurements  of 
the CMB radiation
of its predecessors  {\em COBE} 
(Cosmic Background Explorer, 1989)\footnote{\url{http://lambda.gsfc.nasa.gov/product/cobe}}
and {\em WMAP}  (Wilkinson Microwave Anisotropy Probe, 2001).\footnote{\url{http://map.gsfc.nasa.gov}}
\, Within the  NASA HEASARC web archive,  the data center for \, CMB  research,  
the \, Legacy Archive for Microwave Background Data  
Analysis \, ($\mathrm{\Lambda}$AMBDA),\footnote{\url{http://lambda.gsfc.nasa.gov}}
provides information on the missions and data. 

The in-flight calibrations of {\em WMAP}  and {\em Planck} 
have been done by comparing the raw data to the 
temperature signal expected from the known dipole anisotropy.
This dipole anisotropy is resulting from the motion of the spacecraft relative to the Sun,
as well as  by motion of the solar 
system with respect to the CMB rest frame. 

{\em Planck} carried two 
scientific instruments: the High Frequency Instrument, or HFI, 
and the Low Frequency Instrument, or LFI. 
%
%
LFI 
has been designed to produce high-sensitivity, multi-frequency 
measurements of the microwave sky in the frequency range of 
27~GHz to 77~GHz (wavelength range 11.1~mm to 3.9~mm). The instrument 
consists of an array of 22 tuned radio receivers located in the 
focal plane of the telescope.
%
The    radiometric 
calibration is based on the combination of the orbital dipole 
plus the solar dipole. 
The solar dipole 
provides a signal of a few millikelvin with 
the same spectrum as the CMB anisotropies that is visible throughout the 
mission. 
%
%
An internal reference load at 4~K is used for monitoring 
the calibration stability of the 30~GHz radiometers. 
%
Additionally,  the brightness temperature of Jupiter is compared. 
%
The calibration uncertainties 
of the 2015 LFI/{\em Planck} data release are given as 
0.20~\% at 70~GHz, 0.26~\% at 44~GHz, 
and 0.35~\% at 30~GHz. These updated results agree very well with the 
data from the HFI instrument and within 1~$\sigma$ with {\em WMAP} results 
\citep{2015arXiv150508022P}.\footnote{Preliminary 
results had indicated some discrepancy to {\em WMAP}, which could 
be eliminated by improved data analysis; for updated information see the 
webpages of the {\em Planck} Collaboration 
\url{http://www.cosmos.esa.int/web/planck/publications}.}


%

Similarly, 
HFI 
has been designed to produce high-sensitivity, multi-frequency measurements 
of the diffuse radiation permeating the sky in all directions in the 
frequency range of 
100~GHz to 857~GHz (wavelength range 3.0~mm to 0.35~mm). 
The instrument consists of an array of 52 bolometric detectors placed 
in the focal plane of the telescope.
%
The components and subsystems of HFI have been calibrated and tested on the ground, the 
focal plane units have been characterised in the Saturne 
cryostat of the calibration 
facility of the Institut d'Astrophysique Spatiale in 
Orsay and,  after being integrated,
on the
satellite in a  cryogenic vacuum chamber at the CSL  
(Centre Spatial de Li\`ege).
A preliminary absolute
response was estimated during the focal plane calibration
with an uncertainty 
of 10~\% (and a relative pixel to pixel calibration
of 3~\%). 
The pre-launch calibration of  HFI had been carried out between
September 2004 and August 2008  \citep{2010A&A...520A..10P}. 
%
The in-flight calibration of the submillimetre channels
of HFI relies on the calibration of the FIRAS experiment on the
{\em COBE} satellite, which had provided the most accurate 
radiometric calibration for extended sources in the millimetre and 
submillimetre wavelength range to date. FIRAS
used an absolute black body to provide a flux calibration with an
accuracy superior to 1~\% below 400~GHz and 3~\% above 
\citep{1999ApJ...512..511M}.     
To obtain the best calibration accuracy over the large range from 
100~GHz to 857~GHz, 
two different radiometric calibration schemes have to be 
used. The higher frequency data 
are calibrated  by use of models of planetary atmospheric
 emission (Uranus and Neptune).   
The lower frequencies (from 100~GHz to 353~GHz) are calibrated by use of the 
time-variable cosmological microwave background dipole, i.e., the 
dipole produced by the proper motion of {\em Planck} with respect to
the rest frame of the CMB. 
This source of calibration only depends on the satellite 
velocity with respect to the solar system and permits an independent 
measurement of the amplitude of the CMB solar dipole, 
(3364.5 $\pm 0.8$)~\textmu K,  
which is $1~\sigma$ higher than the {\em WMAP} measurement with a direction that 
is consistent between both experiments  \citep{2015arXiv150201587P}.
%
%
%
%
Uncertainties in the absolute radiometric calibration have been 
found to lie in the range from 
0.54~\% to 10~\% from 100~GHz to 857~GHz. 
Details of the subsequently improved post-launch calibration can be 
found on the {\em Planck} webpages and especially in the publications of the  
{\em Planck}  
collaboration.\footnote{\url{http://www.cosmos.esa.int/web/planck/publications}}

\subsection{Solar spacecraft}

Solar physicists have striven towards 
spectroradiometric measurements of the solar spectrum 
from space since the 1960s.  A series of eight Orbiting Solar Observatories ({\em OSO})
was launched by NASA between 1962 and 1975. 
Instrumentation included spectrometers and radiometers for EUV and X-ray measurements. 
The telescope mirror and the spectrometer of the photoelectric 
spectroheliometer  on 
{\em OSO-6} were separately calibrated before launch, the latter by use of  
 transfer-standard 
photodiodes \citep{1973ApJ...183..291H}. 
However, spatial non-uniformities in the 
performance of a concave grating used in the spectrometer 
calibration  to produce a 
monochromatic beam of test radiation limited the pre-launch radiometric 
uncertainty 
of {\em OSO-6} to 10~\%.  Such optical non-uniformities also affected the 
pre-launch 
calibration of CDS on  the Solar and Heliospheric Observatory 
{\em SOHO} and the in-orbit performance of UVCS, 
also on {\em SOHO}; both instruments will be discussed later in this section.

The calibration of the S-055 spectrometer on {\em Skylab} was similar to that 
of the {\em OSO-6} instrument mentioned above \citep{1977ApOpt..16..837R}, 
and performance was monitored 
during the mission by means of underflights.  Nevertheless, the ultimate 
S-055 radiometric accuracy was insufficient to disentangle 
variations in the VUV output of the Sun from 
changes in the instrument responsivity (cf., Figure~\ref{calroc}).

\subsubsection{Total irradiance monitors}

A number of spaceborne 
instruments dedicated to total solar irradiance measurements have 
in-orbit radiometric 
calibration capabilities that are directly traceable to laboratory 
standards, i.e., 
no models of astronomical objects 
are required.  
Because these instruments do not employ telescopes nor have spectral 
resolving capability, 
their calibration is not entirely relevant to that of astronomical 
telescope-spectroscopic combinations, which are the focus of this paper. 
Nevertheless, we discuss  
total solar irradiance monitors in order to demonstrate the advantages 
of using laboratory standards in orbit.

The most accurate radiometric instruments used for this purpose are electrical 
substitution radiometers, such as the Active 
Cavity Irradiance Monitors \citep[][]{1999mfs..conf...19W} 
on the Solar Maximum Mission 
\citep[{\em SMM}, 1980 to 1989,][]{1980SoPh...65....5B, 1981ApJ...244L.113C}, 
 %
the Upper Atmosphere Research Satellite 
\citep[{\em UARS}, 1991 to 2005,][]{1993JGR....9810643R}, 
%
and on the Active Cavity Radiometer Irradiance Monitor Satellite 
\citep[{\em ACRIMSAT}, launched in 1999,][]{2014Ap&SS.352..341W}, 
TIM on {\em SORCE} 
\citep[launched 2003,][]{2005SoPh..230..111K}, 
the Differential Absolute Radiometers DIARAD and PMO6-V instruments 
that are part 
of the Variability of Solar Irradiance and Gravity Oscillations 
package 
\citep[VIRGO,][]{1997SoPh..175..267F} 
on {\em SOHO} \citep{1995SoPh..162.....F}  
and of the SOLAR package on the International Space Station  {\it ISS}. 
These ESRs operate at about 300~K and, in principle, have 
measurement uncertainties 
far below the percent level. 
For the primary calibration, radiometers operating at liquid helium temperatures 
can reach an SI uncertainty 
of approximately $5\times 10^{-4}$ 
\citep[500~ppm,][]{2014Ap&SS.352..341W}. 
TIM/{\em SOURCE} states an accuracy of $1\times 10^{-4}$, with a precision of 
better than $1\times 10^{-6}$. 
Details on the ground and flight calibrations of TIM are described by 
\citet{2005SoPh..230..111K}. 
%
%
However, in order to make the results of various 
measurements agree with the space absolute radiometric reference scale   
\citep[SARR,][]{1995AdSpR..16...17C}, 
significant adjustments, depending not only 
on the specific instrument optics and geometry 
(such as, e.g., apertures, filters, and slits) 
but also  
on    
apparent instrument degradations and changes in 
observing parameters, have been required.  
\cite{Froehlich13}, \citet{2014Ap&SS.352..341W} and others have 
repeatedly discussed such adjustments, and the various measurements 
of the total solar irradiance from spacecraft have 
been combined into composite time series by several authors.

Recently, 
Fr\"ohlich 
 has undertaken a thorough re-analysis of the test measurements performed 
in the laboratory on the flight spare instruments VIRGO~1 and VIRGO~4 
in order to revise the 
characterisation of the PMO6V-A and PMO6V-B instruments which are 
still operating 
on {\em SOHO}. Similarly, he revised the characterisation of the 
DIARAD radiometer and thus obtained a new absolute value of the 
VIRGO TSI.\footnote{The corresponding references are accessible 
through the webpage 
\url{http://www.pmodwrc.ch/pmod.php?topic=tsi/composite/SolarConstant}.}

\begin{figure}[ht]
\centering \includegraphics[width=\textwidth]{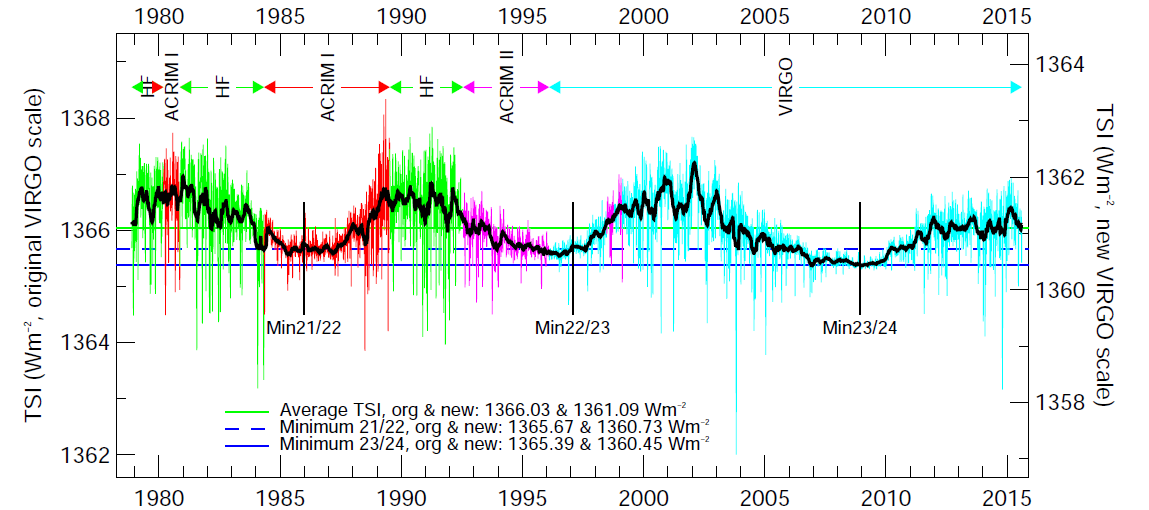} 
\caption{
The PMOD composite of total solar irradiance (TSI). Daily values (1978 to present) 
are plotted in different colours for the different 
experiments. 
The new and original VIRGO scales are shown on the right- and 
left-hand ordinates, respectively. Three numbers  
characterising the  solar-activity cycles that have so far 
been covered by measurements from space are listed. 
(Courtesy of C. Fr\"ohlich.)}
%
\label{TSI_VIRGOnew}
\end{figure}

The resulting   composite  reaching back to 
1978 \citep{FrLea98,Frohlich2015a,Frohlich2015b}
is shown in Figure~\ref{TSI_VIRGOnew}. 
It is noteworthy that the discrepancy 
between the scales of TIM/{\em SORCE} and VIRGO/{\em SOHO} is now 
such that the TSI averaged over the last minimum 
determined by VIRGO,  (1359.66 $\pm$ 2.47)~W\,m$^{-2}$, deviates from 
the 1360.52~W\,m$^{-2}$  measured by TIM 
 by less than half its  uncertainty. 
Preliminary calculations of a composite time series 
combining wavelet methods and Bayesian statistics 
\citep[e.g.,][]{DudokdeWit2011,2014EAS....66...77D,2014bpt..book.....V} 
have led to similar findings. 


%
%
%


%


\subsubsection{Spectral irradiance measurements}

Spaceborne spectrometers with true, on-board,  spectral irradiance 
calibration capabilities that permit efficiency changes to be tracked from
the time of laboratory calibration through integration, launch, and years of
use in orbit have been the Solar Ultraviolet Spectral Irradiance Monitors 
\citep[SUSIM;][]{1988ApL&C..27..163V,1993JGR....9810695B}. 
These have monitored with modest spectral resolving power and
without a telescope 
the full-disk VUV and UV solar spectral irradiance in 
the wavelength 
band that drives the photochemistry of the Earth's ozone layer. 
The SUSIMs, which 
have observed the Sun several times from the Space Shuttle and the 
{\em UARS}, comprised several spectrometers and a number of D$_2$-lamp transfer 
standards.  The initial spectrometer calibrations were established in the 
laboratory by direct comparison with synchrotron radiation.  During orbital 
operations, one of the spectrometers is not used to view the Sun, a procedure 
that could contribute to rapid degradation in performance, but only to monitor 
relative changes in the output of the D$_2$ lamps, which have significantly 
different duty cycles, again to preserve the laboratory calibration 
on the lamp by less frequent operation.

The Solar Stellar Irradiance Comparison Experiment SOLSTICE 
\citep{1994SPIE.2266..317R} 
has  also been part of the {\em UARS} instrument complement 
and monitored the full-disk solar spectral 
irradiance with approximately the same resolution and spectral range as SUSIM.  
SOLSTICE was thoroughly calibrated by use of synchrotron radiation 
before launch, but 
had no on-board calibration capability per se.  Instead  SOLSTICE 
tracked changes 
in its detection efficiency by comparing solar irradiances to those 
of hot stars.  The set of 
these is large enough that changes in the instrument can be 
disentangled from unexpected stellar 
variability, which is thought to be negligible in the stars chosen, 
by comparing each star to 
the ensemble average. 
Difficulties with  SOLSTICE measurements, as with similar other instruments, 
are: ($i$), 
possible undetected change in the instrument efficiency between pre-launch 
calibration and 
in-orbit observations, and, ($ii$), the large dynamic range, about 10$^8$, 
 over which accurate 
radiometry is required.  This range is accommodated by using different 
exposure times and a 
wide range of spectrometer entrance and exit apertures.  However, 
the latter differences mean 
that the solar and stellar observations use different fractions of 
the spectrometer optics, 
which may be non-uniform in performance  
across their surface  
or which may change properties at different rates as a function of time.
The {\em UARS} SUSIM and SOLSTICE instruments  monitored the VUV solar 
irradiance regularly and simultaneously from  1992 to 2005.  
\citet{1996JGR...101.9541W}  
compared the 
results and those of several other VUV solar irradiance monitors.  

A next generation of solar spectral irradiance instruments is represented by 
the Solar EUV Experiment 
\citep[SEE,][]{1998SPIE.3442..180W} 
on the NASA Thermosphere 
Ionosphere Mesosphere Energetics and Dynamics  ({\em TIMED})
spacecraft and the Spectral Irradiance Monitor 
\citep[SIM,][]{1998Metro..35..707R, 1998SPIE.3427..477L}
on the {\em SORCE} mission.  
{\em TIMED} has been launched in December 2001 and  provides VUV 
irradiance measurements 
using  grating spectrographs and silicon photodiodes to cover the
wavelength range 0.1~nm to 200~nm. 
The EUV grating spectrograph (EGS) as well as the XUV (astro)photometer 
system (XPS) are advanced versions of earlier instruments used 
in rocket flights. 
%
%
%
%
%
Calibration as well as comparison with other spectral irradiance monitors  
is documented by  
\citet{2005SoPh..230..345W}.\footnote{\url{http://lasp.colorado.edu/see/calibration.htm}}  
Continued independent measurements with prototype instruments on rockets, 
scheduled approximately annually, provide data for in-flight calibration. 

%
The Solar Radiation and Climate Experiment ({\em SORCE}) spacecraft was 
launched in January 2003, 
and, besides the Total Irradiance Monitor (TIM), carries  the Spectral Irradiance 
Monitor (SIM), the Solar Stellar Irradiance Comparison Experiment 
(SOLSTICE),  and the XUV Photometer System (XPS).  Together they measure 
the solar spectral irradiance from 1~nm to 2~\textmu m, accounting for 95~\% of 
the spectral contribution to TSI, i.e., the so-called ``solar constant''. 
%
%
%
SIM/{\em SORCE} is 
particularly interesting because it incorporates a very sensitive ESR as a 
detector in its focal plane and a dual spectrometer arrangement that 
allows degradation 
in orbit to be tracked. The spectral range covers 310~nm to 
2.4~\textmu m, thus the visible and near 
infrared (Vis/NIR) range. An additional channel covers the 
200~nm to 300~nm
 ultraviolet spectral region in order to overlap with the SOLSTICE instrument. 
SIM's accuracy amounts to 2~\%.\footnote{\url{http://lasp.colorado.edu/home/sorce/instruments/sim/}}
%
%
SOLSTICE/{\em SORCE} 
provides precise daily measurements of solar spectral irradiance 
at ultraviolet wavelengths, 
covering 115~nm to 320~nm with a spectral resolution of 0.1~nm, 
an absolute accuracy of 
better than 6~\%, and a relative accuracy (precision) of 0.5~\% per 
year.\footnote{\url{http://lasp.colorado.edu/home/sorce/instruments/solstice/}} 
The SOLSTICE preflight calibrations included 
component-level calibrations in the LASP 
calibration laboratory and 
system-level calibrations primarily at the NIST Synchrotron 
Ultraviolet Radiation Facility  
SURF~III, 
similar to those used for the SOLSTICE/{\em UARS}  calibration programme 
\citep{1993JGR....9810679W}. 
Unit-level calibrations included characterisation of individual 
optical and instrument elements, 
including gratings, detectors, and mirrors.
The system-level calibrations included detailed wavelength and 
radiometric calibrations 
of the fully assembled instrument. Wavelength calibration used 
platinum and mercury lamps providing well-spaced emission lines. 
The in-flight calibration for the SOLSTICE instruments includes 
long-term multi-wavelength 
observations of stable early-type O-B stars 
and observations using redundant channels with 
strongly reduced duty cycles. 
The wavelength calibration for the in-flight 
data uses 
prominent solar emission features. 
%
%
%
The XPS 
\citep{2005SoPh..230..375W} 
evolved from earlier versions flown on 
the Student Nitric Oxide Explorer 
\citep[{\em SNOE};][]{1999GeoRL..26.1255B} 
and {\em TIMED}.
It measures the solar soft X-ray (XUV) irradiance 
from 1~nm to 34~nm and the bright 
hydrogen emission at 121.6~nm (\hi Lyman-alpha). 
The photodiodes were calibrated at NIST SURF-III with reference Si diodes 
(uncertainty of 2~\%). In-flight calibration is achieved 
via redundant channels with lower (weekly) duty cycles and rocket flights, 
the latter resulting in an absolute irradiance uncertainty 
of about 12~\% to 24~\%.\footnote{\url{http://lasp.colorado.edu/home/sorce/instruments/xps/in-flight-calibration/}} 
%
%
Additionally, any 
 other available solar XUV irradiance measurements, such as from {\em SOHO} 
and SEE/{\em TIMED}, 
are included in the XPS validation programme.


\citet{yeo2015} 
summarise current solar irradiance studies, measurements as well 
as models, and also come to the conclusion that despite a vast 
number of open questions remaining, the combined effort has 
led to major improvements in our understanding of the solar 
cycle and its variation.  
%
%
%
%
%

\subsubsection{The SOLAR instruments on ISS}
The SOLAR package has been 
mounted in 2008 (after being carried into orbit aboard the shuttle
mission {\em STS-122}) as one of the external payloads 
of the Columbus laboratory on the International Space Station ({\em ISS}). 
 The total and spectral irradiance is recorded simultaneously in 
the 16~nm to 3.1~\textmu m  range 
by three instruments. 

\begin{figure}[!htb]
\centering \includegraphics[width=0.8\textwidth]{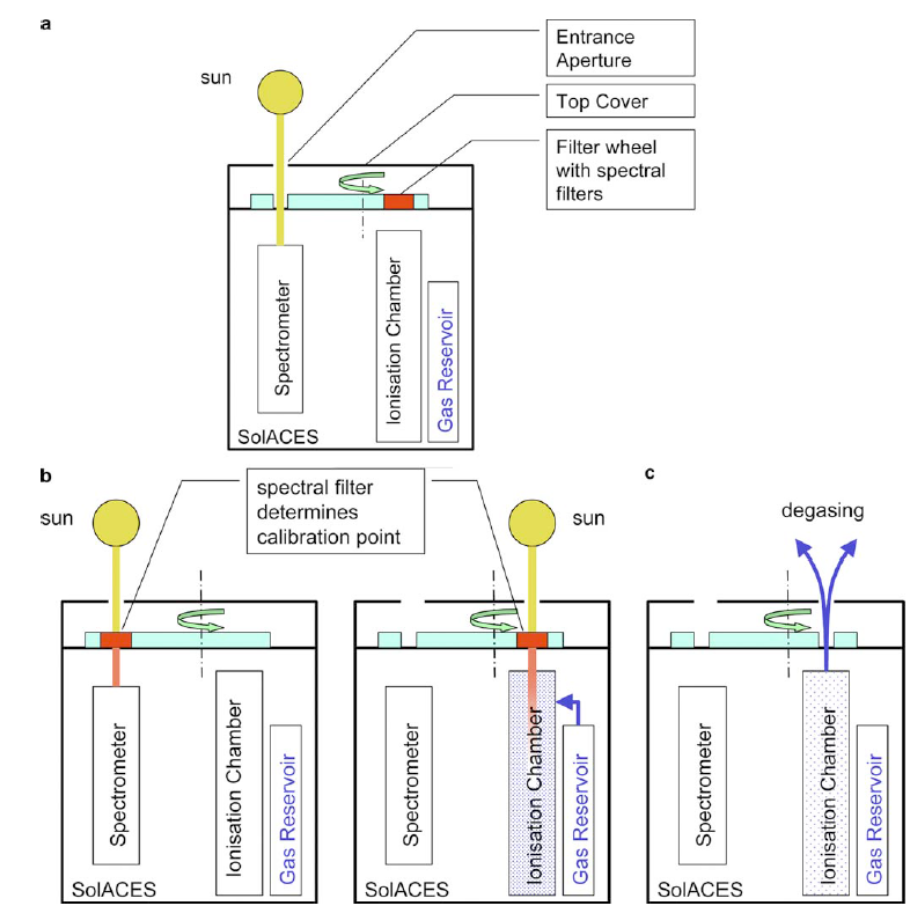} 
\caption{
(a): The scheme of the SolACES spectrometer - filter wheel - ionisation 
chamber arrangement. 
(b)--(c): The sequence of the calibration mode. 
(After Figure~6 in \citet{2006AdSpR..37..273S},   
with permission).}
\label{solaces_ionch}
\end{figure}

The Solar Auto-Calibrating EUV/UV Spectrophotometer (SolACES) 
operates   
  three grazing-incidence plane gratings to determine the absolute fluxes 
from 16~nm to 150~nm \citep{2014SoPh..289.1863S}. 
  Its fourth spectrometer stopped operating shortly after launch.  
The absolute calibration is  achieved by primary standards, 
namely two rare-gas ionisation chambers (cf., Figure~\ref{solaces_ionch}). 
%
%
%
%
%
%
%

%
%
During the mission, the instrument responses steadily declined due to  
ageing  of the channel electron multiplier sensors and contamination. 
A special calibration sequence with various gases and filter settings 
is repeated regularly, 
at a higher frequency than originally foreseen, in order to  monitor 
these changes. 
The overall uncertainty has been determined to be 10~\%. 
The instrument description, its performance and results, as well as comparisons with 
 SEE/{\em TIMED} 
and EVE/{\em SDO} \citep{2012SoPh..275..115W} 
in the EUV band between 16~nm to 25~nm 
  are given by \citet{2014SoPh..289.1863S}. 
The agreement of all three data sets is well within  
20~\% throughout the years 2008 to 2013. 

SOVIM  \citep{2010AdSpR..45.1393M}   
is combining two types of absolute radiometers and 
three-channel filter radiometers.
The SOVIM package is composed of three PMO6 absolute radiometers, 
two solar photometers, one pointing sensor 
and one DIARAD radiometer. 

SOLSPEC \citep{2009SoPh..257..185T}   
is composed of three double spectrometers using 
concave gratings, covering the wavelength range 
  from 165~nm to 3.1~\textmu m.
Like SolACES it has been calibrated by use of primary standard 
sources of the PTB. 
Internal calibrations have been performed regularly, by measuring the 
spectrum of internal deuterium lamps (UV) or tungsten lamps 
(visible and IR).\footnote{
\url{http://solspec.projet.latmos.ipsl.fr/SOLSPEC_GB/Home.html}}

All irradiance measurements in the UV range point out the importance of  
rigorous cleanliness and contamination  
control at instrument and spacecraft levels. Moreover,  
the use of novel materials, such as 
wide-bandgap semiconductors enhances the robustness of detectors 
against proton radiation 
\citep{2015ExA....39...29B}. 


\subsubsection{Solar irradiance spectra}

In addition to the measurements discussed above, which were mainly focussed on the UV 
and VUV wavelength ranges, calibrated solar spectra have been obtained from the 
ground over the past few decades 
\citep{1969ApOpt...8.2215A, 1974ApOpt..13..518T,1984SoPh...90..205N,1992ApJ...390..668L, 1995SoPh..157...51B, 2004GMS...141..171T, 2004AdSpR..34..256T}. 
%
Most of this 
work is not directly relevant to the subject of this paper, but these 
measurements do contribute to the development of the solar reference spectrum 
 and can be used to assess measurements from space in spectral 
regions where there are overlapping data.



During March/April 2008 (Carrington rotation 2068) and still 
within the extended last solar minimum  
(transition between solar cycles 23 and 24), 
solar irradiance reference spectra (SIRS) 
from 0.1~nm to 2.4~\textmu m have been 
generated using a combination of satellite
and sounding rocket observations  \citep{2009GeoRL..36.1101W}. 
These reference spectra
include daily satellite observations from SEE/{\em TIMED} and  
{\em SORCE} instruments. Data on the 
extreme ultraviolet range have also been improved with higher
spectral resolution observations by use of the prototype  {\em SDO}
Extreme ultraviolet Variability Experiment (EVE) aboard a
sounding rocket launched on 14 April 2008.
Comparisons with the \mbox{ATLAS-3} 
reference spectrum\footnote{taken during the Atmospheric 
Laboratory for Applications and Science (ATLAS) {\em Space Shuttle} 
missions ATLAS~1, March 1992, and 
 ATLAS~3, November 1994}  \citep{2004GMS...141..171T, 2004AdSpR..34..256T} 
showed a significant overall improvement in 
uncertainties (10~\% to 15~\% in the EUV and X-ray range, compared to the earlier 30~\% to 50~\%; 
2~\% to 5~\% compared to 3~\% to 10~\% in the FUV-MUV; around 1~\% to 3~\%,  
both in the VIS and NIR) thanks to improved calibration and monitoring.   

\subsubsection{Spectroradiometric telescopes on SOHO}


The next serious efforts at accurate solar spectroradiometric measurements from 
space following the {\em OSO} series and {\em Skylab} were associated with 
the {\em SOHO} mission 
\citep{1995SoPh..162.....F}.   
The {\em SOHO} project required extreme cleanliness in construction, 
integration, and 
launch operations so that changes in performance in orbit would be  minimised.  
The spacecraft builders worked with a cleanliness requirement of a few hundred
nanograms of condensable and particulate contamination per square centimetre, 
while the instrument teams aimed for even less. The minimal deterioration 
in performance 
of these instruments over the entire course of the mission is attributable,  
probably to a large extent,  to the 
cleanliness achieved  \citep{2002ISSIR...2...91T}.  

\begin{figure}[t]
\centering \includegraphics[width=0.8\textwidth]{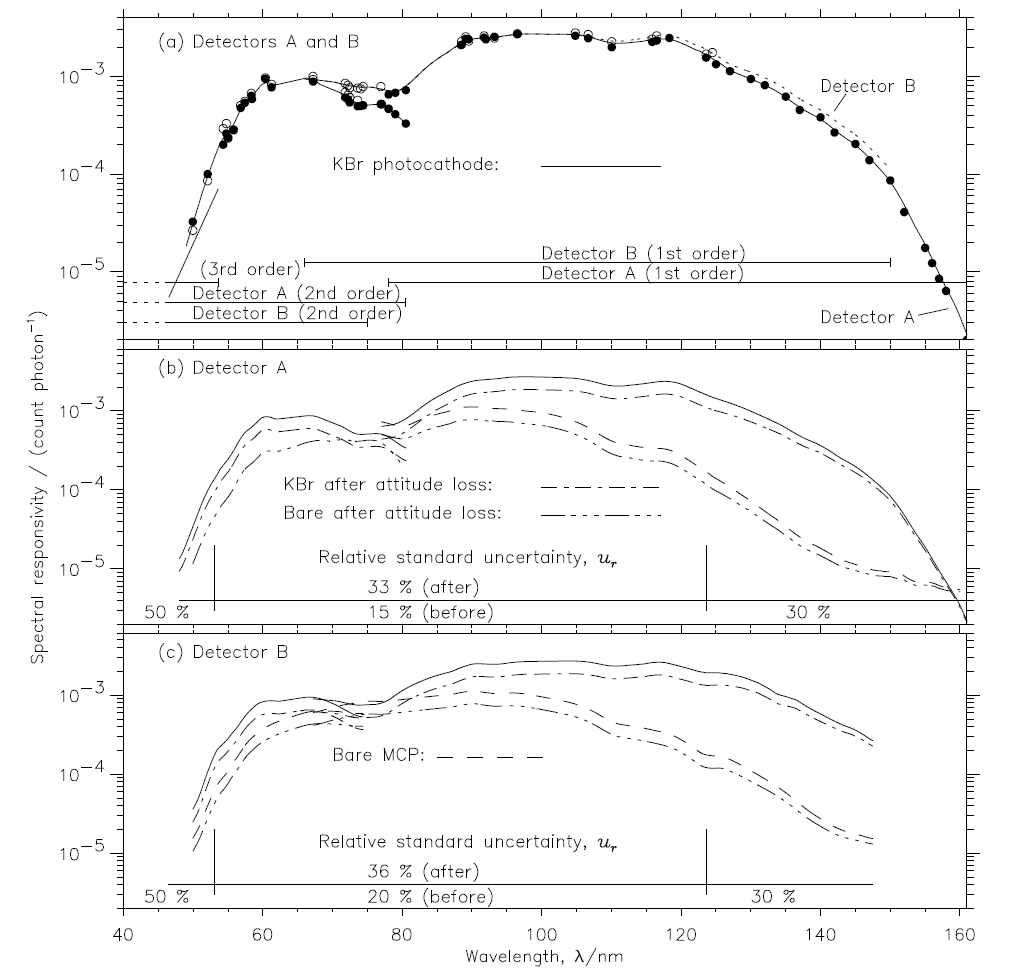} 
\caption{
Spectral responsivities of the SUMER spectrometer 
and the corresponding relative uncertainties.
(a) First-order, second order, and third-order responsivities
for both detectors (interpolated between calibration
points); (b) Independent detector A assessment; and (c) results for detector B. The
relative uncertainties in the central wavelength range (indicated by vertical bars at
53~nm and 124~nm) are derived from the laboratory calibration
and are smaller than those of the in-flight calibration
extensions obtained from calculated line-ratios (i.e., branching fractions) 
and stellar observations.
(After Figure~1 in \citet{Wilhelm02},   
with permission).}
\label{SUMER_responsivity}
\end{figure}
\begin{figure}[h]
\centering \includegraphics[width=0.8\textwidth]{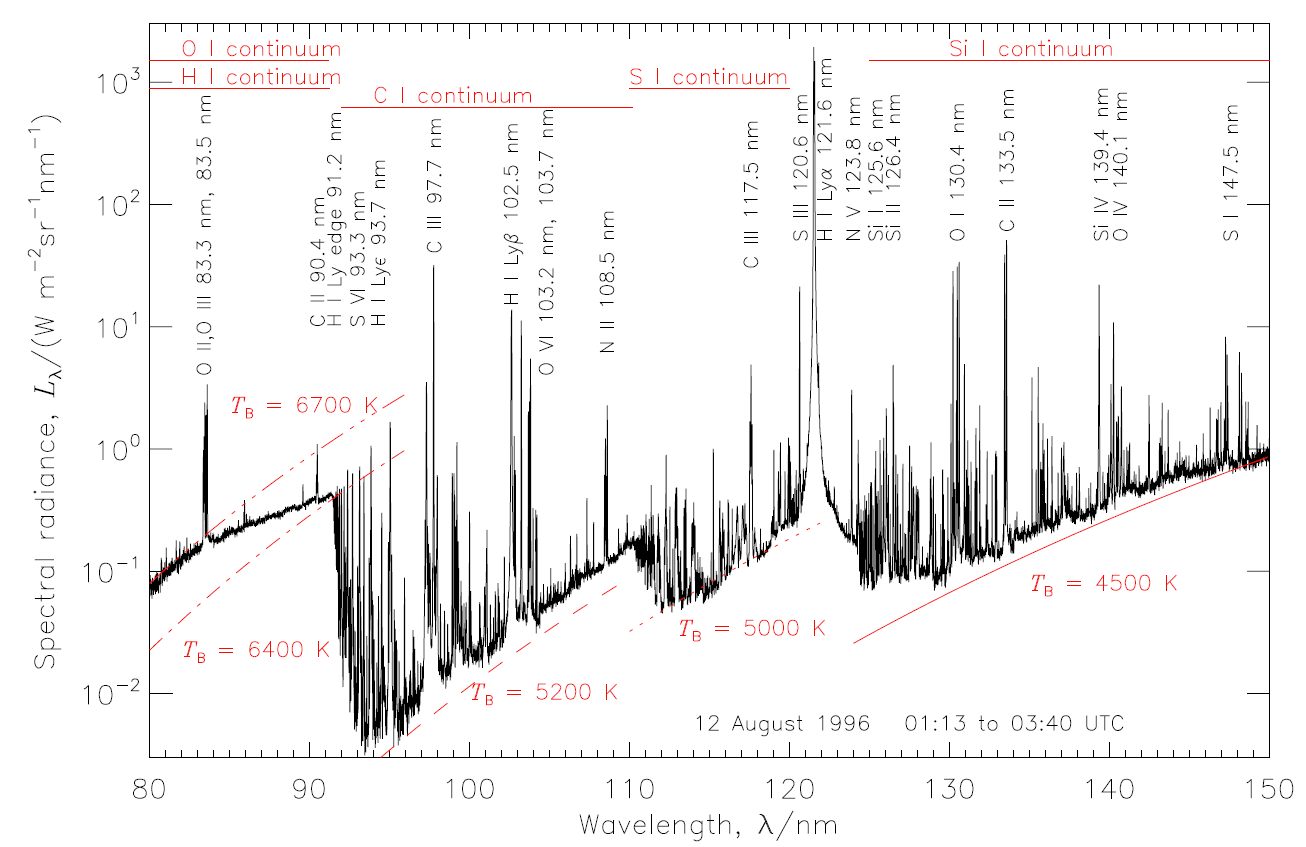} 
\caption{
Spectral radiance of the quiet Sun in the VUV range 
from a region near the centre of the disk, 
measured with the SUMER detector on {\em SOHO}. 
The spectral radiances of some continua expected for 
certain brightness temperatures, $T_{\rm B}$, are also shown. The relative uncertainty
below 124~nm is 15~\% and above this limit 30~\%.
(After Figure~4 in \citet{2010AN....331..502W},   
with permission).}
\label{SUMER_spectrum}
\end{figure}

The spectroradiometric efficiency of the Solar Ultraviolet 
Measurements of Emitted 
Radiation instrument on {\em SOHO} \citep[SUMER,][]{1995SoPh..162..189W}   
was determined 
prior to launch by employing a source standard that had been 
calibrated by comparison to 
synchrotron radiation    \citep{1996A&AS..115..561H}. 
The source standard consisted of a 
hollow cathode which emitted a line spectrum, and a spherical normal-incidence 
collimating mirror. The collimated beam had a 10~mm diameter and a divergence of 
$\pm$1\arcmin.  In the laboratory calibration of SUMER, the image of the 
flux-limiting aperture of the hollow cathode underfilled the aperture as seen 
from the 
focal plane of the telescope, which is, at the same time, the entrance 
slit plane 
of the spectrometer.  An unobstructed observation of the entire source 
radiation 
was thus achieved and spectroradiometric responsivities that included the 
reflectivities of all optical surfaces, the grating efficiencies, 
and the detector performance could be established  
by an appropriate scan of the entrance aperture.\footnote{Note however 
that apertures and stops in the instrument
      itself were not taken into account in this way.}
The pre-launch calibrations have subsequently been further refined under 
operational conditions \citep{1997ApOpt..36.6416W, 2000ApOpt..39..418S}. 
%
SUMER has been measuring with two detectors, A and B,  in almost overlapping wavelength 
bands. 
The central part of  both detectors' photocathode 
microchannel plate (MCP) has a coating of potassium bromide
(KBr) and the other parts use the bare plate. 
The spectral responsivity characterisation of the instrument is shown in 
Figure~\ref{SUMER_responsivity}, and the VUV radiance spectrum of a
quiet-Sun region with 
prominent emission lines and continua
in the wavelength range from 80~nm to 150~nm is presented in Figure~\ref{SUMER_spectrum}. 
%
%
%

The spectroradiometric efficiency of the Ultraviolet Coronagraph Spectrometer 
on {\em SOHO}  \citep[UVCS,][]{1995SoPh..162..313K}  
was determined before launch at 
selected wavelengths by use of transfer-standard photodiodes 
\citep{1996SPIE.2831....2G}. 
The results, which had an estimated uncertainty of 20~\%, have 
been confirmed for 
the instrument parameters used in the laboratory by underflights with the 
SPARTAN~201 Shuttle-borne spectrometer \citep{2002ISSIR...2..249F}.  
Annual 
observations of stars that pass through the UVCS field of view have shown that 
there has been no detectable (at the 10~\% to 15~\% level) change 
of responsivity over the 
course of the mission.
Because of limited resources and instrument limitations, it was not possible 
before launch to calibrate UVCS over the full range of instrument parameters 
used in science observations.  In-orbit studies and laboratory tests using 
replicas of the flight gratings have shown that, because the reflectivity 
of the grating is not uniform across its surface, the UVCS O\,{\sc vi} channel 
effective area is not a linear function of the width of the 
aperture\footnote{
        UVCS is a coronagraph with external and internal occulters. 
        Adjustment of the latter, which is done to reduce the level
        of scattered light, changes the aperture.}
\citep{2000SPIE.4139..362G}. 

Although UVCS did not rely on extensive modelling to determine 
instrument performance parameters, this experience is instructive to 
those who attempt to model the performance of complex instruments from 
first principles:  
It 
is often difficult to predict the behaviour of state-of-the-art optical 
components, including detectors, and coatings (and the 
inevitable contaminants) over the complete range of 
physical dimensions, optical angles, wavelengths, etc., expected in use.  The 
experience 
of the {\em OSO-6} and CDS (discussed below) laboratory calibration teams, 
who 
had to deal with non-uniform beams reinforces this point.  Thus 
performance models should be verified with benchmark measurements over 
as much of the parameter space as possible. 
%

The spectroradiometric efficiency of CDS, the Coronal Diagnostic 
Spectrometer \citep{1995SoPh..162..233H},  
was measured prior to launch 
in a manner similar to that used for SUMER: the source standard used was also 
calibrated by comparing it to synchrotron radiation 
\citep{1996Metro..32..647H}, 
but in this instance, a 
Wolter type-II telescope served as the collimator  
and the collimated beam was limited to 5~mm in 
diameter when it left the Wolter telescope. 
The nominal divergence was $\pm$30\arcsec. 
Inside its vacuum tank, the CDS instrument could be moved perpendicular to the 
beam (while the optical axis was maintained), so that the instrument apertures 
that illuminated the grazing- and normal-incidence gratings could be mapped. The 
laboratory calibration of CDS  \citep{2000JOptA...2...88L} 
turned out to be a 
much more complex undertaking than that of SUMER. Several reasons contributed to 
this, not the least being the fact that the collimated beam showed structure in 
its cross section and exhibited an angular divergence exceeding its nominal 
value by nearly a factor of four. 
Rocket underflights (a NASA/LASP rocket carrying a photoelectric 
EUV Grating Spectrograph, EGS,  
and SERTS-97, the Solar EUV Rocket Telescope and Spectrograph)  
and line-ratio techniques were used to reassess the initial 
calibration at early stages of the mission \citep{2000ApJ...536..959B,2001A&A...379..708D}. 
Later on, an update was performed using data from intercalibration 
campaigns and from two Extreme-Ultraviolet Normal Incidence 
Spectrograph (EUNIS) rocket flights 2006 and 2007. 
As described  by \citet{2011ApJS..197...32W},    
the EUNIS spectrographs had been calibrated at RAL against 
the same transfer standard 
(cross-calibrated against the BESSY~II storage ring as primary source) that  
 had been used for the calibration of CDS and the 
EIS on {\em Hinode} (see below). 

%
%
The Extreme ultraviolet Imaging Telescope EIT has been designed to provide 
full-disk solar images of the lower corona in four EUV passbands 
\citep{1995SoPh..162..291D}.   
It consists of a normal-incidence Ritchey-Chr\'etien telescope with multilayer 
coating and a 
 back-side illuminated thinned CCD sensor at its focus. 
The pre-flight spectral response had been calibrated at the 
Institut d'Astrophysique 
Spatiale (IAS) in Orsay using synchrotron light from the SuperACO positron 
storage ring \citep{1995SPIE.2517...29D}.       
Several parameters and calibration components could only be determined 
in-flight using special operation protocols and sequences, such as 
determining the point spread function (PSF) via flare observations 
and a Mercury transit 
\citep{2002ISSIR...2..121C}.   
Furthermore, the EIT has been suffering from time-varying 
response degradation, partially induced by contaminants on the 
CCD and on optical surfaces, and 
also by radiation damage to the CCD. Some non-polymerized contaminants, 
such as water, could 
successively be eliminated via heating of the components; 
however, as the water re-condensed 
at the instrument's housing wall, 
this led to oscillatory behaviour, especially during the 
first years of the mission. The flat-field  of the CCD and its variation are modelled comparing
and correlating it to flat-fields in the visible range measured by use 
of calibration lamps 
\citep[see, e.g.,][and references therein]{2013SoPh..288..389B}. 
%

%
The  Solar EUV Monitor \citep[SEM,][]{1995SoPh..162..441H} 
comprises a free-standing,  
5000~lines/mm transmission grating, aluminum-coated silicon photodiodes, and 
aluminum filters that define the bandpass.  It has been designed to measure the 
He~{\sc ii} 30.4~nm output from the Sun as well as the 
integrated flux between 17~nm and 70~nm.
The SEM showed steady degradation of the first-order signal over the first seven years of
operation. After that the responsivity has remained almost constant. 
This result followed from 
a series of dedicated calibration-rocket underflights 
\citep{2002ISSIR...2..135M, 2013SoPh..288..389B}.
%

The LASCO experiment \citep{1995SoPh..162..357B} 
has been designed as a set of three independent 
coronagraph telescopes to 
record white-light images of the solar corona from 1.1~R$_{\odot}$ through 
30~R$_{\odot}$. After {\em SOHO}'s temporary loss of attitude in 1998, the C1 
instrument observing the innermost range stopped functioning.  
The C2 instrument covers the geometric range from (2.5 to 6.5)~R$_{\odot}$, and 
the C3 field of view extends from (3.8 to 30)~R$_{\odot}$. 
Pre-flight calibration of the LASCO instruments has been performed at 
the Naval Research Laboratory (NRL). 
The radiometric in-flight calibration is based 
on on-board calibration lamps as well as 
on a set of stars that have to be observed against the background of the 
solar corona    
\citep{2004SPIE.5171...26L, 2006SoPh..233..155T, 2006SoPh..233..331M}. 

%
Intercalibration data sets, taken on a regular basis for 
SUMER and CDS, in selected areas and 
conditions turned out to be extremely valuable, 
in particular 
to re-evaluate calibrations after  {\em SOHO}'s loss of attitude and 
subsequent recovery. 
Various detector parts of CDS (normal incidence, NIS, and grazing incidence, GIS) 
and SUMER (detctors A and B)  
had been compared in dedicated  observing sequences of overlapping solar areas 
 and in several EUV lines.  
Differences of 2~\% to 50~\% were found, and could be used to  re-evaluate the 
calibrations \citep{1999ApOpt..38.7035P, 2001ApOpt..40.6292P, 2003ApOpt..42..657P}. 
In-flight calibration studies were also made employing atomic data 
bases and the line-ratio 
technique \citep{2001A&A...379..708D, 2010A&A...518A..49D},   
and, subsequently,  
rocket underflights were used to update the CDS calibration \citep{2011ApJS..197...32W}. 
 %
For various instruments, such as CDS and EIT, the spectrally summed radiances 
could be compared to irradiance monitors, such as SEM 
\citep{2002ISSIR...2..211T}. 
More details are given in the ISSI report dedicated to {\em SOHO's} radiometric calibration 
\citep{2002ISSIR...2.....P}  
and in the article by \citet{2013SoPh..288..389B} 
which summarises the experience with 
stability of solar space radiometric instrumentation in great detail.

\subsubsection{Spectroradiometric telescopes on TRACE} 
The Transition Region and Coronal Explorer 
\citep[{\em TRACE},][]{1999SoPh..187..229H} 
was a NASA SMEX mission operating from 1998 to 2010 
that imaged the solar photosphere, transition region, and corona with angular  
resolution of 1\arcsec. 
To provide continuous coverage of solar phenomena, {\em TRACE} was 
located in a sun-synchronous 
polar orbit. 
The instrument featured a 30~cm Cassegrain telescope,  
multilayer optics, and a lumogen-coated CCD detector to record three EUV and 
several UV wavelengths. 
The pre-flight calibration had been performed partially in the 
laboratory. 
Measurements of the quantum efficiencies of the CCDs together with the 
lumogen coating, for example, 
were made at the Lockheed Martin ATC\footnote{The Lockheed Martin Advanced Technology 
Center recently changed its name to Space Technology Advanced Research and Development 
Laboratories, or STAR Labs.} facility. 
The subsequent on-orbit characterisation of the filters and  
the consecutive updates and monitoring of the 
flat fields has been described by \citet{1999SoPh..187..229H}. 
Reference spectra from SOLSTICE and the shuttle- or rocket-borne 
High Resolution Telescope and Spectrograph 
\citep[HRTS,][]{1998SoPh..183...29H}, 
as well as comparisons with SUMER/{\em SOHO} observations \citep{2006A&A...456..747K}, 
were used to ``clean'' the spectrum, i.e.,  
correct for overlapping lines and contaminations, 
in particular for the 121.6~nm Ly-$\alpha$ images which contain a 
contribution from the 
UV continuum at wavelengths beyond 150~nm and for the 
C\,{$\scriptstyle {\rm IV}$ doublet which needs to be deconvolved from 
several contributions.


\subsubsection{Telescopes on STEREO} 

The {\em STEREO} mission, launched in 2006 \citep{2008SSRv..136....5K}, 
consists of two nearly 
identical observatories,
one ahead of the Earth on its orbit ({\em \mbox{STEREO-A}, Ahead}), 
the other trailing behind ({\em STEREO-B, Behind}). 
Each spacecraft carries four instrument packages for measuring plasma 
properties of solar energetic particles, the magnetic field, and radio 
disturbances, as well as a set of telescopes to investigate 
the characteristics of coronal mass ejections (CME). 
Since February 2011, the two spacecraft  have been providing  
unprecedented views of the 
far side of the Sun, and, together with data from the Solar Dynamics 
Observatory ({\em SDO}), 
collected 360\gr\ maps of the Sun.  
From late 2014 to early 2016, the  
{\em STEREO} spacecraft will be in a position in space where the 
Sun will affect full data downloads and the communication will be restricted. 
At least one spacecraft, however, will always be collecting data.

{\em STEREO's} Sun Earth Connection Coronal and Heliospheric Investigation (SECCHI) package 
%
comprises five telescopes for imaging   
the solar corona from the solar disk to 
beyond 1~au. These telescopes are: an extreme ultraviolet imager 
(EUVI: (1 to 1.7)~R$_{\odot}$), 
two 
 Lyot coronagraphs (COR1: (1.5 to 4)~R$_{\odot}$  and COR2: (2.5 to 15)~R$_{\odot}$) and two 
heliospheric imagers (HI-1: (15 to 84)~R$_{\odot}$  and HI-2: (66 to 318)~R$_{\odot}$ ). 
All instruments use $2048 \times 2048$ pixel CCD arrays in a backside-illuminated mode. 
Their backside surface has been specially processed for EUV sensitivity, 
while the other has an anti-reflection coating applied. 
The EUVI mirrors 
were calibrated at the synchrotron of the 
IAS in Orsay 
and its CCDs  at the Brookhaven synchrotron and the LMSAL XUV calibration facility 
\citep{2008SSRv..136...67H}.\footnote{see also \url{http://secchi.lmsal.com/EUVI/}} 

The Heliospheric Imager (HI) is a wide-angle, visible-light, imaging 
system for the detection of 
CME events in interplanetary space and, in particular, 
of events directed towards the Earth. 
Large parts of its radiometric  calibration and the monitoring thereof have 
been made by use of 
stars of known minor variability 
as reference\footnote{\url{http://www.stereo.rl.ac.uk/Documents/InstrumentPapers.html}} 
\citep{2009SoPh..254..185B, 2010SoPh..264..433B, 2012SoPh..276..491B}. 
No significant deterioration has been 
found \citep{2013SoPh..288..389B}. 

The paired COR1 telescopes, the innermost coronagraphs of the SECCHI 
instrument suite,
 observe the white-light K-corona from (1.4 to 4)~R$_{\odot}$ in a 
waveband 22.5~nm wide, centered on the H$\alpha$ line at 656~nm. 
Observations of the planet Jupiter are used to establish absolute 
radiometric calibrations for
each telescope. \citet{2008SoPh..250..443T} have  described 
the radiometric and pointing calibration.\footnote{\url{http://cor1.gsfc.nasa.gov/publications/}} 


\subsubsection{Spectroradiometric telescopes on Hinode} 

The Japanese {\em Hinode} mission, launched in 2006,  supports three 
solar telescopes, 
co-funded 
by JAXA-ISAS, NASA, ESA, and the United Kingdom Space Agency (UKSA). 
They cover  
the optical range (Solar Optical Telescope, SOT),  
the EUV range (EUV Imaging Spectrometer, EIS) 
and the X-ray range (X-Ray Telescope, XRT). 
%
%
The EIS pre-launch radiometric calibration has been performed at RAL 
in the same way as that of CDS  
\citep{2004ApOpt..43.1463S, 2006ApOpt..45.8689L,2007SoPh..243...19C, 2008ApJS..176..511B}. 
The calibration is regularly monitored in orbit; this includes 
flat-fields from 
LED exposures, dark images, quartz-crystal microbalance (QCM) measurements, 
as well as dedicated 
observation campaigns. The latter have included 
 EUNIS rocket flights that were also used for the CDS/{\em SOHO} calibration 
\citep{2011ApJS..197...32W}. 

\subsubsection{Spectroradiometric telescopes on SDO} 


The Solar Dynamics Observatory ({\em SDO}), which has been extensively 
described by \citet{2012SoPh..275....3P} in a special 
edition of Solar Physics,  
has been launched in February 2010. 
{\em SDO} carries three 
scientific experiments; the Atmospheric Imaging Assembly 
AIA \citep{2012SoPh..275...17L}, 
the EUV Variability Experiment EVE \citep{2012SoPh..275..115W},  
and the Helioseismic and Magnetic Imager HMI \citep{2012SoPh..275..229S}. 
%

%
EVE measures the 
solar spectral irradiance from  from 0.1~nm to 105~nm at 0.1~nm resolution.  
 EVE's high-resolution irradiance instruments are the Multiple EUV Grating Spectrographs: 
MEGS-A, a grazing-incidence spectrograph 
that measures the solar EUV irradiance in the 6~nm to 37~nm range,\footnote{The MEGS-A  
had to be turned off due to a failure of its CCD electronics in May 2014.} 
and  MEGS-B, 
a normal-incidence, dual-pass spectrograph 
that measures the solar EUV irradiance in the 35~nm to 105~nm range. 
To provide their in-flight calibration, an EUV spectro-photometer (ESP) 
 measures the solar 
EUV irradiance in broad bands between 0.1~nm and 39~nm, and a  
photometer measures the Sun's bright  Ly-$\alpha$ emission at 121.6~nm. 
The ESP instrument on {\em SDO} was calibrated at NIST's SURF~III 
before and 
after integration with the EVE package, and the 
ground calibration results are available via the Laboratory for Astronomy and
Space Physics' (LASP) 
webpage.\footnote{\url{http://lasp.colorado.edu/home/eve/data/ground-calibration-results/}}   
A nominally identical 
instrument is periodically flown on a sounding rocket and makes observations of the Sun 
simultaneously with its {\em SDO} analogue. The rocket instrument is calibrated at 
SURF~III's EUV detector radiometry beamline  
before and after each rocket flight, and the NIST calibration is transferred to 
the {\em SDO} instrument.
In this way, EVE  maintains an absolute calibration accuracy of better than 25~\% 
\citep{2012SoPh..275..115W, 2012SoPh..275..145H, 2012SoPh..275..179D}. 

The AIA instrument is an array of 
four normal-incidence 
reflecting telescopes that image the Sun in seven EUV, two VUV, and 
one visible wavelength channels. 
%
%
The initial pre-flight radiometric calibration of AIA was based on measurements 
of the response of all the telescope components (mainly at LMSAL and the 
 Advanced Light Source (ALS) synchrotron at Lawrence Berkeley National 
Laboratory),  
which were then combined analytically 
into a model of  the overall system performance. 
The dominating terms are the uncertainty in
the contamination thickness and the CCD quantum efficiency determination. 
The accuracy of the preflight calibration has been 
estimated to be of the order of 25~\% 
\citep{2012SoPh..275...41B}.    
For tracking degradation (e.g., by contamination) and to refine and 
maintain the calibration, cross-calibration with EVE is routinely performed. 
Comparisons and cross-calibrations have also been performed between AIA 
and EVE/{\em SDO}, 
EIS/{\em Hinode}  
and  SOLSTICE/{\em SORCE} \citep{2014SoPh..289.2377B}.  
They found that most channels are in agreement within the uncertainties. 
The 30.4~nm channel showed a 
degradation by a factor of three from May 2010 through September 2011, very likely due to 
contamination effects.  
Some other inconsistencies could be ascribed to errors in the wavelength response function 
either of AIA or of EVE. 
Additionally, it was found that the version of the CHIANTI spectral lines data base 
then in use was missing some lines and underrepresenting emission in others.   

\subsubsection{Instruments on  IRIS}

The Interface Region Imaging Spectrograph 
\citep[{\em IRIS},][]{2014SoPh..289.2733D} SMEX  
 spacecraft has been launched in June 2013 and provides simultaneous 
spectra and images of the photosphere, 
chromosphere, transition region, and corona with 0.33\arcsec\ to 0.4\arcsec\ spatial 
resolution, two-second temporal resolution, and 1~km/s velocity 
resolution over a field-of-view of up to 175\arcsec $\times$ 175\arcsec. 
The {\em IRIS} instrument is a multi-channel imaging spectrograph with a 
20~cm UV telescope, 
feeding ({\em i}) two far-UV channels of 
133.2~nm to 135.8~nm and 138.9~nm to 140.7~nm with 2.6~pm resolution and 
effective areas of 1.6~cm$^2$ and 2.8~cm$^2$, respectively, and ({\em ii}) 
a near-UV channel of 
278.3~nm to 283.5~nm with 5.3~pm resolution and an effective area of 0.2~cm$^2$.
%
A large fraction of the radiometric calibration 
as well as tracking longterm contamination and degradation is achieved via 
observation of a selected set of UV-bright stars. 
As described in Section~\ref{sec3}, the 
spectral radiances of such stars 
are given with an uncertainty of 10~\% to 15~\%.  
{\em IUE} spectra of
these stars can be viewed on the website of the  
MAST.\footnote{\url{http://archive.stsci.edu/}} 
Stellar comparisons and quiet-Sun studies are planned for  
long-term monitoring of the calibration.  
The relevant information is also found on the corresponding 
webpages.\footnote{\url{http://iris.lmsal.com/documents.html}}


%


\subsubsection{Solar X-ray spectroscopy}
%
%
{\em Yohkoh} 
\citep[before launch called {\em SOLAR-A},][]{1991SoPh..136....1O} 
was a solar observatory operated from 1991 to 2001 by  the  Institute of Space and 
Astronautical Science (ISAS, Japan) 
in collaboration with the space agencies of the United States and the United Kingdom. 
It carried four instruments: a Soft X-ray Telescope (SXT), a Hard X-ray Telescope (HXT), 
a Bragg Crystal Spectrometer (BCS), and a Wide Band Spectrometer (WBS). 
SXT used already a CCD for readout. 
Part of the calibration had been performed at Rutherford Appleton Laboratory 
\citep{RAL-93-Lang}.  

%
%
The Reuven Ramaty High-Energy Solar Spectroscopic Imager 
\citep[{\em RHESSI},][]{2002SoPh..210....3L} 
is a NASA SMEX mission designed to investigate 
particle acceleration and energy release in solar flares, through 
imaging and spectroscopy of hard X-ray and gamma-ray continua emitted by 
energetic electrons, and of gamma-ray lines produced by energetic ions. 
The single instrument consists of an imager, made up of nine bi-grid 
rotating modulation collimators (RMC), in front of a spectrometer 
with nine cryogenically-cooled germanium detectors, one behind each RMC.
{\em RHESSI} observes solar photons over three orders of magnitude in energy (3~keV to
17~MeV) 
with  high energy resolution. In the range 
from 3~keV to 100~keV, the resolution is 
 about 1~keV FWHM;  in the range  over 1~MeV, it is about 3~keV FWHM  
\citep{2002SoPh..210...33S}.  
%
%
The gain calibration of the detector segments 
  is re-calculated  on a regular basis 
using 
background data collected over a full orbit. The channel-to-energy 
conversion is represented as a purely linear fit over each energy scale. 
The effective area and response of the spectrometer has been simulated and reproduced 
by GEANT3, a high-energy photon and particle transport code that is 
used to generate the response matrix of the instrument. 
Before flight, the GEANT3 performance was
calibrated by taking spectrometer data in the laboratory using radioisotopes 
with
lines from 3.7~keV to 6.1~MeV, placed in many positions above and around the
spectrometer. The laboratory data were compared with GEANT3 simulations of
the same configurations, and the detectors’ internal segmentation boundaries in
GEANT3 were adjusted until the GEANT3 results matched the data.
For monitoring the detector response over time and noting 
radiation damage, {\em RHESSI} carries a weak onboard radioactive 
source (5~nanocuries of $^{137}$Cs)
emitting at 662~keV, far from any line expected to occur in flares or in 
the variable background.
Background is usually high in the instrument, due to limited shielding 
(a consequence of weight constraints) and 
the chosen orbit, crossing the South Atlantic Anomaly (SAA). 
At most energies the continuum is dominant,
which is typical for an unshielded instrument. 
%

{\em Hinode}'s  
X-Ray Telescope (XRT)\footnote{\url{http://hinode.nao.ac.jp/}, \url{http://xrt.cfa.harvard.edu/}}
was designed and developed by the Japan-US collaboration between the Smithsonian Astrophysical 
Observatory (SAO), the NASA Marshall Space Flight Center MSFC, 
JAXA, and the National Astronomical Observatory of Japan. 
The XRT CCD camera was tested and calibrated 
in X-rays at the Advanced Technology Center of the NAOJ with JAXA. 
The telescope calibration has been, among other locations, performed 
at the X-Ray Calibration
Facility (XRCF) of the MSFC, and filters for the 
XRT were tested at the X-Ray Astronomy
Calibration and Testing (XACT) facility of the 
Osservatorio Astronomico di Palermo   
\citep{2007SoPh..243...63G}.   
The detector calibration is described by \citet{2008SoPh..249..263K}.  




%% file: concl5.tex
\section{Conclusions and Outlook}                                     
Ultimately, it must be realised that an unambiguous deconvolution 
of observed signals into valid, scientifically interesting, time-dependent 
data is possible only if an instrument's spectral detection effciency 
can be completely characterised at all times.
This requires either direct on-board calibration systems or 
highly-sophisticated component and system 
performance models supplemented with regular benchmark observations 
using equipment that 
is traceable to laboratory primary standards.

For the sake of completeness, we should mention here that this applies 
not only to the radiometric calibration of a space telescope. An overall 
calibration of a spectrometric space telescope requires that a 
considerable number of additional quantities be determined. These include: 
the spacecraft reference frame, the telescope's pointing accuracy and 
stability and its vignetting function, the point-spread function 
(PSF, both on- and off-axis), flat-field maps, stray light and 
the occurrence of ghost images and other backgrounds, the plate-scale,
radiometric non-linearities owing to pulse pile-up or charge-transfer 
inefficiency,  
as well as the timing of the observations. As some of these quantities 
are prone to change during orbital operations -- sometimes also as a 
consequence 
of particular events like spacecraft eclipses or large solar 
flares -- in-orbit monitoring 
at appropriate intervals and following such events is necessary. 
Curiously, today still, more effort is often spent on such monitoring 
than on a proper radiometric laboratory calibration. 
These in-orbit comparisons are extremely vital and by no means 
rendered unnecessary
by a solid pre-launch calibration, yet a thorough characterisation and 
understanding
of the instrument before launch is a necessity for obtaining optimal 
measurements.
There are, fortunately, indications that the ``launch-now-calibrate-later'' 
attitude is
changing, not least fostered by the demanding requirements of some current 
and many future missions. 

The terrestrial atmosphere has always hampered ground-based observations, 
be it by its turbulence or its tendency to absorb and even hide the radiation 
of cosmic objects arriving at Earth. Balloon and space astronomy have 
successively given access to more and more regions of the electromagnetic spectrum 
and with an image quality that lies beyond what can be attained on the ground. 
Past and
current astrometric space missions have provided accurate distance measurements 
for a large number of stars; modern ground-based interferometry, moreover, 
has provided 
stellar radii. In principle, thus, it should now be possible to determine 
comprehensive properties of a large number of stars and other cosmic objects 
with high accuracy.

The radiometric calibration of space telescopes, a prerequisite to this goal 
has, 
however, turned out to be far from trivial. The environment to which an 
instrument 
after its laboratory calibration is exposed during storage and integration into 
the spacecraft, during transportation, preparations for launch as well as, 
finally, 
upon exposure to the environment of space, has a varying influence on its 
responsivity -- and both improvement and degradation are possible 
(cf., Fig.~\ref{HSTstis}). 
To overcome this problem, an extensive network of standard stars has 
been established. 
This approach becomes problematic, however, if one wants to achieve 
accuracies below a few per cent. When testing stellar 
models with this kind of calibration, there is a danger of circular 
conclusions.

For high-energy astronomy, standard candles, which here take over the role of 
stable standard stars, are rare, and changes of the radiometric calibration 
during operation in space occur as well. 
Under these circumstances an ``International Astronomical Consortium for 
High Energy Calibration'' whose members perceive radiometric calibration 
as a dynamic process has led progress. IACHEC encourages consideration of the 
derived constraints in the data exploitation and in scientific
publications. They now see ``success after years of diverging results on 
calibrations, and lack of
communication among the teams''. They emphasise on their web site that no 
important change in
effective area of an instrument can take place devoid of other 
instruments' results.

In the sub-sections above, we have separately discussed spectroradiometry 
and associated accuracies for different wavelength ranges. We need to 
emphasise, however, that the radiometric calibration of telescopes 
should cover the entire spectral range accessible in space with 
similar accuracies. 
Investigations in multi-wavelength astronomy may address different 
phenomena, but such phenomena tend to be linked -- be it by gravitation 
or by radiative, magnetic and mechanical energy or by particle transport. 
The solar corona, for example, is best observed through VUV and soft X-ray 
radiation, while the photosphere's characteristics are accessible 
through observations in the visible and infrared. A meaningful study of 
the links between phenomena observed in these diverse wavelength regions 
does require or, at least, would benefit from similar radiometric accuracies.

A move away from involving models of celestial standards in calibration to 
physics-based radiometric calibrations of space telescopes now seems to be 
underway, at least in the visible and near-infrared.
Both the use of calibration rockets and calibrated ground-based telescopes 
supplemented 
by measurements of the prevailing properties of the terrestrial atmosphere are 
foreseen. 
This promises a future with one-per-cent and even sub-per-cent accuracies, 
traceable to Earth-based laboratory standards.

The recent turnaround in attitude towards the importance of an accurate 
spectroradiometric calibration, which has to some extent also been 
motivated by the discovery of dark energy, recognises the influence 
of calibration on the progress of science in hindsight. Looking forward, 
the asset of an advanced calibration is now recognised as well, specifically for 
{\em HST} it is pointed out that ``Hubble's power to revolutionise has never 
been greater than now. The instruments are calibrated better than ever, and 
we are using them in ways we had not anticipated at the time of the last 
servicing mission'' \citep{sembach15}.   

Nearly a decade ago \citet{2007ASPC..364...77M}    
incited the participants at a conference on  photometry to 
``turn from practical astrophysics to astronomical metrology.''  
If this were to take hold for general spectroradiometry, we may look 
forward to sound progress in all fields of astronomy.

%% file: abbrev_cal.tex

\let\abbrev\nomenclature   \renewcommand{\nomname}{List of acronyms}
\addcontentsline{toc}{section}{List of acronyms}
\nomenclature{RCSS}{Radiometric calibration spectral source}
\nomenclature{LECS}{Low Energy Concentrator Spectrometer}
\nomenclature{MECS}{Medium Energy Concentrator Spectrometer}
\nomenclature{HPGSPC}{High Pressure Gas Scintillator Proportional Counter}
\nomenclature{PDS}{Phoswich Detection System} 
\nomenclature{GRBM}{Gamma-ray Burst Monitor}
\nomenclature{HIFI}{Heterodyne Instrument for the Far Infrared}
\nomenclature{ACRIM}{Active Cavity Radiometer Irradiance Monitor}
\nomenclature{HF}{Hickey-Frieden radiometer on {\em NIMBUS-7}} %
\nomenclature{ACS}{Advanced Camera for Surveys}
\nomenclature{IRC}{InfraRed Camera}
\nomenclature{SOT}{Solar Optical Telescope}
\nomenclature{ADS}{Astrophysics Data System}
\nomenclature{APS}{Active pixel sensor}
\nomenclature{ATM}{Apollo Telescope Mount}
\nomenclature{HETG}{High Energy Transmission Grating}
\nomenclature{LETG}{Low Energy Transmission Grating}
\nomenclature{ACIS}{Advanced CCD Imaging Spectrometer}
\nomenclature{\emph{AXAF}}{Advanced X-ray Astrophysics Facility, now \emph{Chandra}}
\nomenclature{BESSY}{Berlin Electron Storage ring for 
Synchrotron radiation}
\nomenclature{BIPM}{Bureau International des Poids et Mesures}
\nomenclature{PTB}{Physikalisch-Technische Bundesanstalt}
\nomenclature{CCD}{Charge-coupled device}
\nomenclature{CDS}{Coronal Diagnostic Spectrometer}
\nomenclature{\emph{Chandra}}{X-ray observatory, formerly \emph{AXAF}}
\nomenclature{\emph{CXO}}{\emph{Chandra} X-ray observatory}
\nomenclature{\emph{ORFEUS}}{Orbiting Retrievable Far and Extreme Ultraviolet 
Spectrometer}
\nomenclature{\emph{Skylab}}{NASA space station} 
\nomenclature{\emph{Voyager}}{NASA planetary and interstellar mission 
(two spacecraft, \emph{Voyager-1} and \emph{Voyager-2})} 
\nomenclature{COS}{Cosmic Origins Spectrograph}
\nomenclature{EGS}{Extreme-ultraviolet Grating Spectrograph}
\nomenclature{EIT}{Extreme-ultraviolet Imaging Telescope}
\nomenclature{ESA}{European Space Agency}
\nomenclature{EUV}{Extreme ultraviolet}
\nomenclature{\emph{EXOSAT}}{European X-ray Observatory Satellite}
\nomenclature{FOV}{Field of view}
\nomenclature{FUV}{Far ultraviolet}
\nomenclature{FWHM}{Full width at half maximum}
\nomenclature{\emph{GALEX}}{Galaxy Evolution Explorer}
\nomenclature{GI}{Grazing incidence}
\nomenclature{GIS}{Grazing incidence spectrometer}
\nomenclature{NIS}{Normal incidence spectrometer}
\nomenclature{GSFC}{Goddard Space Flight Center}
\nomenclature{HRTS}{High Resolution Telescope and Spectrograph}
\nomenclature{HRI}{High Resolution Imager}
\nomenclature{HUT}{Hopkins Ultraviolet Telescope}
\nomenclature{\emph{HST}}{Hubble Space Telescope}
\nomenclature{IAU}{International Astronomical Union}
\nomenclature{ISSI}{International Space Science Institute}
\nomenclature{SolACES}{Solar Auto-Calibrating EUV/UV Spectrophotometer}
\nomenclature{LASP}{Laboratory for Atmospheric and Space Physics}
\nomenclature{MCP}{Microchannel plate}
\nomenclature{MPG}{Max-Planck-Gesellschaft}
\nomenclature{NASA}{National Aeronautics and Space Administration (US)}
\nomenclature{NBS}{National Bureau of Standards, now NIST}
\nomenclature{NI}{Normal incidence}
\nomenclature{TCF}{Telescope Calibration Facility (NIST)}
\nomenclature{NIST}{National Institute of Standards and Technology (US)}
\nomenclature{NUV}{Near ultraviolet}
\nomenclature{\emph{OSO}}{Orbiting Solar Observatory}
\nomenclature{PSF}{Point spread function}
\nomenclature{RCSS}{Radiometric calibration spectral source}
\nomenclature{\emph{ROSAT}}{R\"ontgensatellit}
\nomenclature{\emph{Rosetta}}{ESA mission to comet 67P/Churyumov-Gerasimenko}
\nomenclature{\emph{SDO}}{Solar Dynamics Observatory}
\nomenclature{SED}{Spectral energy distribution} 
\nomenclature{SEM}{Solar Extreme-ultraviolet Monitor}
\nomenclature{SERTS}{Solar Extreme-ultraviolet Research Telescope and Spectrograph}
\nomenclature{\emph{SMM}}{Solar Maximum Mission}
\nomenclature{\emph{SOHO}}{Solar and Heliospheric Observatory}
\nomenclature{\emph{Suzaku}}{Japanese X-ray astronomy mission, 
formerly  \emph{Astro-E2}} 
\nomenclature{\emph{Yohkoh}}{Solar X-ray observatory}
\nomenclature{\emph{Hinode}}{Solar observatory, formerly \emph{Solar-B}}
\nomenclature{SOLSTICE}{Solar-Stellar Irradiance Comparison Experiment}
\nomenclature{\emph{STS}}{Space Transportation System}
\nomenclature{SUMER}{Solar Ultraviolet Measurements of Emitted Radiation}
\nomenclature{SURF}{Synchrotron Ultraviolet Radiation Facility}
\nomenclature{SIM}{Spectral Irradiance Monitor}
\nomenclature{SUSIM}{Solar Ultraviolet Spectral Irradiance Monitor}
\nomenclature{\emph{STEREO}}{Solar Terrestrial Relations Observatory}
\nomenclature{SXT}{Soft X-ray Telescope}
\nomenclature{HXT}{Hard X-ray Telescope}
\nomenclature{STIS}{Space Telescope Imaging Spectrograph}
\nomenclature{\emph{TIMED}}{Thermosphere, Ionosphere and Mesosphere Energetics and Dynamics mission}
\nomenclature{\emph{TRACE}}{Transition Region and Coronal Explorer}
\nomenclature{\emph{UARS}}{Upper Atmosphere Research Satellite}
\nomenclature{UVCS}{Ultraviolet Coronagraph Spectrometer}
\nomenclature{UV}{Ultraviolet}
\nomenclature{VUV}{Vacuum ultraviolet}
\nomenclature{MUV}{Medium ultraviolet}
\nomenclature{FUV}{Far ultraviolet}
\nomenclature{NISP}{Near-infrared Spectrograph and Photometer on \emph{Euclid}}
\nomenclature{\emph{Euclid}}{Future ESA mission to map the geometry of the dark Universe}
\nomenclature{XUV}{Extreme ultraviolet}
\nomenclature{XPS}{XUV Photometer System}
\nomenclature{BCS}{Bragg Crystal Spectrometer and bent crystal spectrometer} 
\nomenclature{CME}{Coronal mass ejection} 
\nomenclature{\emph{COBE}}{Cosmic Background Explorer}
\nomenclature{\emph{FUSE}}{Far-Ultraviolet Spectroscopic Explorer}
\nomenclature{HRTS}{High Resolution Telescope and Spectrograph}
\nomenclature{\emph{IUE}}{International Ultraviolet Explorer}
\nomenclature{LASCO}{Large Angle Spectroscopic Coronagraph}
\nomenclature{LASP}{Laboratory for Atmospheric and Space Physics}
\nomenclature{MPS}{Max-Planck-Institut f\"ur Sonnensystemforschung, 
formerly Max-Planck-Institut f\"ur Aeronomie (MPAE)}
\nomenclature{MPE}{Max-Planck-Institut f\"ur extraterrestrische Physik}
\nomenclature{STScI}{Space Telescope Science Institute}
\nomenclature{IAS}{Institut d'Astrophysique Spatiale}
\nomenclature{NRL}{Naval Research Laboratory}
\nomenclature{SI}{Syst\`eme International d'Unit\'es, International System of Units}
\nomenclature{PMOD/WRC}{Physikalisch-Meteorologisches Observatorium Davos / World Radiation Center} 
\nomenclature{SSI}{Solar spectral irradiance} 
\nomenclature{SEE}{Solar EUV Experiment}
\nomenclature{VIRGO}{Variability of Solar Irradiance and Gravity Oscillations}
\nomenclature{\emph{RHESSI}}{Reuven Ramaty High Energy Solar Spectroscopic Imager}
\nomenclature{AIA}{Atmospheric Imaging Assembly}
\nomenclature{EIS}{EUV Imaging Spectrometer}
\nomenclature{ESP}{EUV spectro-photometer}
\nomenclature{\emph{IRIS}}{Interface Region Imaging Spectrograph} 
\nomenclature{SN}{Supernova} 
\nomenclature{QE}{Quantum efficiency}
\nomenclature{RMC}{Rotation Modulation Collimator}
\nomenclature{JAXA}{Japan Aerospace Exploration Agency}
 \nomenclature{WBS}{Wide Band Spectrometer}
\nomenclature{\emph{HST}}{Hubble Space Telescope} 
\nomenclature{\emph{ANS}}{Astronomical Netherlands Satellite} 
\nomenclature{\emph{JWST}}{James Webb Space Telescope}
\nomenclature{EUVI}{EUV Imager} 
\nomenclature{$\mathrm{\Lambda}$AMBDA}{Legacy Archive for Microwave Background Data Analysis}
\nomenclature{\emph{OAO}}{Orbiting Astronomical Observatory} 
\nomenclature{SAO}{Smithsonian Astrophysical Observatory}
\nomenclature{MIT}{Massachusetts Institute of Technology}
\nomenclature{MAST}{Mikulski Archive for Space Telescopes}
\nomenclature{IACHEC}{International Astronomical Consortium for High Energy Calibration}
\nomenclature{HEASARC}{High Energy Astrophysics Science Archive Research Center}
\nomenclature{ESR}{Electrical substitution radiometer} 
\nomenclature{FOC}{Faint Object Camera}
\nomenclature{FOS}{Faint Object Spectrograph}
\nomenclature{HEXTE}{High Energy X-ray Timing Experiment}
\nomenclature{ASM}{All-Sky Monitor}
\nomenclature{PCA}{Proportional Counter Array}
\nomenclature{STIS}{Space Telescope Imaging Spectrograph} 
\nomenclature{HRC}{High Resolution Camera} 
\nomenclature{WFC}{Wide Field Camera}
\nomenclature{\emph{EUVE}}{Extreme Ultraviolet Explorer} 
\nomenclature{\emph{WFIRST}}{Wide-Field Infrared Survey Telescope} 
\nomenclature{SM1(2,3,4)}{\emph{HST} Servicing Missions}
\nomenclature{CIPM}{Comit\'{e} International des Poids et Mesures}
%
%
\nomenclature{\emph{IRTS}}{Infrared Telescope in Space} 
\nomenclature{ACCESS}{Absolute Color Calibration Experiment for Standard Stars} 
\nomenclature{2MASS}{Two Micron All Sky Survey}
\nomenclature{NIR}{Near infrared}
\nomenclature{MIR}{Mid infrared}
\nomenclature{IR}{Infrared}
\nomenclature{CALSPEC}{Calibration data base for the {\em HST} and the {\em JWST}, maintained by the STScI}
\nomenclature{\emph{MSX}}{Midcourse Space Experiment}
\nomenclature{FIRAS}{Far Infrared Absolute Spectrophotometer}
\nomenclature{CMB}{Cosmic Microwave Background}
\nomenclature{HFI}{High Frequency Instrument} 
\nomenclature{LFI}{Low Frequency Instrument}
\nomenclature{\emph{WMAP}}{Wilkinson Microwave Anisotropy Probe}
\nomenclature{CNES}{Centre National d'Etudes Spatiales}
\nomenclature{HRMA}{High Resolution Mirror Assembly} 
\nomenclature{KAO}{Kuiper Airborne Observatory} 
\nomenclature{SECCHI}{Sun Earth Connection Coronal and Heliospheric Investigation}
\nomenclature{HI}{Heliospheric Imager}
\nomenclature{\emph{ISO}}{Infrared Space Observatory} 
\nomenclature{FTS}{Fourier transform spectrometer} 
%
\nomenclature{LTE}{Local thermodynamic equilibrium}
\nomenclature{NLTE}{Non local thermodynamic equilibrium}
\nomenclature{SAA}{South Atlantic Anomaly} 
\nomenclature{GEANT}{Geometry and Tracking, a high-energy photon and particle transport code}
\nomenclature{SDD}{Silicon drift detector} 
\nomenclature{\emph{NuSTAR}}{Nuclear Spectroscopic Telescope Array} 
\nomenclature{CTE}{Charge transfer efficiency}
\nomenclature{MOS}{Metal oxide semiconductor} 
\nomenclature{FIS}{Far Infrared Surveyor}
\nomenclature{MLS}{Metrology Light Source} 
\nomenclature{TSI}{Total solar irradiance}
\nomenclature{ECR}{Electrically calibrated radiometer} 
\nomenclature{\emph{SORCE}}{Solar Radiation and Climate Experiment}
\nomenclature{\emph{Spacelab}}{Laboratory for use on \emph{Space Shuttle} flights}
\nomenclature{TIM}{Total Irradiance Monitor}
\nomenclature{CMB}{Cosmic Microwave Background}
\nomenclature{\emph{ISS}}{International Space Station}
\nomenclature{\emph{IRAS}}{Infrared Astronomy Satellite} 
\nomenclature{\emph{Spitzer}}{Space Infrared Telescope Facility}
\nomenclature{QCM}{Quartz-crystal micro balance}
\nomenclature{IR}{Infrared} 
\nomenclature{\emph{AKARI}}{Japanese space mission for infrared astronomy, formerly \emph{Astro-F}} 
\nomenclature{\emph{SMEX}}{Small Explorer Mission}
\nomenclature{FIR}{Far infrared}
\nomenclature{\emph{Swift}}{NASA Gamma-Ray Burst Mission}
\nomenclature{\emph{BeppoSAX}}{Italian-Dutch satellite for X-ray astronomy}
\nomenclature{\emph{SPARTAN}}{\emph{Shuttle}-launched satellites for solar studies}
\nomenclature{\emph{RXTE}}{Rossi X-ray Timing Explorer} 
\nomenclature{\emph{XMM-Newton}}{X-ray Multi-Mirror Mission}
\nomenclature{CHIANTI}{An atomic database for spectroscopic diagnostics of 
astrophysical plasmas}
\nomenclature{EPIC}{European Photon Imaging Camera}
\nomenclature{HMI}{Helioseismic and Magnetic Imager for \emph{SDO}}
\nomenclature{ISAS}{Institute of Space and Astronautical Science, Japan}
\nomenclature{DIARAD}{Differential Absolute Radiometer}
\nomenclature{\emph{Herschel}}{ESA infrared and sub-millimetre telescope mission}\nomenclature{\emph{Planck}}{ESA mission for microwave astronomy}
\nomenclature{PACS}{Photodetector Array Camera and Spectrometer for \emph{Herschel}}
\nomenclature{SPIRE}{Spectral and Photometric Imaging Receiver}
\nomenclature{NIRSpec}{Near infrared multiobject dispersive spectrograph to be 
flown on the \emph{JWST}} 
\nomenclature{PSI}{Paul Scherrer Institut}
\nomenclature{LMSAL}{Lockheed Martin Solar and Astrophysics Laboratory}
\nomenclature{RAL}{Rutherford Appleton Laboratory}  
\nomenclature{UKSA}{United Kingdom Space Agency} 
\nomenclature{XRT}{X-ray Telescope}  
\nomenclature{XRS}{X-ray Spectrometer} 
\nomenclature{XIS}{X-ray Imaging Spectrometer} 
\nomenclature{HXD}{Hard X-ray Detector} 
\nomenclature{PSPC}{Position Sensitive Proportional Counters}
\nomenclature{PANTER}{X-ray test facility near M\"unchen, Germany}
\nomenclature{MSFC}{Marshall Space Flight Center}
\nomenclature{XRCF}{X-ray and Cryogenics Facility, NASA MSFC}
\nomenclature{XACT}{X-ray Astronomy Calibration and Testing, Palermo, Italy}
\nomenclature{RaMCaF}{Rainwater Memorial Calibration Facility}
\nomenclature{WD}{White Dwarf}
\nomenclature{CSL}{Centre Spatial de Li\`ege}
\nomenclature{\emph{Gaia}}{ESA astrometry mission}
\nomenclature{\emph{TD-1}}{ESA UV mission} 
\nomenclature{\emph{SNOE}}{Student Nitric Oxide Explorer}
\nomenclature{SARR}{Space absolute radiometric reference scale}
\nomenclature{EVE}{Extreme ultraviolet Variability Experiment}
\nomenclature{SIRS}{Solar irradiance reference spectra} 
\nomenclature{ATLAS}{Atmospheric Laboratory for Applications and Science, 
on the \emph{Space Shuttle}} 
\nomenclature{EUNIS}{Extreme-Ultraviolet Normal Incidence Spectrograph}
\nomenclature{MEGS}{Multiple EUV Grating Spectrograph}
\nomenclature{ALS}{Advanced Light Source} 